\documentclass[british,english,10pt]{article}

\usepackage[a4paper,top=2cm,bottom=2cm,left=3cm,right=3cm]{geometry} 

\usepackage{amsmath}
\usepackage{amssymb}
\usepackage{graphicx}

\usepackage[usenames]{color}

\def\ba{\begin{eqnarray}}
\def\ea{\end{eqnarray}}
\def\be{\begin{equation}}
\def\ee{\end{equation}}
\def\nn{\nonumber}
\def\bc{\begin{center}}
\def\ec{\end{center}}
\def\bmp{\begin{minipage}}
\def\emp{\end{minipage}}

\def\calL{\ensuremath{\mathcal{L}}}
\def\Ncq{\ensuremath{N_{35}}\ }
\def\Nqc{\ensuremath{N_{23}}\ }
\def\bfv{\ensuremath{\mathbf{v}}}
\def\e{\ensuremath{\mathrm{e}}}
\def\i{\ensuremath{\mathrm{i}}}
\def\I{\ensuremath{\mathcal{I}}}
\def\R{\ensuremath{\mathbb{R}}}
\def\erf{\ensuremath{\mathrm{erf}}}
\def\epsilon{\ensuremath{\varepsilon}}
\def\Im{\ensuremath{\mathrm{Im}}}
\def\v{\vspace{0.5cm}}

\numberwithin{equation}{section}

\usepackage{babel}
\begin{document}

\centerline{\Large The Swift--Hohenberg equation with a nonlocal nonlinearity}

\v\v

\centerline{\large David Morgan and Jonathan H. P. Dawes}

\v

\centerline{\it Department of Mathematical Sciences, University of Bath,}
\centerline{\it Claverton Down, Bath BA2 7AY, UK}

\v\v

\centerline{\today}

\begin{abstract}
It is well known that aspects of the formation of localised states in a
one-dimensional
Swift--Hohenberg equation can be described by Ginzburg--Landau-type
envelope equations. This paper extends these multiple scales analyses
to cases where an additional nonlinear integral term, in the form of a
convolution, is present. The
presence of a kernel function introduces a new lengthscale into the
problem, and this results in additional complexity in both the
derivation of envelope equations and in the bifurcation
structure.

When the kernel is short-range, weakly nonlinear
analysis results in envelope equations of standard type but whose
coefficients are modified in complicated ways by the nonlinear
nonlocal term. Nevertheless, these computations can be 
formulated quite generally in terms of properties of the Fourier transform of
the kernel function.
When the lengthscale associated with the kernel is longer,
our method leads naturally to the derivation of two different, novel,
envelope equations that describe aspects of the dynamics in these
new regimes. The first of these contains additional bifurcations,
and unexpected
loops in the bifurcation diagram. The second of these captures the
stretched-out nature of the homoclinic snaking curves that
arises due to the nonlocal term.
\end{abstract}


\section{Introduction}

Motivated by fluid mechanics, reaction-diffusion chemistry, and biological
systems, pattern forming nonequilibrium systems continue
to attract significant research interest \cite{Hoyle06,Cross09}.
They form a broad class
of dissipative continuum nonlinear systems that describe important
processes in nature. It is well known that pattern forming, or Turing,
instabilities can give rise to solution branches corresponding to
regular spatial structures \cite{Dawes10},
which typically emerge from a bifurcation point as a system parameter
is varied. In some cases one observes localised patches of pattern
rather than regular structure that fills the entire domain. These
are often referred to as \emph{localised states}; the building blocks
for such solutions can be considered to be isolated regions in which
a front exists between a number of periods of the underlying pattern
and a homogeneous background state, as illustrated in Figure 1.

Localised states form near subcritical Turing instabilities, in which
case two features of the system combine to result in the existence
of stable localised states. Firstly, the system is \emph{bistable},
meaning that both the homogeneous and patterned states are linearly stable
over an open interval of parameter values. Secondly, a \emph{pinning}
mechanism exists so that the configuration of fronts between pattern
and homogeneous states is stable.

Physical pattern-forming systems are most commonly modelled through
systems of parabolic partial differential equations that describe
local interactions: the time evolution of the fields at a point $x$
in the domain depends only on the values
of the variables and their derivatives at $x$ itself. However, there
are a number of situations in which models containing nonlocal terms
emerge as natural descriptions of the dynamics; for example Firth et al.
\cite{Firth07}
propose a nonlocal model for a nonlinear optical system, Purwins et al.
\cite{Purwins10}
present a nonlocal model for dielectric gas discharge dynamics, and
Plaut and Busse \cite{PB02} derive a nonlocal Complex Ginzburg--Landau
equation for the dynamics of thermal convection in a rotating
annulus, in a particular parameter regime of low Prandtl numbers.
The substantial mathematical biology literature on neural field models
contains many examples of nonlocal model equations \cite{HA05,VCM07}.
The paper by Coombes, Lord and Owen \cite{CLO03} includes numerical
evidence for homoclinic snaking in a nonlocal neural field model, although
the nonlocality is of a different kind to that considered in this paper.

In this paper we extend the well-known multiple scales asymptotic
treatment of the standard 1D Swift--Hohenberg equation
\be
\partial_{t}u=\left[r-(1+\partial_{x}^{2})^{2}\right]u+N(u), \label{eq:she}
\ee
for a scalar variable $u(x,t)$, where $N(u)$ denotes nonlinear terms
in $u$ and $r$ is a real parameter, to cases where the right-hand side
of~(\ref{eq:she}) is augmented by a nonlinear nonlocal term
of the form $u (K \ast u^2)$, i.e. $u(x,t)$ multiplied by a convolution of
$u^2$ with a kernel $K(x)$. Such a form includes the example discussed
by Firth et al. \cite{Firth07}.
Importantly, this choice of nonlocal
term also maintains the variational structure of the problem. The variational
nature of the standard Swift--Hohenberg equation has been exploited
in much recent work since the variational character guarantees, for 
example, that there are no oscillatory instabilities. For similar
reasons of simplification we choose to work with a variational nonlocal term.
It is then apparent that (as we show in section~\ref{sec:nonlocal}),
proposing a nonlocal term in the free energy
function that depends only on
$u^2$, so that the sign of $u$ is immaterial, and which preserves
the linear part of the Swift--Hohenberg equation, we are led to the (still
rather general) form that we consider here.

We find that consideration of even the `weakly nonlocal' case,
for which the kernel of the nonlocal integral term decays
extremely rapidly in space, introduces considerable
complexity to the multiple scales analysis. Further complexity arises in
the bifurcation structure of the localised states as the width of the
kernel increases.

Given the level of complexity we uncover, it is clear that many aspects
of this problem deserve a fuller treatment than we are able to give here.
In particular, details of the
snaking bifurcation structure are left to be the subject of future work.
In this paper our focus is on the extension of the multiple scales analysis and the existence of three asymptotic regimes for amplitude equations to operate in.

The paper is organised as follows. In
section~\ref{sec:local} we summarise briefly the relevant aspects of
the behaviour of the local Swift--Hohenberg equation and the
formation of localised states.
Section~\ref{sec:nonlocal} introduces the nonlocal term into the Swift--Hohenberg
model and makes general remarks on the modified equation.
Sections~\ref{sec:short3} and \ref{sec:short5} present the main results of the paper:
these extend the standard asymptotic analyses to the weakly nonlocal
case in which the kernel function decays on the lengthscale
that is asymptotically short in the multiple-scales setup. In
section~\ref{sec:longrange} we briefly comment on the two
other natural distinguished asymptotic limits of the problem, in
which the characteristic lengthscale of the kernel is considered to be
longer: these result in the derivation of 
different Ginzburg--Landau-like equations.
Section~\ref{sec:discussion} concludes.

\section{Localised states in the 1D Swift--Hohenberg equation}
\label{sec:local}

The standard Swift--Hohenberg equation (SHE) given in~(\ref{eq:she})
is a one dimensional PDE for the
scalar field $u\left(x,t\right)$ posed on the domain
$x\in\Omega\subseteq\mathbb{R}$.
It has often been viewed as a near-threshold approximation for
Rayleigh-B\'enard
convection, although it is frequently used in its own right as a canonical
model equation for pattern formation, and most recently, as a canonical
example of a scalar PDE that has localised solutions and
homoclinic snaking \cite{Avitabile10}. Given suitable nonlinear terms
$N(u)$, the base state $u\left(x,t\right)\equiv0$
undergoes a pattern-forming instability as $r$ passes through zero.
Typically, the nonlinear
term incorporates a second parameter whose value determines the
criticality of the bifurcation at $r=0$.
In the remainder of this section we
review the typical choices of $N(u)$ and the three types of solution
to~(\ref{eq:she}) prior to discussing the nonlocal form.

The two commonly used variants of the SHE are known
as the quadratic-cubic and cubic-quintic cases due to the
nonlinear terms they include. We label the two corresponding choices of
$N(u)$ by
\begin{equation}
N_{23}(u) = bu^2 -u^3, \qquad \mathrm{and} \qquad N_{35}(u)=su^3 -u^5.
\label{eq:Swift Hohenberg Nonlinear Terms}
\end{equation}

The \Ncq version is a less generic choice
because it is symmetric under sign changes in $u$.
However, the presence of this symmetry makes sense in some physical
situations, for example thermal convection in a Boussinesq fluid
with identical upper and
lower boundary conditions. Certainly, algebraic computations are
often much more straightforward with $N_{35}$ compared to $N_{23}$.
In both the $N_{23}$ and $N_{35}$ cases,
the term with the smaller exponent controls
the extent to which the system exhibits subcritical behaviour at
small amplitude, while the term with the larger exponent re-stabilises
solutions at large amplitude.

Analysis shows that for $b^2>\frac{27}{38}$, and for $s>0$, the Turing instability at
$r=0$ is subcritical; if the reverse inequalities hold then it is supercritical.
Both choices of $N(u)$ lead to an equation that is invariant under the
spatial reflection $\left(x,u\right)\rightarrow\left(-x,u\right)$.
Moreover, changing the sign of $b$ in the \Nqc equation is equivalent
to changing the sign of $u(x,t)$. We therefore consider only the cases $b>0$
and $s>0$ in what follows.

The solution $u(x,t)\equiv 0$ is linearly stable in $r<0$ and
undergoes a pattern-forming instability at $r=0$
from which emerges a branch of stationary spatially-periodic solutions. For
small $r$ these solutions have wavenumber $k$ close to unity.
Other spatially constant branches bifurcate from $u=0$ at
$r=1$; since these branches lie well away from the initial instability at $r=0$
their behaviour will not be considered further.

\begin{figure}[!ht]
\bc
\includegraphics[width=14.0cm]{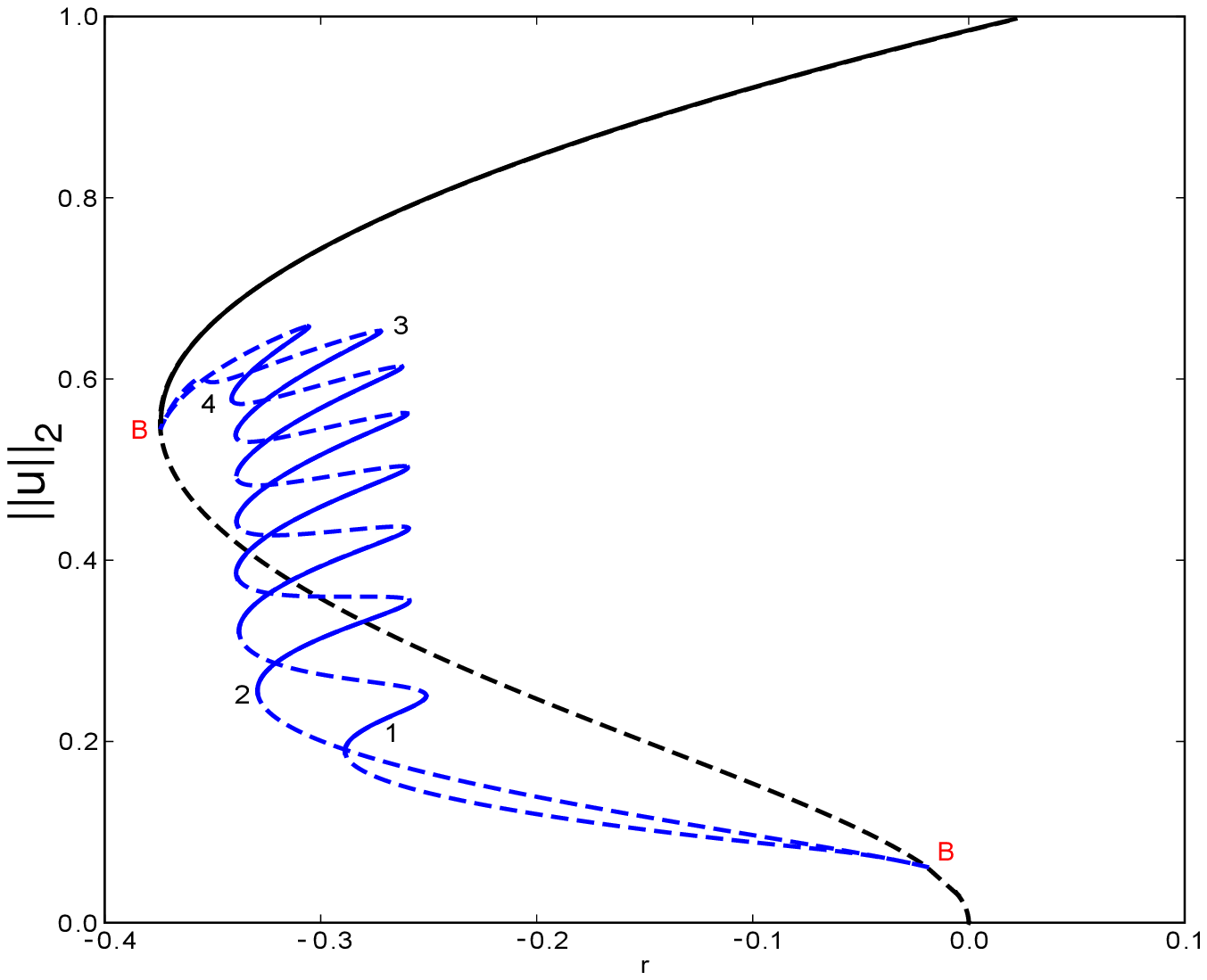}
\includegraphics[width=6.5cm]{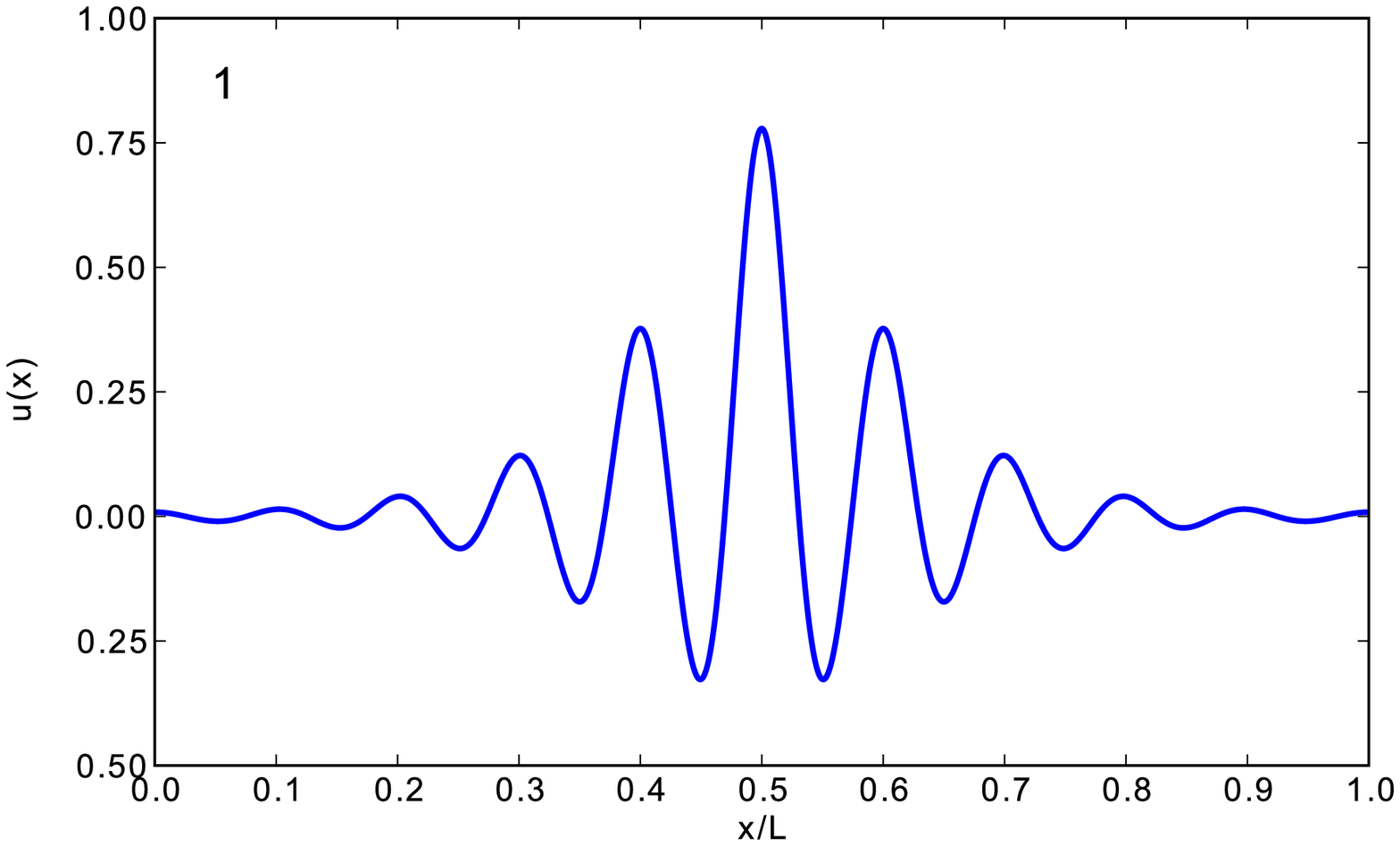}
\includegraphics[width=6.5cm]{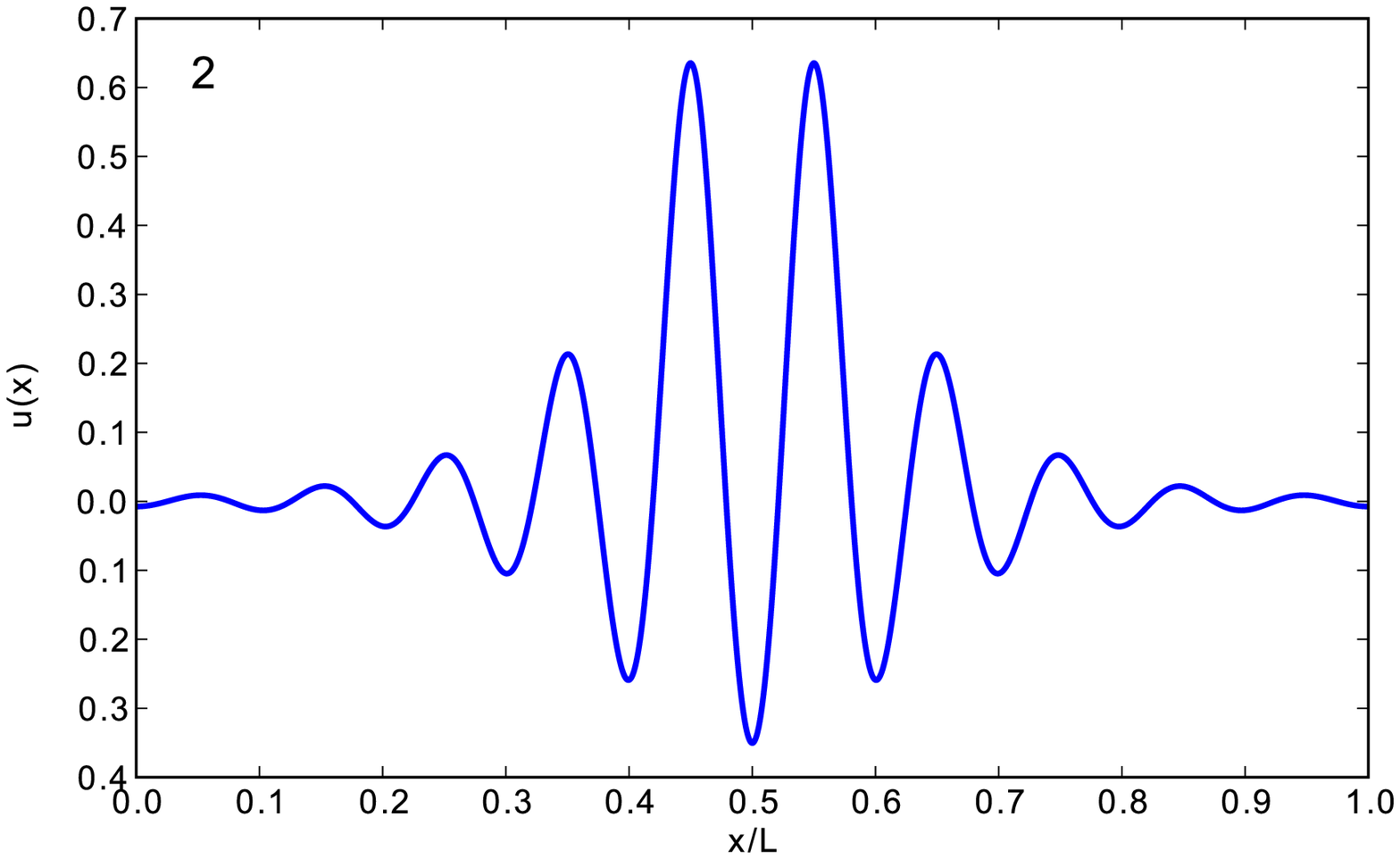}
\includegraphics[width=6.5cm]{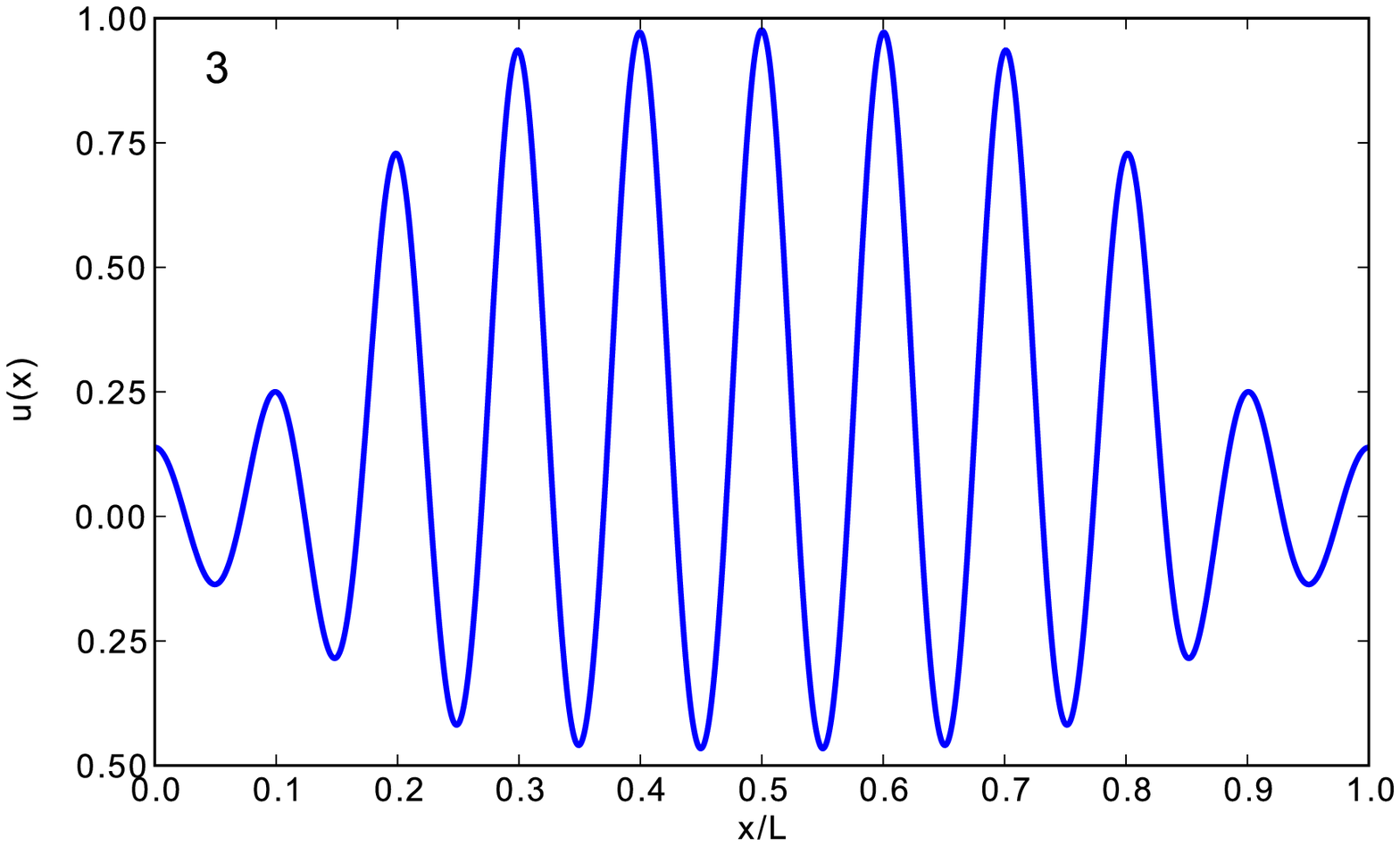}
\includegraphics[width=6.5cm]{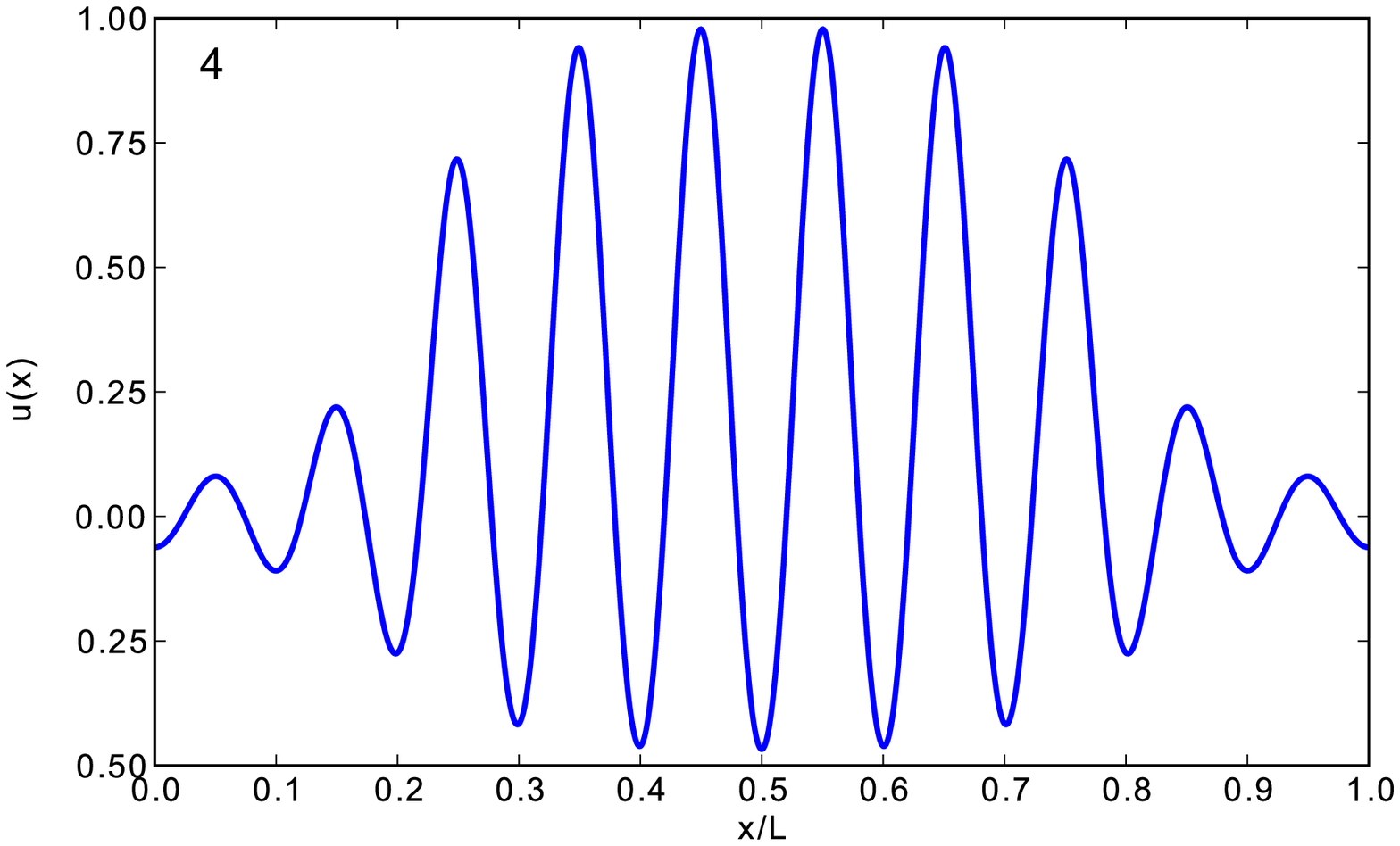}
\ec
\caption{\label{fig:snaking}\small Homoclinic snaking for the SHE.
Upper panel:
bifurcation diagram showing branches of periodic
patterns (black) and localised states (blue) in the $(r,||u||_2)$-plane for
the \Nqc SHE, for $b=1.8$. Solid (dashed) lines
indicate stable (unstable)
solutions. Points marked `B' are the bifurcations at which the localised
states emerge from the periodic states and are a consequence of the finite
computational domain.
Lower panels: illustrative solution
profiles $u(x,t)$ at the four numbered locations on the bifurcation diagram.
The domain length used is $0 \leq x \leq 20\pi$, with periodic boundary conditions.
This figure, and several others, was computed using the continuation software
AUTO \cite{AUTO}.}
\end{figure}

When the initial bifurcation at $r=0$ is subcritical, the patterned
branch undergoes a saddle-node bifurcation at finite amplitude
in $r<0$ in both versions of the SHE, illustrated for the \Nqc case in
figure~\ref{fig:snaking}. In a finite domain,
additional bifurcation points (labelled `B' in the figure) arise
along the branch. In a finite domain these bifurcation points are connected
by intertwining branches of modulated solutions
that become increasingly localised as the
domain size increases. The formation of these intertwining curves is therefore
often referred to as `homoclinic snaking',
an (at first sight) unusual
structure that has a rich theoretical background as discussed
by many authors \cite{Champneys98,CRT00,Burke06}. 
Figure~\ref{fig:snaking} also shows typical solution profiles at the
top and bottom of each snaking curve. The snaking curves
characterise the set of stationary solutions where $u\left(x\right)$
describes an orbit that is spatially homoclinic to the state $u(x)\equiv 0$
after exactly one excursion near the periodic pattern.
The oscillations in each smooth curve are
connected through a set of saddle-node bifurcations
that `snake' backwards and forwards between limiting values of $r$.
For the parameter values of figure~\ref{fig:snaking} this is
approximately the interval $-0.34<r<-0.26$.

Within this snaking region lies a Maxwell point at which the periodic
state is energetically equal to the background state, and therefore one
would expect stationary fronts between them to be possible. Such a Maxwell
point can only be defined for systems that are variational in structure.
This is the case for the SHE, and the corresponding free energy quantity is
\ba
\mathcal{F}[u] = \int_\Omega \left( 
\frac{1}{2}u_{xx}^2 - u_x^2 + \frac{1-r}{2}u^2
-\int^u N(s) \ ds 
\right) \ dx , \nn
\ea
so that $u_t =-\delta \mathcal{F} / \delta u$ and so
\ba
\frac{d \mathcal{F}}{dt} = \frac{\partial u}{\partial t}
\frac{\delta \mathcal{F}}{\delta u} = -\left( u_t \right)^2 \leq 0, \nn
\ea
which guarantees convergence to equilibrium states, as long as $\mathcal{F}[u]$
is bounded from below.
As one moves up the snake, each additional crossing back-and-forth
on the bifurcation diagram is associated
with the creation of an additional pair of large amplitude peaks,
one at each edge of
the localised pattern; solutions at the top of the snake are wider than those
at the bottom and the number of turns on the snake is proportional to the size
of the domain. 
In addition to the curves shown in figure~\ref{fig:snaking} there
are stationary asymmetric states that exist on `rungs' that
link the two intertwining curves together. The rungs emerge in pitchfork
bifurcations near to 
the saddle-node bifurcation points on the snaking curves.
The rung states closely resemble the symmetric localised states on the
snaking curves but they have a different phase relation between the
small-scale periodic spatial oscillations and their overall envelope.
For the SHE they are always unstable. The detailed description of
the snaking curves and rung branches demands a close examination of
the role of the relative phase between the small-scale and the envelope
and this can only be done, in an asymptotic approach,
by considering terms that are formally exponentially small.
The form of these exponentially small terms has
been studied using both exponential asymptotics \cite{KC06,Chapman09,Dean11} and
the variational approximation method \cite{Susanto11,MS11,Dawes13}.

\subsection{Multiple scales asymptotics for the SHE}
\label{sec:multi}

In this section we review the results of the well-known
multiple scales computations for the \Nqc and \Ncq cases of the SHE.
To make progress we consider a perturbation expansion that considers
small amplitude solutions, introducing the parameter $\epsilon$ to describe
the solution amplitude:
\begin{equation}
u\left(x,t\right)=\varepsilon u_{1}+\varepsilon^{2}u_{2}
+\varepsilon^{3}u_{3}+\varepsilon^{4}u_{4}+\varepsilon^{5}u_{5}
+\cdots . \label{eq:u_ansatz}
\end{equation}
In addition, the scalar field $u(x,t)$ is modelled as
evolving on two different length
scales near the bifurcation at $r=0$: $u(x,t)$ comprises
a periodic function on the short lengthscale $x$, modulated
by an unknown envelope $A(X)$ that is a function of a long
lengthscale $X$. In general one would also assume
that the two components evolve
on different timescales: the underlying pattern varying on the short
timescale $t$, and the envelope evolving on a long timescale $T$. Since
the instability at $r=0$ is steady, the dependence on $t$ is in fact trivial.
We therefore look for leading-order solutions in the form
\begin{equation}
u_1\left(x,X,t,T\right) = A\left(X,T\right)\e^{ix}
+\bar{A}\left(X,T\right)\e^{-ix}, \label{eq:u1_ansatz}
\end{equation}
where we add the complex conjugate so that $u_1$ is real even though
$A(X,T)$ may be complex-valued.
Our last requirement is that the bifurcation parameter $r$ is small which
in turn demands that $r$ is rescaled: the rescaled parameter is denoted $\mu$.

We will consider two different sets of parameter scalings. For the
SHE the results of the multiple scales calculations are well-known and
we give only the results. These provide useful comparisons with the
results for the nonlocal SHE presented in section~\ref{sec:nonlocal}.
The first set of scalings leads to a solvability
condition being imposed at $O(\varepsilon^3)$ in the
multiple scales expansion, and results in an envelope equation that
can be solved analytically using Jacobi elliptic functions. The
second set of scalings does not require a solvability condition being imposed before
$O(\varepsilon^5)$. Although this analysis generates
a PDE that does not, in general, have closed form equilibrium solutions,
it does provide a leading order estimate of the location of the snaking
curves in the bifurcation diagram.

\subsection{Derivation of a Ginzburg--Landau equation at third order}
\label{sec:gl3}

The first set of scalings considered is:
\ba
X=\varepsilon x;      \qquad \qquad 
T=\varepsilon^{2}t;   \qquad \qquad 
r=\varepsilon^{2}\mu.  \label{eq:scaling3}
\ea
Proceeding with the standard multiple scales technique of
substituting~(\ref{eq:u_ansatz}) into~(\ref{eq:she}) and
solving for $u_n(x,X,t,T)$ at successive orders in $\epsilon$ 
we derive evolution equations for $A(X,T)$,
in both the \Nqc and \Ncq cases, through the solvability conditions
that arise at $O(\varepsilon^3)$. For the \Nqc case we obtain
\begin{equation}
A_{T}=\mu A+\frac{38}{9}\left(b^{2}-\frac{27}{38}\right)A\left|A\right|^{2} 
+ 4A_{XX},
\label{eq:gl3-23}
\end{equation}
and for the \Ncq case we obtain
\begin{equation}
A_{T}=\mu A + 3sA\left|A\right|^{2} + 4A_{XX}.
\label{eq:gl3-35}
\end{equation}
The behaviour and solutions of these cubic Ginzburg--Landau equations
are discussed in many places, for example \cite{Vega05} and \cite{Hoyle06}.
The cubic GL equations have explicit solutions in
terms of Jacobi elliptic functions.
Of more interest for our purposes is that the coefficient of the
nonlinear term vanishes in the cases
$b^2=\frac{27}{38}$ and $s=0$; at these parameter values the pattern-forming
instability switches between supercritical and subcritical. Examination of
the codimension-two points $(r,b)=(0,\sqrt{27/38})$ and $(r,s)=(0,0)$
allows us to extend the multiple scales analysis to include
the restabilisation of periodic patterns at larger amplitudes; this is 
considered in the next section.

\subsection{Derivation of a Ginzburg--Landau Equation at fifth order}
\label{sec:gl5}

To examine the codimension-two points at which the instability at $r=0$
switches from supercritical to subcritical we rescale the coefficient of the
nonlinear term with the lower exponent, writing
\begin{equation}
b=\sqrt{\frac{27}{38}}+\varepsilon^{2}b_{2}; \qquad \mathrm{or} \qquad 
s=\varepsilon^{2}s_{2},  \label{eq:b scaling (4.4)}
\end{equation}
in the \Nqc and \Ncq cases, respectively. To balance the linear terms
in~(\ref{eq:gl3-23}) and~(\ref{eq:gl3-35})
at a higher order in the expansion, we
introduce the alternative set of scalings
\begin{equation}
X=\varepsilon^{2}x; \qquad \qquad
T=\varepsilon^{4}t; \qquad \qquad
r=\varepsilon^{4}\mu.
\label{eq:scaling5}
\end{equation}
No secular terms are generated
before $O\left(\varepsilon^{5}\right)$. At $O(\epsilon^5)$ we
obtain cubic-quintic Ginzburg--Landau equations as follows.
For the \Nqc case we obtain
\begin{equation}
A_{T}=\mu A +\frac{2}{3}\sqrt{114}b_{2}A\left|A\right|^{2}
-\frac{8820}{361}A\left|A\right|^{4}+i\frac{16}{19}A_{X}\left|A\right|^{2}
+4A_{XX},
\label{eq:gl5-23}
\end{equation}
and for the \Ncq case we obtain
\begin{equation}
A_{T}=\mu A +3s_{2}A\left|A\right|^{2} -10A\left|A\right|^{4}
+4A_{XX}.
\label{eq:gl5-35}
\end{equation}
The linear terms are identical to those in~(\ref{eq:gl3-23})
and (\ref{eq:gl3-35}), but the different scalings bring additional nonlinear terms
into the asymptotic balance.

\begin{figure}[!ht]
\bc
\includegraphics[width=7.0cm]{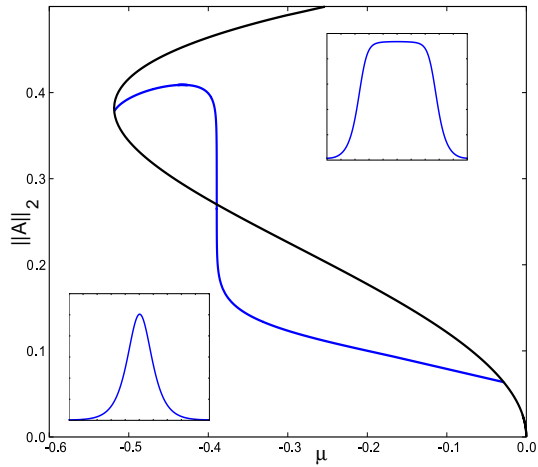}
\caption{\label{fig:gl5-23}\small Bifurcation diagram in the $(\mu,\| A\|_2)$ plane
for the cubic-quintic Ginzburg--Landau
equation~(\ref{eq:gl5-23}) for $b_2=1$ showing
the homogeneous (black) and modulated
(blue) branches of solutions. The inset figures indicate the general shape of the
modulus of the
amplitude $|A(X)|$ at two points low down ($\mu \approx -0.1$) and
higher up ($\mu \approx -0.4$) on the blue curve.
The domain used is $0 \leq X \leq 20\pi$, with periodic boundary conditions.}
\ec
\end{figure}

The cubic-quintic Ginzburg--Landau equations~(\ref{eq:gl5-23})
and~(\ref{eq:gl5-35}) derived at $O(\varepsilon^5)$
are able to capture the bistability through their more complicated
collection of competing nonlinear terms. This is illustrated in
figure~\ref{fig:gl5-23}.

Figure~\ref{fig:gl5-23} is clearly similar in structure to figure~\ref{fig:snaking}
in that a modulated branch (blue)
bifurcates from the primary solution branch (black) at two points, one
close to $\mu=0$ and one 
close to the saddle node bifurcation. The vertical section of the modulated
branch indicates the Maxwell point at which the envelope increases rapidly
over a very small range of $\mu$. We note that the cubic-quintic Ginzburg--Landau equation~(\ref{eq:gl5-23}) and its more general counterparts have been
recently investigated by Kao and Knobloch \cite{KK12} in some detail.

\section{Including a nonlocal term}
\label{sec:nonlocal}

Having summarised the usual asymptotic analysis of the local SHE, we now extend
it to the case in which an additional nonlinear nonlocal term exists.
Specifically we consider the nonlocal Swift--Hohenberg equation
\ba
\partial_{t}u=\left[r-(1+\partial_{x}^{2})^{2}\right]u+N(u)
-\gamma u(x,t) \int_{\Omega}K\left(x-y\right)u\left(y,t\right)^2 \, dy,
\label{eq:nlshe}
\ea
where $K(x)$ is a bounded function defined on $\Omega$ that either
has bounded support, or decays
sufficiently rapidly at large $|x|$,
so that its Fourier transform exists. These properties define what we
mean by `short range': we treat the form of $K(x)$ and any
parameters on which it depends as fixed while
working asymptotically in the limit $\epsilon \ll 1$.

\subsection{Properties of $K(x)$ and the free energy $\mathcal{F}[u]$}

We assume also that (i) $K$ is normalised
so that $\int_\Omega K(x) \ dx =1$; (ii) $K$ is even:
$K(x)=K(-x) \ \ \forall x \in \Omega$, and (iii) $K$ is non-negative:
$K(x) \geq 0 \ \ \forall x \in \Omega$.
The parameter $\gamma$ is the coefficient of the nonlocal term.
The form of the nonlinear nonlocal term proposed in~(\ref{eq:nlshe}) is
motivated naturally from consideration of the free energy
functional
\ba
\mathcal{F}[u] = \int_\Omega \left( 
\frac{1}{2}u_{xx}^2 - u_x^2 + \frac{1-r}{2}u^2
-\int^u N(v) \ dv 
\right) \, dx + \frac{\gamma}{4} \int_\Omega \int_\Omega K(x-y) u(x)^2 u(y)^2 
\, dx \ dy. \nn
\ea
In order to check that the time evolution $u_t = -\delta \mathcal{F} / \delta u$
is well-posed, we need to ensure that $\mathcal{F}$ is bounded below.
In the cubic-quintic case $N(u)=su^3-u^5$ this is true for all real
values of $\gamma$. In the quadratic-cubic case $N(u)=b u^2-u^3$ we require
$\gamma>-1$ for $\mathcal{F}$ to be bounded below.
These conclusions follow from the following estimate on the size of the
nonlocal term in $\mathcal{F}$, deduced by combining the
Cauchy--Schwarz inequality and Young's inequality for convolutions.
Details of these inequalities can be found, for example, in the
textbook by Lieb and Loss \cite{LL10}.
Since the nonlocal term is non-negative we have
\ba
\int_\Omega \int_\Omega K(x-y) u(x)^2 u(y)^2 
\, dx \ dy
& \equiv & \| (K \ast u^2) u^2 \|_1 \nn \\
& \leq   & \| K \ast u^2\|_2  \  \| u^2 \|_2 \nn \\
& \leq & \left\| K \right\|_1  \  \| u^2 \|_2  \  \| u^2 \|_2 \nn \\
& \leq & \int_\Omega u(x)^4 \, dx, \label{eq:estimate}
\ea
where we adopt the usual notation $\| f\|_p := \left(\int_\Omega |f|^p \, dx
\right)^{1/p}$ for the $p$-norm of a function $f(x)$, and
we have used Cauchy--Schwarz to derive the second line,
Young's inequality for the third line, and the fact that $\| K \|_1=1$
(normalisation) for the fourth line.
As a result we can see that in $\mathcal{F}[u]$ in the cubic-quintic
case the nonlocal term is always
dominated by the $\frac{1}{6}u^6$ term in $\mathcal{F}[u]$.
In the quadratic-cubic case with $\gamma>0$
we see that the nonlocal term makes
a non-negative contribution to $\mathcal{F}[u]$.
In the remaining case, of quadratic-cubic terms with $\gamma<0$,
we can estimate that
\ba
\mathcal{F}[u] \geq \int_\Omega
\frac{1}{2}(u_{xx}+u)^2 - \frac{r}{2}u^2
-\frac{b}{3} u^3 + \frac{1}{4}(1-|\gamma|) u^4 \, dx \nn
\ea
which follows after integrating the term $-u_x^2$ by parts and using
the estimate~(\ref{eq:estimate}). Hence
\ba
\mathcal{F}[u] \geq \int_\Omega - \frac{b}{3} u^3 + \frac{1}{4}(1-|\gamma|) u^4 \,
dx, \nn
\ea
which is bounded below, per unit length of the domain $\Omega$, by $-b^4/(12 (1-|\gamma|)^3)$.
The bound $\gamma>-1$ is strict since it is attained by the choice of kernel $K(x)=\delta(x)$ the `Dirac delta' distribution.

\subsection{Limiting choices for $K(x)$}

There are two limiting choices of $K(x)$ for which the behaviour of solutions
to~(\ref{eq:nlshe}) can be determined by inspection.
The first is the case where $K(x) \equiv \frac{1}{|\Omega |}$,
referred to as `global coupling' in \cite{Firth07}. In this case the
bifurcation parameter $r$ is effectively replaced by the new parameter 
$r^{\prime}=r-\gamma\left\langle u^{2}\right\rangle $ where
the angled brackets indicate the domain-averaged integral:
$\langle u^2\rangle:=\frac{1}{|\Omega |}\int_{\Omega} u(x,t)^2 dx
\equiv \frac{1}{|\Omega |}\| u \|_2^2$.
As noted in \cite{Firth07}, in this case there is a correspondence between
solutions $u_{local}$ and $u_{nonlocal}$ of the local and nonlocal problems, 
respectively, given by changing the
parameter value:
\begin{equation}
u_{nonlocal}\left(x,t;r^{\prime},\gamma\right) = 
u_{local}\left(x,t;r-\gamma\left\langle u^{2}\right\rangle \right). \label{eq:shift}
\end{equation}
The second limiting case is where $K(x)=\delta\left(x \right)$, a `Dirac
delta' function. This case amounts to only a change in the nonlinearity
from $N(u)$ to $N(u)+\gamma u^3$, i.e. in the \Nqc and \Ncq cases
the coefficient of the cubic term is increased by $\gamma$.

\begin{figure}[!ht]
\bc
\includegraphics[width=12.0cm]{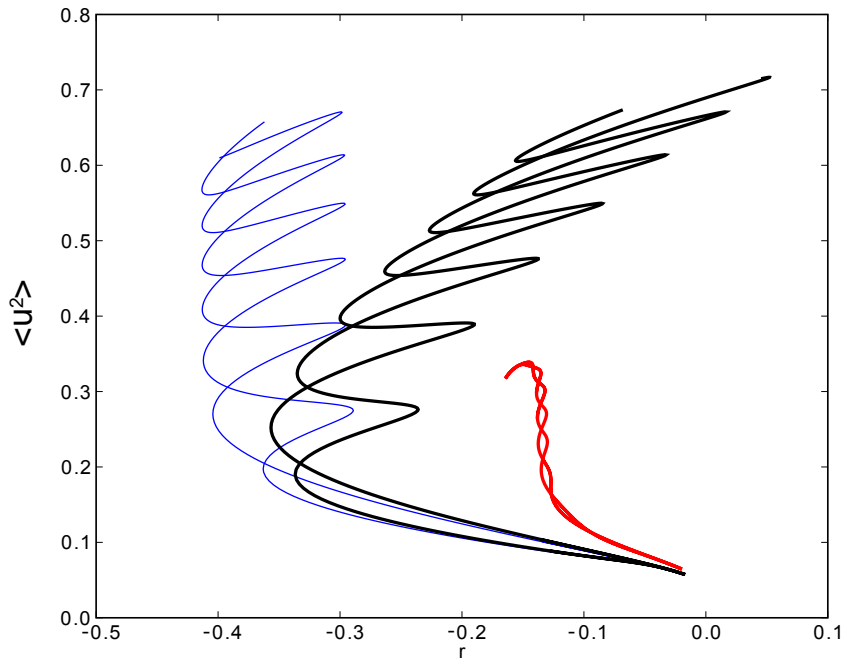}
\caption{\label{fig:extreme}\small Homoclinic snaking in local and global
cases of the (2--3) Swift--Hohenberg equation~(\ref{eq:nlshe}).
Homoclinic snaking curves in the $(r,\| u \|_2)$ plane
are shown for the local case (blue, thin line), and for the nonlocal case~(\ref{eq:nlshe})
in the limiting cases $K(x)=\delta\left(x\right)$ (red, right-hand curves),
and $K=\frac{1}{L}$
(black, sloping curves). Parameter values are $\gamma=0.7$,
$b=1$.
A domain size $L=20\pi$ and periodic boundary conditions were used. Stability
is not indicated.
}
\ec
\end{figure}

Figure~\ref{fig:extreme} illustrates the snaking curves for the \Nqc case, for
fixed values of the parameters $b$ and $\gamma$, varying the choice of kernel function. It confirms the observations made in the paragraph above: the global coupling
case results in a slanted version of the snake from the local problem: they
can be mapped onto each other via the relation~(\ref{eq:shift}). The
choice of the Dirac delta function for the kernel produces the smaller snake
at $r$ closer to zero: it is vertical rather than slanted since effectively
only the coefficient of $u^3$ has been altered.
Since for a general kernel $K(x)$, the nonlocal problem cannot
be related directly to a simple transformation of the local problem, and
detailed analysis is necessary, the remainder of this paper can be
thought of as understanding how the snake shifts between these two
limiting cases.

\subsection{Homoclinic snaking in the nonlocal SHE}

\begin{figure}[!h]
\bc
\includegraphics[width=7.4cm]{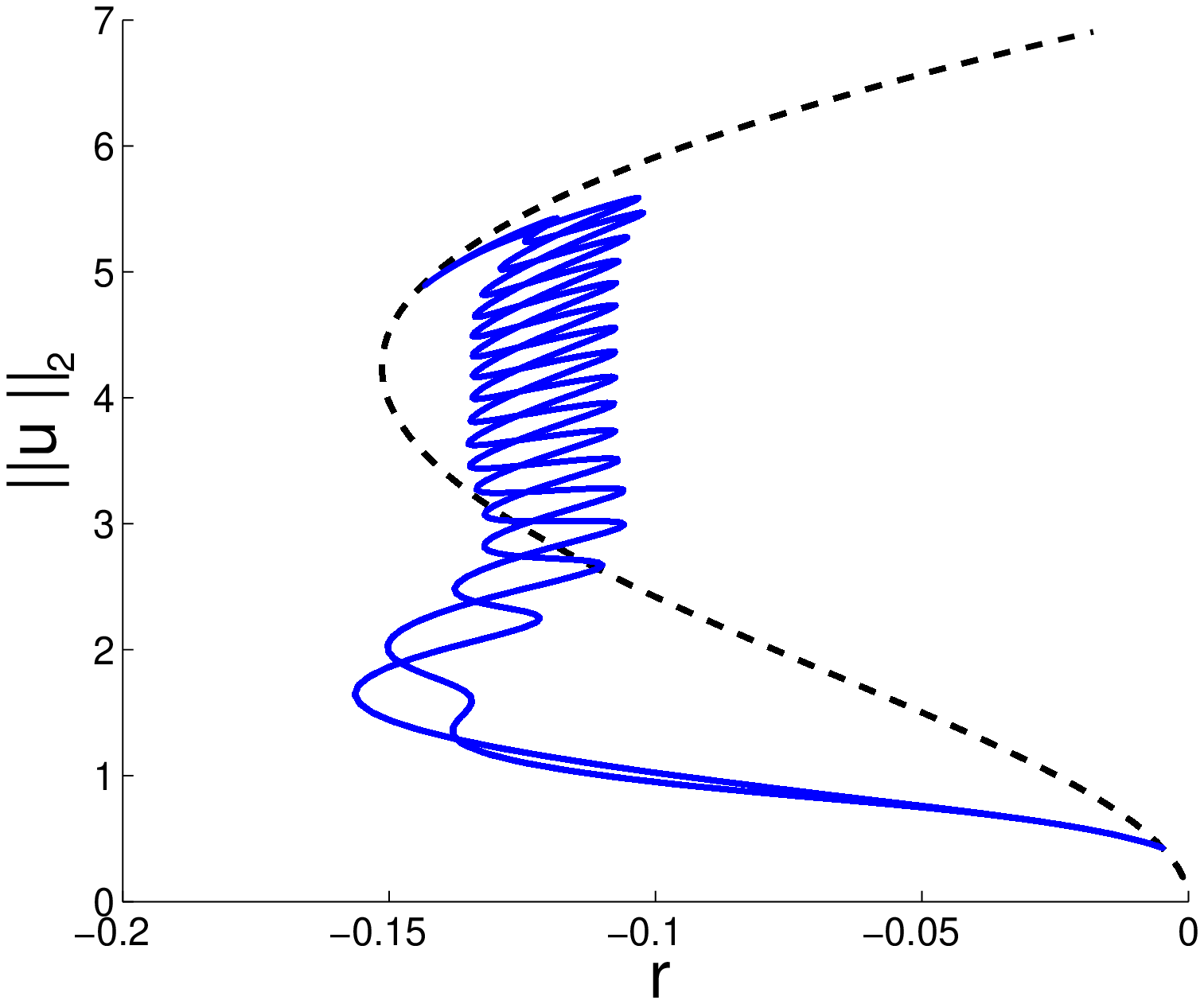}
\includegraphics[width=7.4cm]{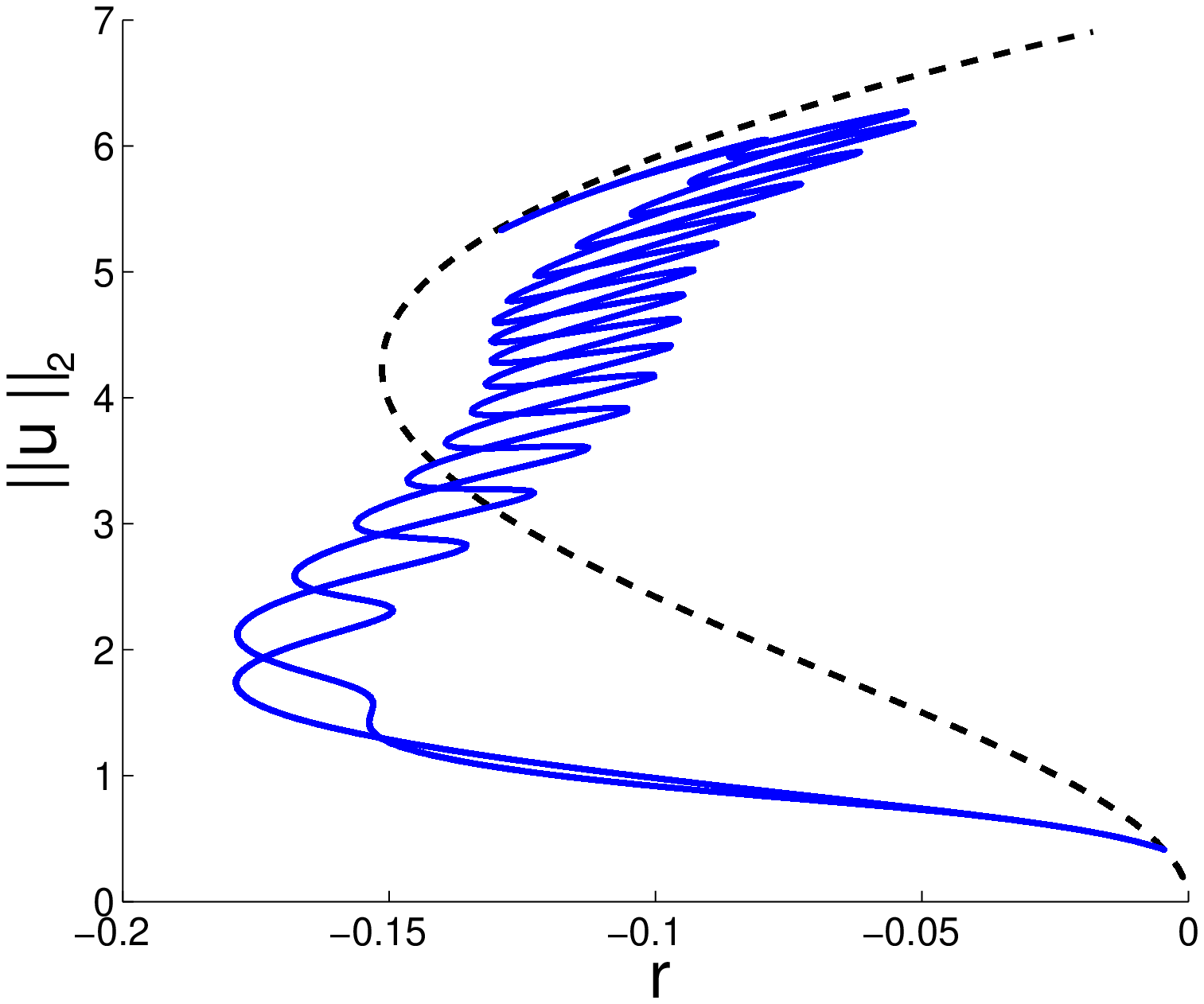}
\caption{\label{fig:nonlocal001}\small Homoclinic snaking in the nonlocal
\Nqc Swift--Hohenberg equation~(\ref{eq:nlshe}) with
the Gaussian kernel $K_G(x)$ defined in~(\ref{eq:g_kernel}).
Homoclinic snaking curves on which the localised states lie are
shown as solid blue lines. Dashed black lines indicate the location of spatially periodic solutions. (a) $\sigma=4\pi$; (b) $\sigma=10\pi$.
Other parameter values are $\gamma=0.5$, $b=1.6$, domain size $L=40\pi$.
Periodic boundary conditions were used, and stability is not indicated.
}
\ec
\end{figure}

When the kernel $K(x)$ is assumed
either to decay very rapidly, or to decay very slowly compared to the
(finite) size of the domain $\Omega$,
we might propose that the effect of the nonlocal term is
only to shift the snaking curves in $r$ a little from the purely
local or purely global problem, respectively.
Figure~\ref{fig:nonlocal001} presents bifurcation diagrams for the snaking curves
in each of these cases, using the Gaussian kernel
\ba
K_{G}\left(x\right) & = & \frac{1}{\sqrt{2\pi\sigma^{2}}}
\e^{-x^2/(2\sigma^2)}. \label{eq:g_kernel}
\ea
In figure~\ref{fig:nonlocal001}(a) the width of
the Gaussian kernel is $\sigma=4\pi$ which is
small compared to the domain size $L=40\pi$. The first few
turns on the snake are shifted to lower $r$ and the last few slightly shifted to
higher $r$, while the centre of the snake remains close to vertical.
Figure~\ref{fig:nonlocal001}(b) contains the corresponding snaking bifurcation diagram for $\sigma=10\pi$ in the same size domain $L=40\pi$. The homoclinic snaking is stretched out in a manner similar to slanted snaking, but with an overall `S' shape that the purely global term cannot generate.

\begin{figure}[!ht]
\bc
\includegraphics[width=7.25cm,angle=0]{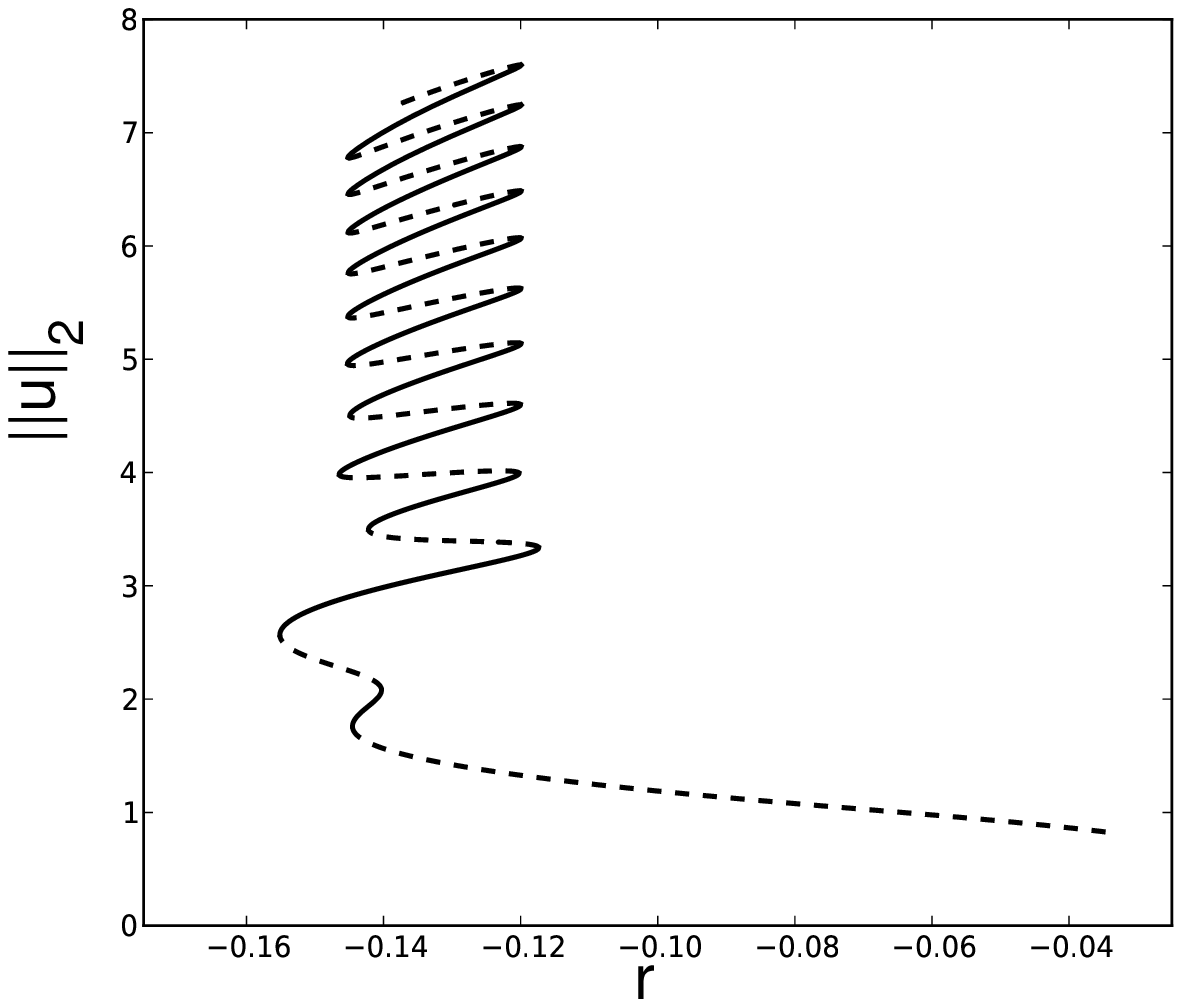}
\includegraphics[width=7.25cm]{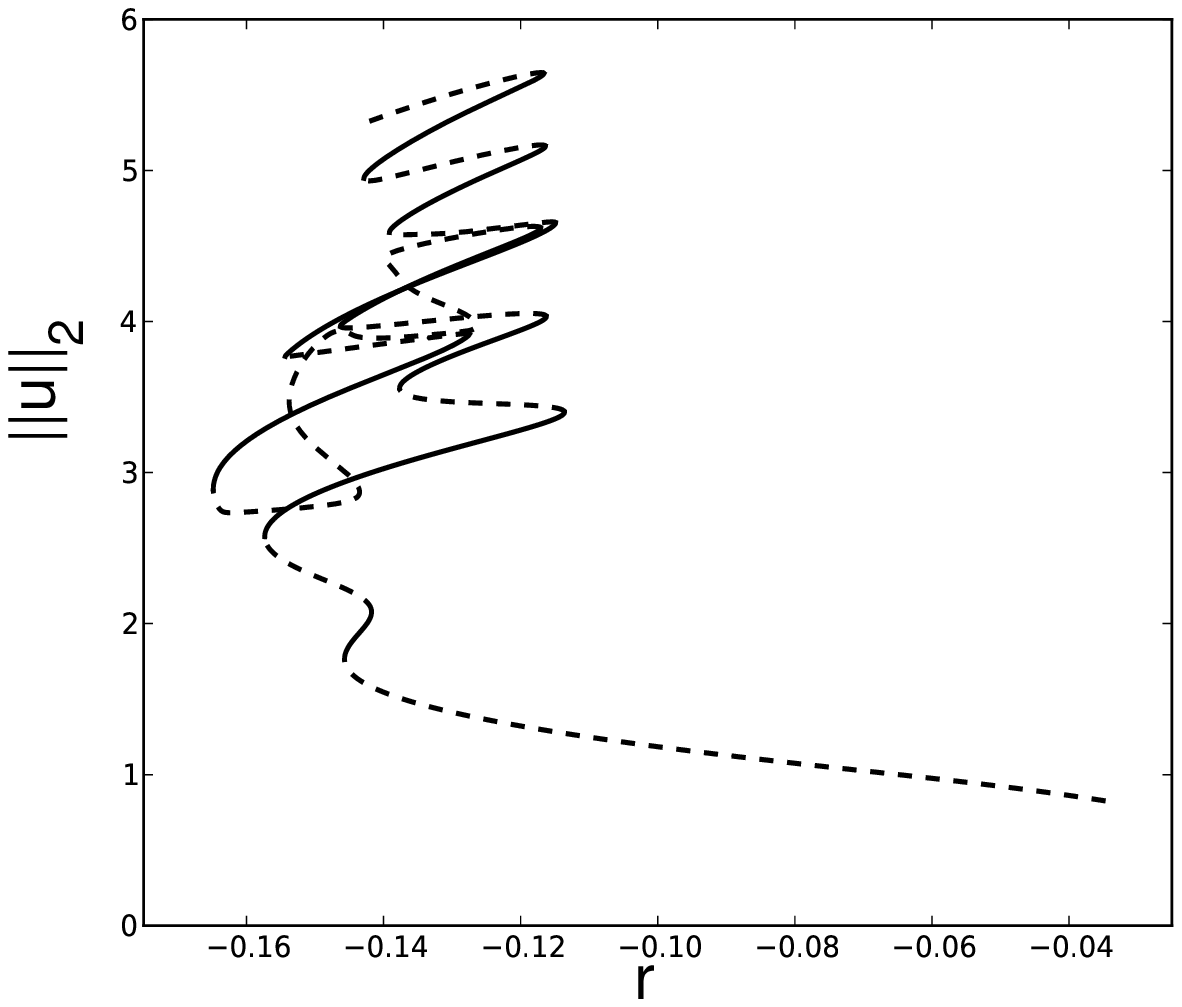}
\caption{\label{fig:th_snake001}\small Bifurcation diagrams in the
$(r,\| u \|_2)$ plane illustrating homoclinic snaking in the nonlocal
\Nqc Swift--Hohenberg equation~(\ref{eq:nlshe}) with
the top hat kernel $K_{TH}(x)$ defined in~(\ref{eq:th_kernel}).
(a) $\delta=36$; (b) $\delta=38$.
Other parameter values are $\gamma=0.5$, $b=1.6$, domain size $L=40\pi$.
Periodic boundary conditions were used, and solid (dashed) lines indicate
stable (unstable) solutions.
}
\ec
\end{figure}

\begin{figure}[!ht]
\bc
\bmp{9.0cm}
\includegraphics[width=9.5cm]{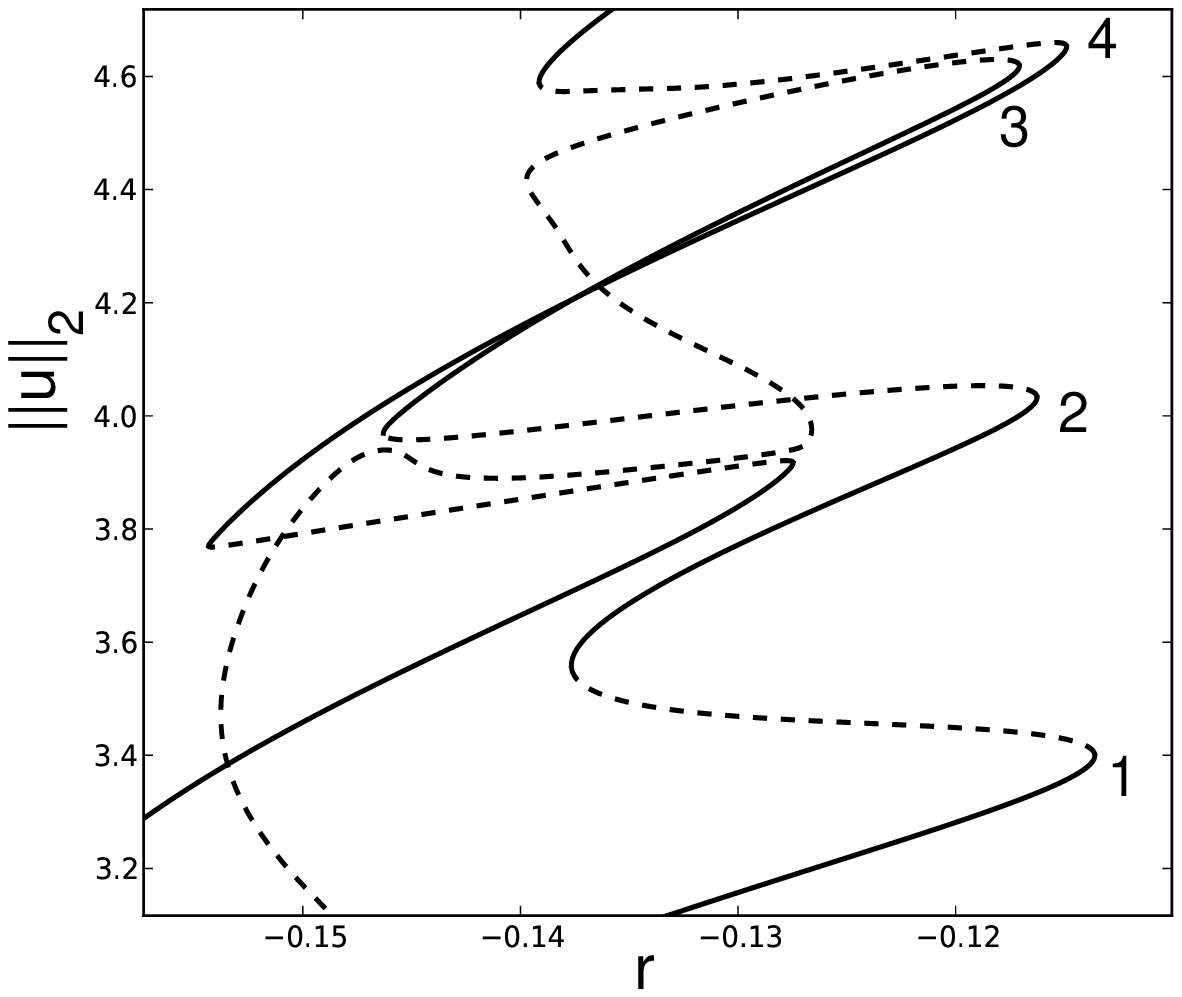}
\emp
\bmp{7.0cm}
\includegraphics[width=6.25cm,height=1.7cm,angle=0]{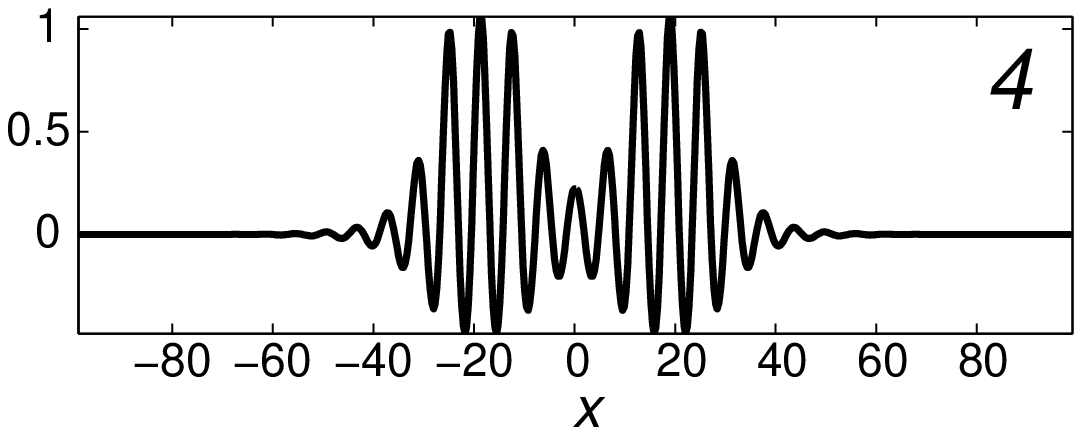}
\includegraphics[width=6.25cm,height=1.7cm,angle=0]{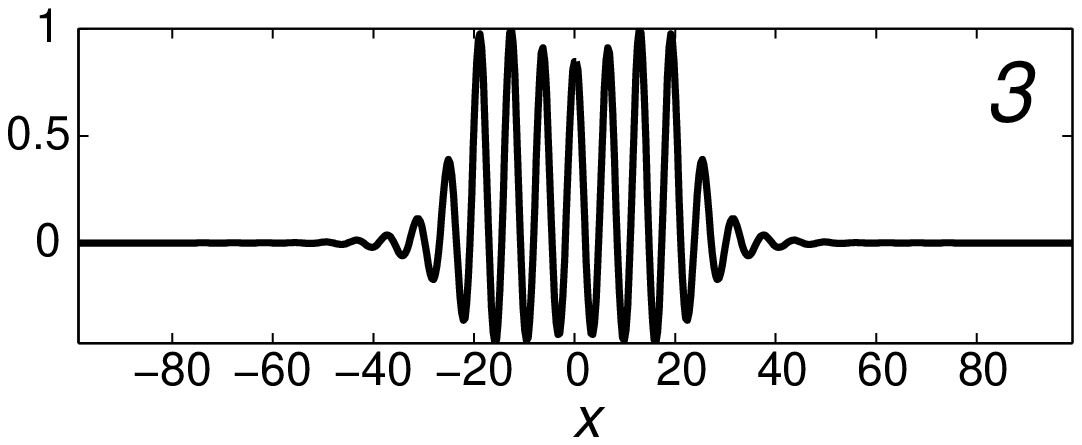}
\includegraphics[width=6.25cm,height=1.7cm,angle=0]{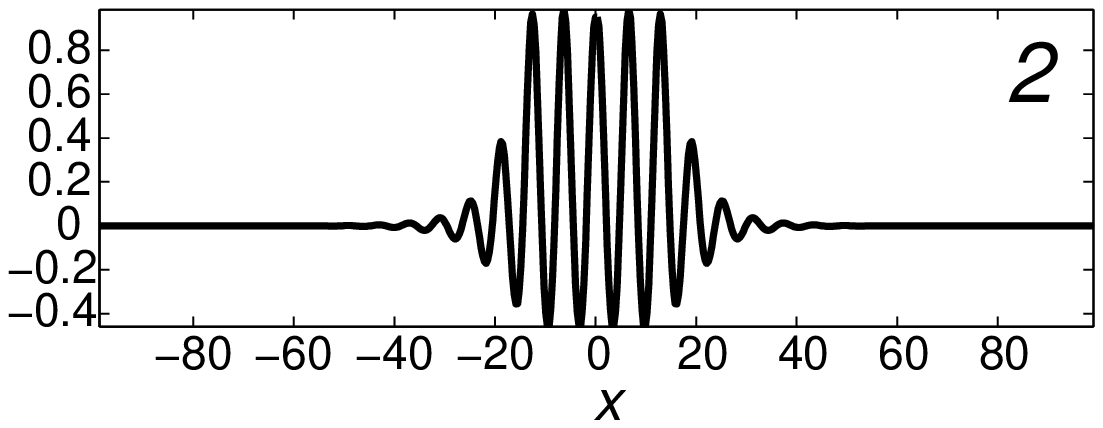}
\includegraphics[width=6.25cm,height=1.7cm,angle=0]{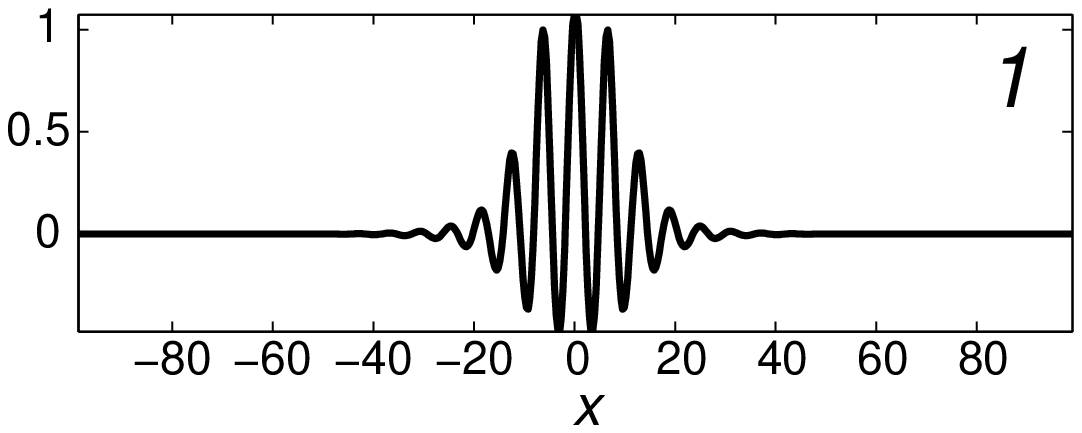}
\emp
\caption{\label{fig:th_snake002}\small Homoclinic snaking in nonlocal versions
of the \Nqc Swift--Hohenberg equation~(\ref{eq:nlshe}) with
a top hat kernel $K_{TH}(x)$.
(a) Enlargement of figure~\ref{fig:th_snake001}(b); (b) solution profiles
at the four points labelled $1$--$4$ in (a).
Other parameter values are $\gamma=0.5$, $b=1.6$, domain size $L=40\pi$.
Periodic boundary conditions were used, and solid (dashed) lines indicate
stable (unstable) solutions.
}
\ec
\end{figure}

It is surprising that the transition between figure~\ref{fig:nonlocal001}(a)
and figure~\ref{fig:nonlocal001}(b) as the width parameter $\sigma$ increases
does not occur in the obvious fashion that one might expect, with the snaking
curves deforming smoothly from one figure to the other. Instead, it appears to
be a complicated process in which the snaking curves collide with other
solution branches, consisting of multipulse states, and re-connect before
pinching off again as $\sigma$ increases further. To illustrate
this complex process, figures~\ref{fig:th_snake001} and~\ref{fig:th_snake002}
present some of the details of the initial changes in bifurcation structure as
the kernel width increases, in this case for the top hat kernel
\ba
K_{TH}\left(x\right) & = & \frac{1}{\delta}\left[H\left(x+\frac{\delta}{2}\right)
-H\left(x-\frac{\delta}{2}\right)\right], \label{eq:th_kernel}
\ea
where $H(x)$ denotes the Heaviside function: $H(x)=0$ if $x \leq 0$ and
$H(x)=1$ if $x>0$.
Figure~\ref{fig:th_snake001}(a) shows the lower end of
one of the snaking branches at $\delta=36$; this value of $\delta$ should be
considered narrow compared to the overall domain width $L=40\pi$. Figure~\ref{fig:th_snake001}(b) is computed in the same way for the slightly increased
value $\delta=38$. The snaking curve has collided with two isolas, leading
to a number of additional loops over which the solution amplitude, as measured
by the $L_2$ norm, decreases. Such loops were referred to as `switchbacks'
by Taylor \& Dawes \cite{TD10}. Figure~\ref{fig:th_snake002}(a) is an
enlargement of part of figure~\ref{fig:th_snake001}(b) that shows the reconnection
of several parts of the snaking curve to link to the isolas. Figure~\ref{fig:th_snake002}(b) presents plots of the solution $u(x)$ at
the four saddle-node bifurcation points on the right-hand side of figure~\ref{fig:th_snake002}(a), in sequence as the $L_2$ norm increases.
Despite the upper pair occurring very close to each other both in $L_2$
norm and values of $r$, the solutions look very different: the upper one
is clearly related to a multipulse localised state.
Very similar behaviour is observed for the Gaussian kernel; the formation
of switchbacks in general appears not to be related to the choice of
kernel.

Figures~\ref{fig:extreme} and~\ref{fig:nonlocal001}
lay out the challenges that we analyse
in detail in the remainder of the paper: developing the well-known
multiple-scale analysis to include nonlocal terms allows us to
determine the location of the branches of localised states and captures
the `S' shaped nature of the branches as shown in figure~\ref{fig:nonlocal001}.
As is also well-known, the regular multiple-scales analysis does not shed light on the
behaviour of individual snaking curves, and so we are unable to say
anything analytic about the breakup of the snaking curves and the collisions
with branches of multipulse states that are clearly shown in
figures~\ref{fig:th_snake001}
and~\ref{fig:th_snake002}. We suspect that the details of
these collisions are strongly dependent on the precise form chosen
for the kernel $K(x)$ and we leave these details to be the subject of
future work.

\begin{figure}
\bc
\includegraphics[width=7.4cm]{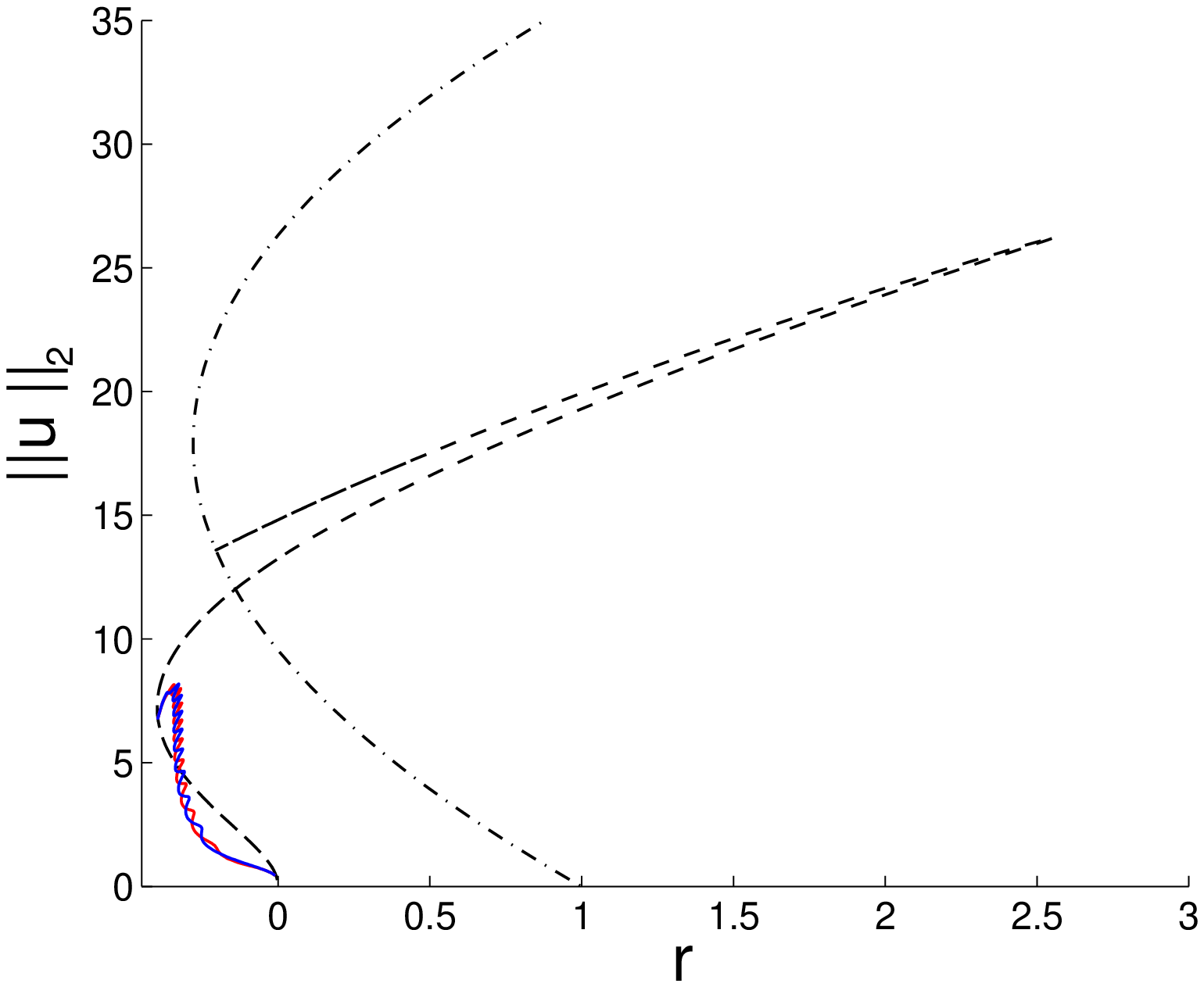}
\includegraphics[width=7.4cm]{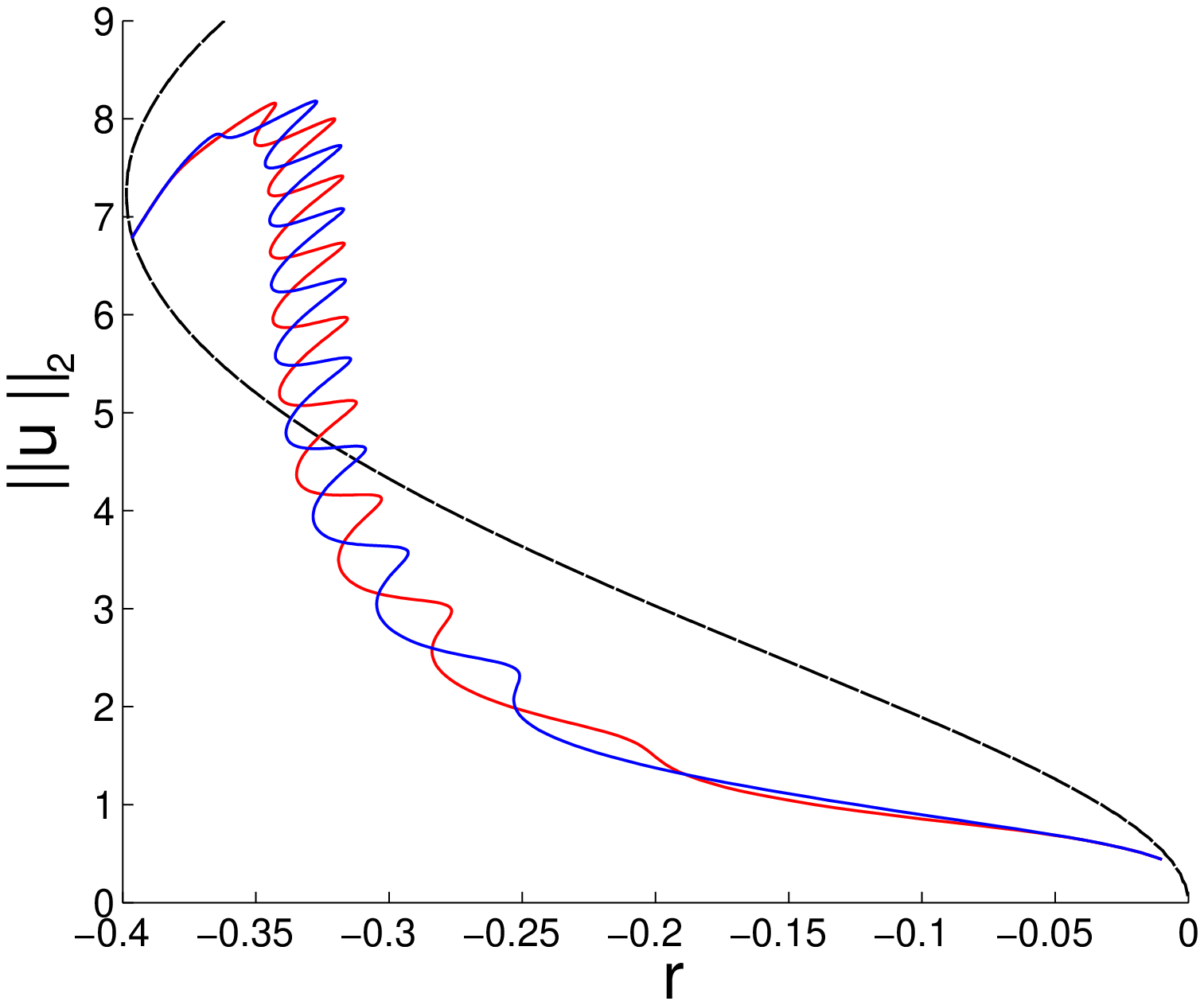}
\caption{\small Left: Bifurcation diagram for solutions of the nonlocal Swift--Hohenberg
equation~(\ref{eq:nlshe}) with a Gaussian kernel,
for $\gamma=-0.5$.
Dashed line indicates the branch of spatially periodic states that bifurcates from the trivial state at $r=0$. Dashed-dotted line indicates the non-zero constant state
that bifurcates from the trivial state at $r=1$. Solid (red and blue) curves are the branches of localised states. Stability is not shown.
Other parameter values are $\sigma=4\pi$, $b=1.6$, domain size $L=40\pi$.
}
\label{fig:gamma_m0_5}
\ec
\end{figure}

\begin{figure}
\bc
\includegraphics[width=7.4cm]{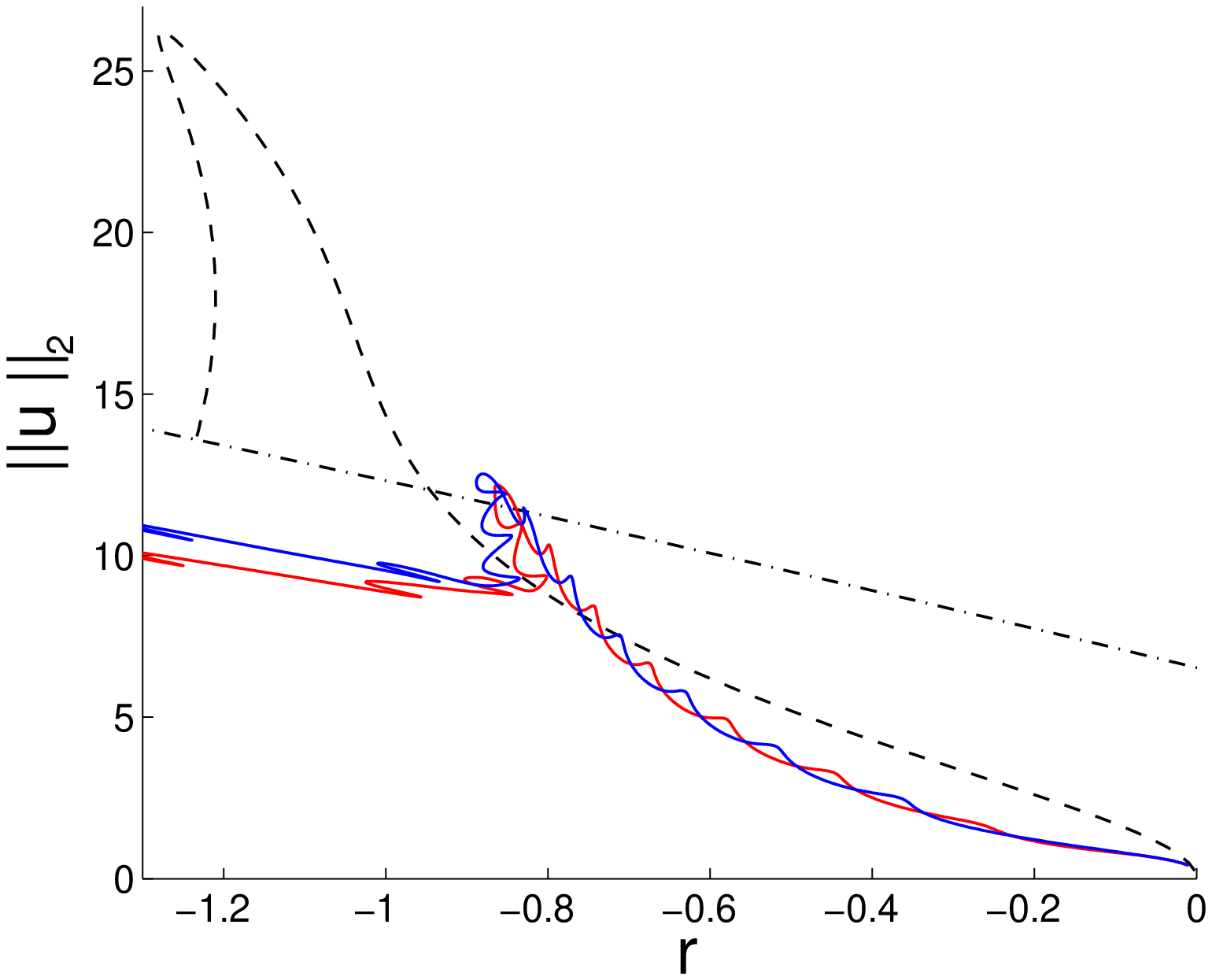}
\includegraphics[width=7.4cm]{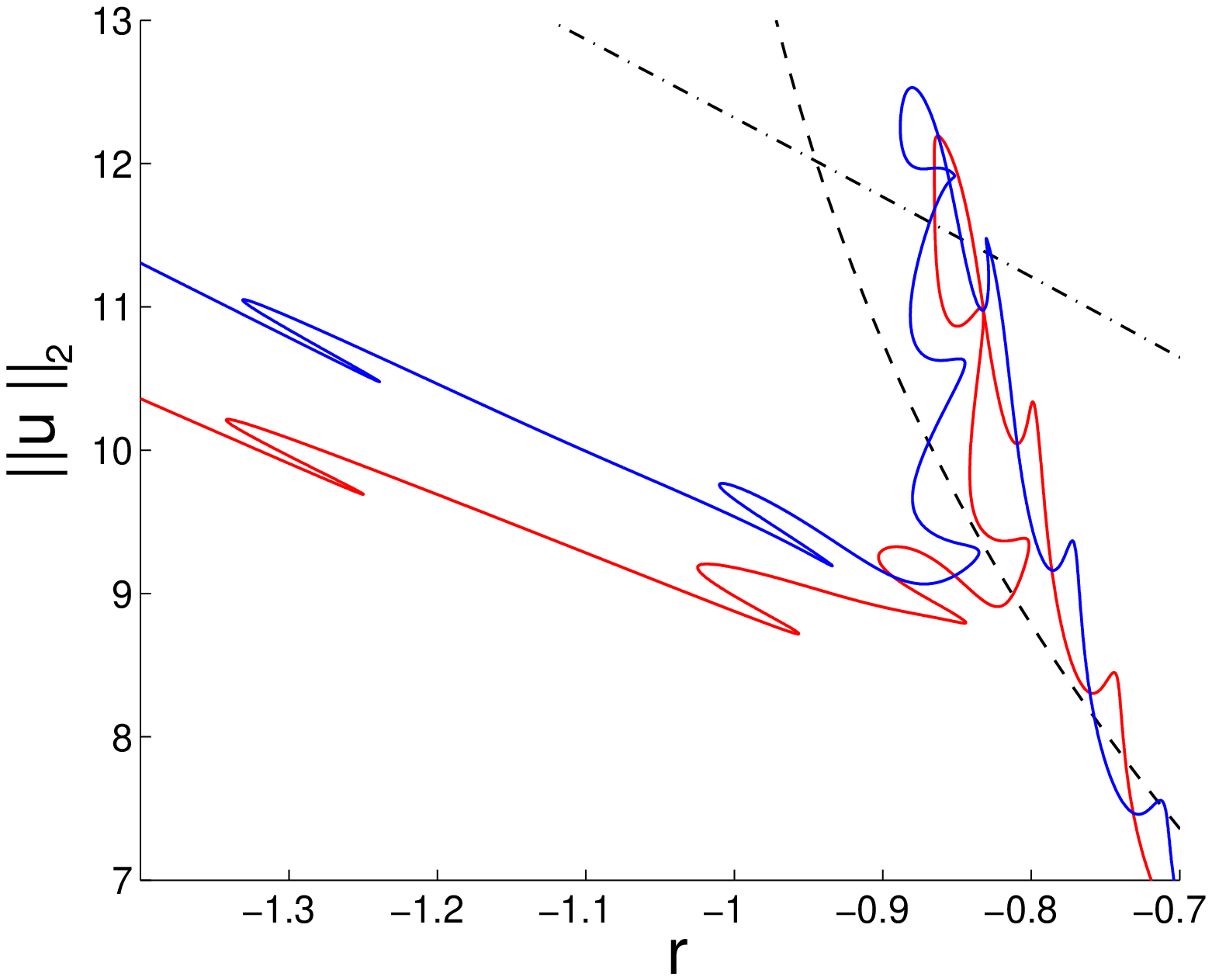}
\caption{\small Left: Bifurcation diagram for solutions of the nonlocal Swift--Hohenberg
equation~(\ref{eq:nlshe}) with a Gaussian kernel,
for $\gamma=-1.2$.
Dashed line indicates the branch of spatially periodic states that bifurcates from the trivial state at $r=0$. Dashed-dotted line indicates the non-zero constant state
that bifurcates from the trivial state at $r=1$. Solid (red and blue) curves are the branches of localised states. Stability is not shown.
Other parameter values are $\sigma=4\pi$, $b=1.6$, domain size $L=40\pi$.
}
\label{fig:gamma_m1_2}
\ec
\end{figure}

Finally, figures~\ref{fig:gamma_m0_5} and~\ref{fig:gamma_m1_2} show results for
$\gamma=-0.5$ and $\gamma=-1.2$, respectively, using the Gaussian kernel $K_G(x)$
with $\sigma=4\pi$. The case $\gamma=-0.5$ shows
standard homoclinic snaking with branches of localised states reconnecting to
the spatially periodic branch near its usual saddle-node bifurcation as the localised
state expands to fill the entire domain. One might expect the snaking curves
to be tilted backwards, which they are at the lower end but they remain vertical
in the central section, as in the case $\gamma=0.5$ shown in figure~\ref{fig:nonlocal001}(a). 
Figure~\ref{fig:gamma_m0_5} shows that the branch of spatially periodic states that
bifurcates from $r=0$ (and shown as a dashed line) extends into $r>0$ before turning round at a second saddle-node
point and then terminating on the branch of non-zero spatially constant states (shown as the dash-dotted line) that
bifurcates from $r=1$. This point is therefore another Turing-type instability, this time of the non-zero constant solution.

Numerical investigations reveal that the usual
homoclinic snaking structure persists, for $\sigma=4\pi$, down to around $\gamma=-1.13$
and certainly into $\gamma<-1$. But below a critical value, the localised states
undergo new instabilities in which the central peaks of the localised state
widen significantly. These snaking curves
do not reconnect to the periodic branch, but instead extend into negative $r$,
undergoing a number of additional twists and turns. This behaviour is
illustrated in figure~\ref{fig:gamma_m1_2}. We remark also that taking $\gamma$
large and negative changes the behaviour of the constant and spatially periodic
branches: in figure~\ref{fig:gamma_m1_2} we observe that the branch of spatially periodic states (dashed line) evolves monotonically until reaching a first saddle-node bifurcation at $r \approx -1.2$, and that the branch of positive constant solutions
(dash-dotted line) extends into $r<0$ and does not undergo a saddle-node bifurcation.

\section{Derivation of cubic Ginzburg--Landau equations}
\label{sec:short3}

In this section we discuss the construction of amplitude equations in the
simplest case, in which
the kernel $K(x)$ varies only on the short length scale. Equivalently,
this is the case where all terms posed on the long length scale
$X$ are assumed to be almost constant over the region where
the kernel is large. 
This makes sense in the case that the envelope function
is smooth and the kernel function decays much more rapidly than the envelope;
hence on functions of $X$, the kernel $K(x)$ acts to leading order
in $\epsilon$, as  a Dirac $\delta$-distribution.
This statement will be made more precise in section~\ref{sec:operator}
below. For the moment we will illustrate the computations involved in
deriving the modifications to the cubic Ginzburg--Landau
equations~(\ref{eq:gl3-23}) and~(\ref{eq:gl3-35}) when the nonlocal
term is included.

Within the weakly nonlinear multiple scales expansion~(\ref{eq:u_ansatz}),
we can see by inspection that the presence of the nonlinear nonlocal term contributes only at $O(\epsilon^3)$
and higher. Direct substitution of the ansatz~(\ref{eq:u_ansatz}) into the nonlocal
term indicates that, at the orders
in $\epsilon$ that are of interest in deriving the solvability conditions at third
or fifth order in $\epsilon$, the following terms need to be evaluated:
\ba
O\left(\varepsilon^{3}\right): \ \ & & \I_3 := u_{1}\left(x\right) J_{11},
\label{eq:int_eps3} \\
O\left(\varepsilon^{4}\right): \ \ & & \I_4 := u_{2}\left(x\right) J_{11}
+2u_1(x) J_{12},
\label{eq:int_eps4} \\
O\left(\varepsilon^{5}\right): \ \ & & \I_5 := u_{3}\left(x\right) J_{11}
+2u_{2}\left(x\right) J_{12},
+ u_1\left(x\right) \left( 2 J_{13} + J_{22} \right),
\label{eq:int_eps5}
\ea
where the integrals $J_{ij}(x)$ for $i,j \in \{1,2,3\}$ are defined to be
\ba
J_{ij}(x) := \int_{\Omega} K\left(x-y\right)u_i\left(y\right)u_j\left(y\right)\, dy.
\ea

To simplify the presentation we will carry out the computations
initially for two explicit kernel functions: the Gaussian $K_G(x)$ defined
in~(\ref{eq:g_kernel}) and the piecewise-constant top hat
function $K_{TH}(x)$ defined in~(\ref{eq:th_kernel}).
Since the scalings~(\ref{eq:scaling3}) generate
secular terms at $O(\varepsilon^3)$ and higher, for the cubic
Ginzburg--Landau equations
we need only to evaluate the leading-order part of the expression~(\ref{eq:int_eps3})
and then incorporate any additional resonant terms (i.e. those having a short-scale
dependence $\sim \e^{\pm \i x}$) into the Ginzburg-Landau equations~(\ref{eq:gl3-23}) and~(\ref{eq:gl3-35}) derived previously. Substituting the ansatz~(\ref{eq:u1_ansatz}) into~(\ref{eq:int_eps3})
we obtain
\ba
J_{11} & = & \int_{\Omega}K\left(x-y\right)\left(2\left|A(Y)\right|^{2}
+A(Y)^{2}\e^{2\i y}+\bar{A}(Y)^2\e^{-2\i y}\right)dy. \nn
\ea
To determine the leading order contribution
we consider the envelope $A(Y)$ to be constant within the integral, using the
implicit decoupling of the short and long length scales $x$ and $X=\epsilon x$
that arises asymptotically as $\epsilon \rightarrow 0$. We therefore
obtain the leading order simplification
\ba
J_{11} & = & 2 \left|A(X)\right|^{2}\int_{\Omega}K\left(x-y\right)dy
+A(X)^2 \int_{\Omega}K\left(x-y\right)\e^{2\i y}dy \nn \\
& & +\bar{A}(X)^2\int_{\Omega}K\left(x-y\right)\e^{-2\i y}dy \ + O(\epsilon).
\label{eq:I3}
\ea
We now evaluate these contributions in the cases of top hat and Gaussian kernels.

\subsection{The top hat kernel $K_{TH}(x)$}

In this subsection we evaluate the integrals in~(\ref{eq:I3}) for the top hat kernel
$K_{TH}(x)$ given in~(\ref{eq:th_kernel}).
The computations are the same regardless of the choice of a finite
domain $\Omega$ or an infinite domain $\Omega=\mathbb{R}$, assuming
that the width $\delta$ of the support of $K_{TH}(x)$ is at most the domain length $L$.
Explicitly we obtain
\ba
J_{11} & = & \frac{2}{\delta}\left|A(X)\right|^{2}\intop_{x-\frac{\delta}{2}}^{x+\frac{\delta}{2}}dy
+\frac{1}{\delta}A(X)^{2}\intop_{x-\frac{\delta}{2}}^{x+\frac{\delta}{2}}\e^{2iy}dy
+\frac{1}{\delta}\bar{A(X)}^{2}\intop_{x-\frac{\delta}{2}}^{x+\frac{\delta}{2}}\e^{-2iy}dy. \nn
\ea
Recall the general result, for integer $m$, that
\begin{equation}
\frac{1}{\delta}\intop_{x-\frac{\delta}{2}}^{x+\frac{\delta}{2}}\e^{imy}dy=\e^{imx}\frac{\sin\left(m\delta/2\right)}{m\delta/2}, \label{eq:e^(imx) integral}
\end{equation}
which we use in the three cases $m=-2,0,2$ to obtain
\ba
J_{11} & = & 2\left|A(X)\right|^{2}+\frac{\sin\delta}{\delta}\left(A(X)^{2}\e^{2ix}+\bar{A(X)}^2 \e^{-2ix}\right).
\nn 
\ea
From $J_{11}$ it is straightforward to compute $\I_3$:
\begin{equation}
\I_3=\left(2+\frac{\sin\delta}{\delta}\right)A\left|A\right|^2 \e^{ix}
+\left(\frac{\sin\delta}{\delta}\right)A^3\e^{3ix} +c.c.,
\nn 
\end{equation}
where $c.c.$ denotes the complex conjugate, so that $\I_3$ is always real-valued.
Comparison with the calculations in the previous sections shows that in
the presence of the nonlocal term with kernel $K_{TH}(x)$
we obtain the Ginzburg--Landau equations
\begin{equation}
A_{T}=\mu A+\left[\frac{38}{9}\left(b^{2}-\frac{27}{38}\right)
-\gamma\left(2+\frac{\sin\delta}{\delta}\right)\right]A\left|A\right|^{2}
+4A_{XX}, \label{eq:gl3_23_th}
\end{equation}
for the \Nqc case, and
\begin{equation}
A_{T}=\mu A +\left[3s-\gamma\left(2
+\frac{\sin\delta}{\delta}\right)\right]A\left|A\right|^{2}
+4A_{XX}, \label{eq:gl3_35_th}
\end{equation}
for the \Ncq case.
Alternatively, the effect of the nonlocal term can be described as
an effective shift in the coefficient of the nonlinear term.
In both cases the effect of the nonlocal term is, when $\gamma>0$, always to make the coefficient of the nonlinear term more negative (since $|\sin \delta| \leq \delta$
for all positive $\delta$), and hence the instability is always
more supercritical in the presence of the nonlocal term. If $\gamma<0$ then
the nonlocal term makes the instability more subcritical.

\subsection{The Gaussian kernel $K_G(x)$}

By substituting~(\ref{eq:g_kernel}) into~(\ref{eq:I3})
and taking the domain $\Omega:=[-L/2,L/2]$, the 
leading order contribution from $J_{11}$ to the
integral term $\I_3$ for the Gaussian kernel $K_G(x)$ 
on the short length scale only can be expressed as
\begin{eqnarray}
J_{11} & = & \int_{\Omega}K\left(x-y\right)\left(2\left|A(Y)\right|^{2}+A(Y)^2\e^{2iy}
+\bar{A}(Y)^2\e^{-2iy}\right)dy +O(\epsilon) \nn \\
& = & \frac{2}{\sqrt{2\pi\sigma^{2}}}\left|A(X)\right|^{2}\hat{I}_1
+\frac{1}{\sqrt{2\pi\sigma^2}}A(X)^2 \hat{I}_2
+\frac{1}{\sqrt{2\pi\sigma^{2}}}\bar{A}(X)^2 \hat{I}_3
\nn 
\end{eqnarray}
where
\begin{equation}
\hat I_1=\int_{-\frac{L}{2}}^{\frac{L}{2}}\e^{-\frac{(x-y)^2}{2\sigma^2}}\ dy,
\qquad
\hat I_2=\int_{-\frac{L}{2}}^{\frac{L}{2}}\e^{-\frac{(x-y)^2}{2\sigma^2}}\e^{2iy}\ dy,
\qquad
\hat I_3=\int_{-\frac{L}{2}}^{\frac{L}{2}}\e^{-\frac{(x-y)^2}{2\sigma^2}}\e^{-2iy}\ dy.
\label{eq:Gaussian Integrals}
\end{equation}
These integrals can be evaluated straightforwardly in terms of the
(complex-valued) error function, which we define as: $\erf\left(z\right)=\frac{2}{\sqrt{\pi}}\int_{0}^{z}e^{-t^{2}}dt$.
We obtain
\ba
\hat I_1 = \frac{\sqrt{2\pi\sigma^2}}{2} G_1(x),
\qquad
\hat I_2 = \frac{\sqrt{2\pi\sigma^2}}{2} \e^{-2\sigma^2}\e^{2ix}G_2(x),
\qquad
\hat I_3 = \frac{\sqrt{2\pi\sigma^2}}{2} \e^{-2\sigma^2}\e^{-2ix}\bar{G_2}(x), \nn
\ea
where
\ba
G_1(x) & := & \erf\left(\frac{x+\frac{L}{2}}{\sqrt{2\sigma^2}}\right) - \erf\left(\frac{x-\frac{L}{2}}{\sqrt{2\sigma^2}}\right), \nn \\
G_2(x) & := & \erf\left(\frac{x+\frac{L}{2}+2i\sigma^2}{\sqrt{2\sigma^2}}\right) - \erf\left(\frac{x-\frac{L}{2}+2i\sigma^2}{\sqrt{2\sigma^2}}\right), \nn
\ea
and in computing $\hat I_3$ we have used the property that
$\erf\left(\bar{z}\right)=\overline{\erf\left(z\right)}$
Putting all this together we obtain the leading order approximation
\begin{eqnarray}
J_{11} = \left|A(X)\right|^{2}G_1(x)
+ \frac{1}{2}\e^{-2\sigma^2}\left(G_2(x) A(X)^2\e^{2ix} 
+ \bar{G}_2(x)\bar{A}(X)^2\e^{-2ix}\right)
+ O(\epsilon).
\nn 
\end{eqnarray}
As with the derivation in the previous subsection for the top hat kernel, we now
multiply by $u_1(x)$ and collect terms in like powers of 
of $\e^{ix}$ to compute $\I_3$. This gives
\begin{equation}
\I_3 =
\left(G_1(x)+\frac{1}{2}\e^{-2\sigma^2}G_2(x)\right)A\left|A\right|^{2} \e^{ix}
+\frac{1}{2}\e^{-2\sigma^2}G_2(x) A^{3}\e^{3ix} +c.c. .
\label{eq:Gaussian Integrals 3}
\end{equation}

Extracting the resonant terms for the Ginzburg--Landau equation
from~(\ref{eq:Gaussian Integrals 3}) by multiplying by $\e^{-ix}$ and 
integrating over $\Omega$ is substantially more difficult in the case of a finite
domain; we therefore consider the case $L \rightarrow \infty$ in order to make
further progress. In this limit we find that the functions
$G_1$ and $G_2$ tend to constant values: $G_1=G_2=2$. For the
\Nqc and \Ncq cases, respectively, we obtain the following Ginzburg--Landau
equations:
\ba
A_{T} & = & \mu A +\left[\frac{38}{9}\left(b^{2}-\frac{27}{38}\right)
-\gamma\left(2+e^{-2\sigma^{2}}\right)\right]A\left|A\right|^{2}+4A_{XX},
\label{eq:gl3_23_g} \\
A_{T} & = & \mu A +\left[3s-\gamma\left(2
+\e^{-2\sigma^{2}}\right)\right]A\left|A\right|^{2} + 4A_{XX}.
\label{eq:gl3_35_g}
\ea
As in the top hat case, the effect of the nonlocal term at this order is therefore
effectively to shift the coefficient of the nonlinear term. The shifts
are in the same directions as in the case of a top hat kernel.
If $\gamma>0$ then the shift makes the bifurcation behaviour more supercritical.
In the limit of a narrow kernel, $\sigma \rightarrow 0$, we see
that the results from the Gaussian case agree with those of the Dirac `delta function'
discussed at the beginning of~section \ref{sec:nonlocal} which shifts the coefficient of the cubic term in $N(u)$ either from $-1$ to $-1+\gamma$ (in the \Nqc case) or from
$s$ to $s+\gamma$, in the \Ncq case. This agreement is found also in the top hat case, as can be seen by considering the limit $\delta \rightarrow 0$ in~(\ref{eq:gl3_23_th}) and~(\ref{eq:gl3_35_th}).

In the opposite limits, of wide kernels, where $\sigma \rightarrow \infty$ in the Gaussian
case,
and $\delta \rightarrow \infty$ in the top hat case, the coefficients again converge
to the same values, although the convergence is oscillatory for the top hat
kernel and monotonic for the Gaussian case.

\section{Derivation of a cubic-quintic Ginzburg--Landau equation}
\label{sec:short5}

In order to go beyond the solvability condition at
$O\left(\varepsilon^{3}\right)$ we need to incorporate higher-order contributions from the nonlocal term. This is a necessary preliminary to the
investigation of the effects of the nonlocal
term on the cubic-quintic Ginzburg--Landau equations near
the codimension-two point at which the initial instability changes from
supercritical to subcritical. In order to organise this calculation clearly, we
first present our approach to the nonlocal term, before using these results
in dealing order by order with the multiple scales computation.

\subsection{Series expansion of the nonlocal term}
\label{sec:operator}

In this section we develop an asymptotic expansion of the nonlocal term as a series
in the envelope $A(X)$ and its derivatives.
The first term in this new expansion is the one that we have relied
on in previous calculations. It turns out that we are able to
derive a compact and general expression in terms of Fourier coefficients
of the kernel: our expansion is therefore applicable to
any kernel function that has a well
defined, continuously differentiable Fourier transform.

The first step is to observe that all the integrals generated by
the asymptotic analysis can be written in the general form
\begin{equation}
J = \int_{\Omega}K\left(x-y\right)F\left[A\left(Y\right)\right]
\e^{imy}\, dy, \label{eq:general}
\end{equation}
where $F$ is a function
of the amplitude $A(X)$. We adopt the rescalings~(\ref{eq:scaling5}), i.e. $X=\epsilon^2 x$, and we consider the
domain $\Omega=\mathbb{R}$. We assume throughout that
there are no technical difficulties with applying the Fourier transform.
We use the following notation for the Fourier transform and its inverse:
\ba
\mathcal{F}\left[f\left(x\right)\right]\left(k\right)
\equiv \hat{f}\left(k\right)
& := &
\int_{-\infty}^{\infty}f\left(x\right)\e^{-ikx}\ dx,
\label{eq:FFT} \\
\mathbf{\mathcal{F}}^{-1}\left[\hat{f}\left(k\right)\right]\left(x\right)\equiv
f\left(x\right)
& := &\frac{1}{2\pi}\int_{-\infty}^{\infty}f\left(k\right)\e^{ikx}\ dk.
\label{eq:invFFT)}
\ea

We expand $J$ asymptotically in a series of straightforward steps which we
now present. First, we
make the substitution $Y=\varepsilon^{2}y$ in~(\ref{eq:general}), and
change the integration variable from $y$ to $z=y-x$ to give
\ba
J = \int_{\R}K\left(x-y\right)F\left[A\left(Y\right)\right]\e^{imy}dy = \int_{\R} K\left(-z\right)F\left[A\left(\varepsilon^{2}x+\varepsilon^{2}z\right)\right]\e^{im\left(x+z\right)}\ dz . \nn
\ea
Now we Taylor expand $A(Y)$ about the point $X$, which gives
\ba
J & = & \int_{\R}K\left(-z\right)\e^{im\left(x+z\right)}\sum_{n=0}^{\infty}\frac{\left(\varepsilon^2 z D_X\right)^{n}}{n!}F\left[A\left(X\right)\right]\ dz,
\label{eq:Fourier Integral 2}
\ea
where $D_X$ stands for $d/dX$. Reversing the change of integration variable yields
\begin{eqnarray}
J & = &
\int_{\R} K\left(x-y\right)\e^{imy}\left[\sum_{n=0}^{\infty}\frac{\left(-\varepsilon^{2}D_X \right)^{n}}{n!}F\left[A\left(X\right)\right]\left(x-y\right)^{n}\right]\ dy .
\nn
\ea
We now take all the $y$-independent terms outside the integral to obtain
\begin{eqnarray}
J & = & \sum_{n=0}^{\infty}\frac{\left(-\varepsilon^2
D_X\right)^n}{n!}F\left[A\left(X\right)\right]\int_{\R}\left(x-y\right)^{n}K\left(x-y\right)\e^{imy}\ dy .
\label{eq:Fourier Integral 4}
\end{eqnarray}
The integral part of this expression is a convolution and so
can be written (very compactly) in terms of Fourier transforms:
\begin{eqnarray}
\int_{-\infty}^{\infty}\left(x-y\right)^{n}K\left(x-y\right)\e^{imy}dy & = & \mathcal{F^{\mathbf{\mathrm{-1}}}}\left[\mathcal{F}\left[x^{n}K\left(x\right)\right]\mathcal{F}\left[\e^{imx}\right]\right] .
\label{eq:Fourier Integral 5}
\end{eqnarray}
This expression can be simplified significantly further using the
well-known Fourier transform properties that (i) the Fourier
transform of a complex exponential is a Dirac delta function and
and (ii) the effect of multiplication by a power of $x$:
\begin{equation}
\int_{-\infty}^{\infty}\e^{i\left(m-k\right)x}\ dx
= 2\pi\delta\left(m-k\right), 
\qquad \mathrm{and} \qquad
\int_{-\infty}^{\infty}x^{n}K\left(x\right)\e^{-ikx}\ dx
= \left(i D_k\right)^n\hat{K}\left(k\right) .
\label{eq:FTfacts}
\end{equation}
On substituting (\ref{eq:FTfacts})
into~(\ref{eq:Fourier Integral 5}), we obtain 
\begin{eqnarray}
\int_{-\infty}^{\infty}\left(x-y\right)^{n}K\left(x-y\right)\e^{imy}dy
& = & 2\pi\mathcal{F^{\mathbf{\mathrm{-1}}}}\left[i^{n} D_k^n\hat{K}\ \delta\left(m-k\right)\right] \nn \\
& = & 
\int_{-\infty}^{\infty}i^{n}D_k^n\hat{K}\ \delta\left(m-k\right)\e^{ikx}dk=\e^{imx} \left(i D_k\right)^n\left.\hat{K}\left(k\right)\right|_{k=m}
\nn
\end{eqnarray}
Returning to~(\ref{eq:Fourier Integral 4}) we see that we have
the expansion:
\begin{eqnarray}
J = \int_{\R} K\left(x-y\right)F\left[A\left(Y\right)\right]\e^{imy}dy =
\e^{imx} \sum_{n=0}^{\infty}\frac{\left(-i\varepsilon^2 D_X D_k\right)^{n}}{n!}F\left[A\left(X\right)\right] \left.\hat{K}\left(k\right)\right|_{k=m} .
\label{eq:Fourier Integral 10}
\end{eqnarray}
For use in the multiple scales analysis, it is convenient to collect the
kernel-dependent parts of the expression into a set of coefficients $I_{mn}$
defined by
\begin{equation}
I_{mn} :=\left(-i D_k\right)^n \left.\hat{K}\left(k\right)\right|_{k=m}, \label{eq:I_mn}
\end{equation}
which can then be evaluated at a later point, for a specific choice of kernel.
Note that since we consider kernels that are even functions
we have the relation $I_{m0}=I_{\left(-m\right)0}$.
In conclusion we have (formally) derived the series representation
\begin{eqnarray}
\int_{-\infty}^{\infty}K\left(x-y\right)F\left[A\left(Y\right)\right]\e^{imy}\
dy = \e^{imx} \sum_{n=0}^{\infty}I_{mn}\frac{\left(\varepsilon^{2} D_X\right)^n}{n!}F\left[A\left(X\right)\right] .
\label{eq:FI_expand}
\end{eqnarray}

\subsection{Asymptotic expansion}

Having found a general expansion for the integrals arising in the
multiple scales analysis, it is now possible to carry out the derivation of a
cubic-quintic Ginzburg--Landau equation where the nonlocal term is
included, using 
the alternative scalings~(\ref{eq:scaling5})
and the expansion (\ref{eq:u_ansatz}).
Previously, under the $O(\varepsilon^3)$ scalings,
the nonlocal term was found to introduce a width dependency into the
coefficient of the $A\left|A\right|^{2}$ term. Using the more complicated
scalings, it is now shown that similar dependencies appear in the
higher order $A\left|A\right|^{4}$ term. As stated, we leave the
kernel dependent coefficients $I_{mn}$ unevaluated in the derivation
and later generate appropriate values for both top hat and Gaussian
kernels through equation (\ref{eq:I_mn}).

The procedure for generating the relevant PDE for $A$ through the
multiple scales analysis is much the same as that discussed
for the local problem in section~\ref{sec:gl5}.
As there are no integral terms at $O\left(\varepsilon\right)$ the
ansatz for $u_{1}$ remains unchanged, as do the scalings for $X$,$T$
and $\mu$. However, with the addition of the nonlocal term the
codimension-two
point in the \Nqc equation is no longer $b=\sqrt{\frac{27}{38}}$ and
the parameter $b$ must instead be expanded about the point
\begin{equation}
b_0=\sqrt{\frac{27}{38}+\frac{9\gamma}{38}\left(2I_{00}+I_{20}\right)},
\label{eq:b0nonlocal}
\end{equation}
which depends on the form of (the Fourier transform of) the kernel $K(x)$.
Explicitly we write
\begin{equation}
b=b_0 + \varepsilon^2 b_2 . \label{eq:b_scaling}
\end{equation}
We remark, for reference, that the codimension two point defined by
$r=0$ and $b=b_0$, given by~(\ref{eq:b0nonlocal}), corresponds to the
codimension-two case of the bifurcation problem in the 
spatial dynamics setting studied by Woods and Champneys \cite{WC99}. In that
paper the condition $b=b_0$ is the condition $q_2=0$ where $q_2$ is one of the
crucial normal form coefficients for determining the qualitative behaviour of
solutions near the bifurcation point.

\subsection{Detailed derivation of the cubic--quintic amplitude equation}

We now present the derivation of the cubic-quintic amplitude equation in detail
in order to show exactly how we make (repeated) use of the series
expansion~(\ref{eq:FI_expand}). 
For clarity, we restate the nonlocal (2--3) Swift--Hohenberg equation
(\ref{eq:nlshe}).
\begin{equation}
\partial_t u = \left[r-\left(1+\partial_{x}^{2}\right)^{2}\right]u+bu^{2}-u^{3}-\gamma u(x,t) \int_{\Omega}K\left(x-y\right)u(y,t)^2 dy . \nn
\end{equation}
We expand as usual in an asymptotic series:
\begin{equation}
u\left(x,X,t,T\right)=\varepsilon u_{1}+\varepsilon^{2}u_{2}+\varepsilon^{3}u_{3}+\varepsilon^{4}u_{4}+\varepsilon^{5}u_{5} + \cdots , \nn
\end{equation}
and solve order by order in $\epsilon$ until we obtain a non-trivial
solvability condition. To simplify notation we define the linear operator
\ba
\calL[u]:=-(1+\partial_x^2)^2 u . \nn
\ea

\paragraph{Terms at $O\left(\varepsilon\right)$ and $O\left(\varepsilon^2\right)$.}
Since the nonlocal term is cubic in $u$, it does not affect the
forms of $u_1$ and $u_2$. Solving at $O\left(\varepsilon\right)$ and
$O\left(\varepsilon^2\right)$ and using~(\ref{eq:b_scaling})
we obtain the expected expressions
\begin{equation}
u_{1}\left(x,X,t,T\right) = A\left(X,T\right)\e^{ix}
+\bar{A}\left(X,T\right)\e^{-ix}, \nn
\end{equation}
and
\begin{equation}
u_2 = b_0\left(2\left|A\right|^{2}
+\frac{1}{9}A^{2}\e^{2ix}+\frac{1}{9}\bar{A}^{2}\e^{-2ix}\right) .
\label{eq:5th Order Nonlocal - u2}
\end{equation}

\paragraph{Terms at $O(\varepsilon^3)$.}
$O(\varepsilon^3)$ is the lowest order
at which a term from the nonlocal expression enters directly. We have
\begin{eqnarray}
\partial_{t}u_{3} & = & \calL\left[u_{3}\right]-4\partial_{x}\partial_{X}\underbrace{\left(1+\partial_{x}^{2}\right)u_{1}}_{=0}+2 b_0 u_{1}u_{2}-u_{1}^{3}-\gamma u_{1}\int_{\Omega}K\left(x-y\right)u_{1}^{2}\left(y\right)dy,
\label{eq:5th Order Nonloca - O(eps^3)}
\end{eqnarray}
and so to find $u_3$ we need to solve
\ba
\calL[u_3]+2b_0 u_{1}u_{2}-u_{1}^{3}-\gamma u_{1}\int_{\Omega}K\left(x-y\right)u_{1}^{2}\left(y\right)dy =0 . \nn
\ea
We expand the integral operator by using the series expansion~(\ref{eq:FI_expand}), noting that all terms in the
expansion where $n>0$ contribute at least an additional factor of
$\varepsilon^{2}$ and hence only the leading order term in $J_{11}$ contributes
at this order:
\begin{eqnarray}
J_{11} \equiv \int_{\Omega}K(x-y)u_{1}^{2}\left(y\right)dy
& = & J_{11}^{(0)} + \epsilon^2 J_{11}^{(2)} + O(\epsilon^4) \label{eq:j11_2} \\
& = & 2\left|A\right|^{2}I_{00} + I_{20}\left(A^2\e^{2ix}+\bar{A}^2\e^{-2ix}\right)
+O(\epsilon^2). \nn
\end{eqnarray}
At this order we ignore $J_{11}^{(2)}$, and,
after multiplying by $\gamma u_1$ we obtain
\begin{equation}
\gamma u_1 \int_{\Omega}K\left(x-y\right)u_{1}^{2}\left(y\right)dy = \gamma \left[\e^{ix}\left(2I_{00}+I_{20}\right)A\left|A\right|^{2}
+\e^{3ix} I_{20} A^{3}\right] +c.c. + O(\epsilon^2)
\label{eq:nonlocal_e3_1}
\end{equation}
The remaining (local) nonlinear terms in (\ref{eq:5th Order Nonloca - O(eps^3)})
are
\begin{eqnarray}
2b_0 u_1 u_2 - u_1^3 & = & \frac{2}{9}b_0^2 \left(A^{3}\e^{3ix}
+\bar{A}^{3}\e^{-3ix} +19A\left|A\right|^{2}\e^{ix}
+19\bar{A}\left|A\right|^{2}\e^{-ix}\right) \nonumber \\
 &  & -\left(A^{3}\e^{3ix} +3A\left|A\right|^2 \e^{ix}
+3\bar{A}\left|A\right|^2 \e^{-ix} +\bar{A}^{3}\e^{-3ix}\right)\nonumber \\
& = & \e^{ix}\left(\frac{38}{9}b_0^2 - 3\right)A\left|A\right|^2
+\e^{3ix}\left(\frac{2}{9}b_0^2 - 1\right)A^{3} +c.c.
\label{eq:nonlocal_e3_2}
\end{eqnarray}

Combining~(\ref{eq:nonlocal_e3_1}) and~(\ref{eq:nonlocal_e3_2}) we have
\ba
2b_0 u_1 u_2 -u_1^3 
-\gamma u_{1}\int_{\Omega}K\left(x-y\right)u_{1}^{2}\left(y\right)dy
& = & \e^{ix}\left(\frac{38}{9}b_0^2 -3 -\gamma\left(2I_{00}+I_{20}\right)\right)A\left|A\right|^{2} \nn \\
 &  & +\e^{3ix}\left(\frac{2}{9}b_0^2 -1 -\gamma I_{20}\right)A^{3}+c.c. .\nn
\ea
The definition~(\ref{eq:b0nonlocal}) of $b_0$ shows that the coefficients
of $\e^{\pm ix}$ vanish, and that the coefficient
of $\e^{3ix}$ can be simplified:
\ba
\frac{2}{9}b_0^2 -1 -\gamma I_{20} & = & \frac{2}{9}\left(\frac{27}{38}+\frac{9\gamma}{38}\left(2I_{00}+I_{20}\right)\right)-1-\gamma I_{20}
= -\frac{16}{19}+\frac{2\gamma}{19}\left(I_{00}-9I_{20}\right) . \nn
\ea
We therefore obtain a solution for $u_3$ in the form
\ba
u_3 = C_2\left(A^3 \e^{3ix} +\bar{A}^3\e^{-3ix}\right),
\qquad
\mathrm{where}
\qquad
C_2 = -\frac{1}{76}+\frac{\gamma}{608}\left(I_{00}-9I_{20}\right) . \nn
\ea

\paragraph{Terms at $O(\varepsilon^4)$.}
At fourth order in $\varepsilon$ we obtain 
\begin{eqnarray}
\partial_{t}u_{4} & = & \calL\left[u_{4}\right] -4\partial_{x}\partial_{X}\left(1+\partial_{x}^{2}\right)u_{2} +b_2u_1^2 +b_0\left(u_2^2 +2u_{1}u_{3}\right) -3u_{1}^{2}u_{2}\nonumber \\
 &  & -\gamma u_{2}\int_{\Omega}K\left(x-y\right)u_{1}^{2}\left(y\right)dy-2\gamma u_{1}\int_{\Omega}K\left(x-y\right)u_{1}\left(y\right)u_{2}\left(y\right)dy \label{eq:nonlocal_e4}
\end{eqnarray}
which includes contributions from two integral terms: $J_{11}$ and $J_{12}$.
We write this schematically in the form
\begin{eqnarray}
\partial_{t}u_{4} & = & \calL\left[u_4\right] + f_{4L} + f_{4NL}, \nn
\end{eqnarray}
collecting the local and integral terms into $f_{4L}$ and $f_{4NL}$
respectively so that we can consider them separately in what follows.

\paragraph{Local terms at $O(\epsilon^4)$.} Expanding the four local
terms we obtain:
\ba
-4\partial_{x}\partial_{X}\left(1+\partial_{x}^{2}\right)u_{2} & = &
\frac{16i}{3}b_0 \left(AA_{X}\e^{2ix}-\bar{A}\bar{A}_{X}\e^{-2ix}\right),
\nn
\\
b_2 u_1^2 & = & b_2\left(2\left|A\right|^{2}+A^2\e^{2ix} +\bar{A}^2\e^{-2ix}\right),
\nn
\\
b_0\left(u_2^2 +2u_1 u_3\right) & = & \frac{326}{81}b_0^3\left|A\right|^4 
+\e^{2ix}\left(\frac{4}{9}b_0^3 +2b_0 C_2\right)A^2\left|A\right|^2
\nn \\
& & +\e^{4ix}\left(\frac{b_0^3}{81} +2b_0C_2\right)A^{4} +c.c. ,
\nn
\\
-3 u_1^2 u_2 & = & -\frac{38}{3}b_0\left|A\right|^4
-\e^{2ix}\frac{20}{3}b_0 A^{2}\left|A\right|^2
-\e^{4ix}\frac{1}{3}C_{1}A^{4} - c.c. .
\nn
\end{eqnarray}
Collecting together all these terms we obtain
\begin{eqnarray}
f_{4L} & = & \frac{16i}{3}b_0\left(AA_{X}\e^{2ix} -\bar{A}\bar{A}_{X}\e^{-2ix}\right) +2b_3\left|A\right|^{2} +b_3 A^2 \e^{2ix}\nonumber \\
 &  & +\left(\frac{326}{81}b_0^3-\frac{38}{3}b_0\right)\left|A\right|^{4} +\e^{2ix}\left(\frac{4}{9}b_0^3 +2b_0C_2 -\frac{60}{9}b_0 \right)A^{2}\left|A\right|^{2}
 \nn \\
 &  & +\e^{4ix}\left(\frac{b_0^3}{81} +2b_0 C_2 -\frac{b_0}{3}\right)A^{4} +c.c. \ . \nn
\end{eqnarray}

\paragraph{Integral terms at $O(\epsilon^4)$.}

We now turn to the nonlocal terms. Using the series expansions, the two integral terms are evaluated to give the following:
\begin{eqnarray}
J_{11}  \equiv \int_{\Omega}K\left(x-y\right)u_{1}^{2}\left(y\right)dy & = & 2\left|A\right|^{2}I_{00}+I_{20}\left(A^{2}e^{2ix}+\bar{A}^{2}e^{-2ix}\right),
\nn \\
J_{12} \equiv \int_{\Omega}K\left(x-y\right)u_{1}\left(y\right)u_{2}\left(y\right)dy
& = & \frac{b_0}{9}\left[ 19I_{10} (A\left|A\right|^2 e^{ix}
+\bar{A}\left|A\right|^{2}e^{-ix}) \right.
\nn \\
& & \left. +I_{30}(A^{3}e^{3ix}+\bar{A}^{3}e^{-3ix})\right], \nn
\end{eqnarray}
ignoring higher-order contributions in $\epsilon$.
The nonlocal terms can then be evaluated to give
\begin{eqnarray}
\gamma u_2\int_{\Omega}K\left(x-y\right)u_{1}^{2}\left(y\right)dy & = & \gamma b_0\left[\left(4I_{00}+\frac{2}{9}I_{20}\right)\left|A\right|^{4} +\e^{2ix}\left(2I_{20}+\frac{2}{9}I_{00}\right)A^{2}\left|A\right|^{2} 
\right. \nn \\
& & \left. +\frac{I_{20}}{9}A^4\e^{4ix} \right] +c.c., \nn
\ea
and
\ba
2\gamma u_{1}\int_{\Omega}K\left(x-y\right)u_{1}\left(y\right)u_{2}\left(y\right)dy & = & \frac{2}{9}\gamma b_0\left[38I_{10}\left|A\right|^{4} +\e^{2ix}\left(19I_{10} +I_{30}\right)A^{2}\left|A\right|^2 \right.
\nn \\
& & \left. +I_{30}A^4\e^{4ix} \right] 
+c.c. \ . \nn 
\end{eqnarray}
These contributions to the right hand side of~(\ref{eq:nonlocal_e4}) can be collected
together to give
\ba
f_{4NL} & = & \frac{\gamma b_0}{9}\left[\left(36I_{00}+76I_{10}+2I_{20}\right)\left|A\right|^{4} +\e^{2ix}\left(2I_{00}+38I_{10}+18I_{20}+2I_{30}\right)A^{2}\left|A\right|^{2}
\right. \nn \\
& & \left. +\e^{4ix}\left(I_{20}+2I_{30}\right)A^{4}\right] +c.c. \nn
\ea

\paragraph{Solution at $O(\epsilon^4)$.}
The equation for $u_4$ therefore reduces to
\begin{eqnarray}
0 = \calL\left[u_{4}\right] +2b_3\left|A\right|^{2} +C_{3}\left|A\right|^{4} +\left(i\frac{16}{3}C_{1}AA_{X}+b_{3}A^{2}+C_{4}A^{2}\left|A\right|^{2}\right)\e^{2ix} +C_{5}A^{4}\e^{4ix} +c.c. \nn \\
\label{eq:nonlocal_e4_soln}
\end{eqnarray}
where we have defined the constants $C_{3}$, $C_{4}$ and $C_{5}$ to be
\ba
C_3 & = & b_0\left[\frac{326}{81}b_0^2 -\frac{38}{3} -2\gamma\left(2I_{00} +\frac{38}{9}I_{10} +\frac{1}{9}I_{20}\right)\right], \nn \\
C_4 & =  & b_0\left[\frac{4}{9}b_0^{2} +2C_2 -\frac{20}{3} -\frac{2\gamma}{9}\left(I_{00}+19I_{10}+9I_{20}+I_{30}\right)\right], \nn \\
C_5 & = & b_0\left[\frac{b_0^2}{81}+2C_{2}-\frac{1}{3} -\frac{\gamma}{9}\left(I_{20}+2I_{30}\right)\right] . \nn
\ea
Equation~(\ref{eq:nonlocal_e4_soln}) can be solved straightforwardly, since there
are no terms in $\e^{\pm i x}$, by acting with $\calL^{-1}$. We therefore obtain
the solution
\begin{equation}
u_4=2b_3\left|A\right|^{2} +C_{3}\left|A\right|^{4} +\left(i\frac{16}{27}C_{1}AA_{X}+\frac{b_{3}}{9}A^{2}+\frac{C_{4}}{9}A^{2}\left|A\right|^{2}\right)\e^{2ix}
+\frac{C_{5}}{225}A^{4}\e^{4ix}+c.c. \ . \nn
\end{equation}

\paragraph{Terms at $O(\varepsilon^5)$.}
Continuing to fifth order in $\varepsilon$ we pick up contributions from
three additional integral
terms as well as higher-order contributions from $J_{11}$ - recall
that at $O(\epsilon^3)$ we considered only the leading order part of $J_{11}$. 
At $O(\epsilon^5)$ we seek only to identify the coefficients of the $\e^{\pm ix}$
terms in order to deduce a solvability condition; we do not need to
solve completely for $u_5$. Considering terms at $O(\epsilon^5)$
in~(\ref{eq:nlshe}) we have
\begin{eqnarray}
\partial_t u_5 +\partial_T u_1 & = & \calL\left[u_5 \right] +\left(\mu+4\partial_{X}^2\right)u_1 -4\partial_{x}\partial_{X}\left(1+\partial_{x}^{2}\right)u_{3}
+2b_2 u_1 u_2 +2b_0\left(u_2 u_3 +u_1 u_4\right) \nonumber \\
& & -3\left(u_{1}u_{2}^{2}+u_{1}^{2}u_{3}\right)
 -\gamma u_3 \int_{\Omega}K\left(x-y\right)u_{1}^{2}\left(y\right)dy
\nn \\
& & -2\gamma u_2 \int_{\Omega}K\left(x-y\right)u_1 \left(y\right)u_2 \left(y\right)dy 
\nn \\
& & -\gamma u_1 \int_{\Omega}K\left(x-y\right)\left(2u_1\left(y\right)u_3\left(y\right)
+u_2^2\left(y\right)\right)dy, \label{eq:nonlocal_e5} \\
& = & \calL\left[u_5\right] + f_{5L} + f_{5NL} . \nn
\end{eqnarray}

\paragraph{Local terms at $O(\epsilon^5)$.}
The local terms in~(\ref{eq:nonlocal_e5}) contribute the following
terms that contain a factor of $\e^{ix}$ (we list only those terms, and
ignore terms containing factors of $\e^{iqx}$ where $q \neq 1$):
\ba
\left(\mu+4\partial_{X}^{2}\right)u_{1} & = & \left(\mu A+4A_{XX}\right)\e^{ix} + \cdots , \nn \\
2b_2 u_1 u_2 & = & \frac{38}{9}b_2 b_0 A\left|A\right|^{2} \e^{ix} +\cdots ,
\nn \\
2b_0 u_2 u_3 & = & \frac{2}{9}b_0^2 C_{2} A\left|A\right|^{4}\e^{ix} + \cdots , \nn \\
2b_0 u_1 u_4 & = &  2b_0 \left[\frac{19}{9}b_2A\left|A\right|^{2}+\left(C_{3}+\frac{C_{4}}{9}\right)A\left|A\right|^{4}+i\frac{16}{27}b_0\left|A\right|^{2}A_{X}\right] \e^{ix} +\cdots , \nn \\
-3\left(u_1 u_2^2 +u_1^2 u_3\right) & = & -\left(\frac{362}{27}b_0^2 +3C_{2}\right)A\left|A\right|^{4}\e^{ix} +\cdots .
\nn
\ea
Collecting these terms together we obtain, for the local terms,
\begin{equation}
f_{5L} = \left( \mu A +4A_{XX} +\frac{76}{9}b_0 b_2A\left|A\right|^2 +C_{6}A\left|A\right|^{4} +i\frac{32}{27}b_0^2A_{X}\left|A\right|^{2}
\right) \e^{ix} + \cdots ,
\nn
\end{equation}
where
\ba
C_{6} = \frac{2}{9}\left(b_0^2 C_{2} +b_0 C_{4}\right) +2b_0C_{3} -3C_{2}-\frac{362}{27}b_0^2 . \nn
\ea

\paragraph{Nonlocal terms that make leading order contributions at $O(\epsilon^5)$.}
We consider the contributions made by
each of the nonlocal terms in $f_{5NL}$ in turn.
As in the case of the local terms, we give only the terms containing a factor
of $\e^{ix}$. For each nonlocal term we first give the leading order
contribution of the integral term, followed by the terms containing
factors of $\e^{ix}$ that contribute to the solvability condition.
For the first nonlocal term in~(\ref{eq:nonlocal_e5})
we have
\ba
\int_{\Omega}K\left(x-y\right)u_{1}^{2}\left(y\right)dy
& = & 2\left|A\right|^{2}I_{00}+I_{20}\left(A^{2}e^{2ix}+\bar{A}^{2}e^{-2ix}\right)
+ O(\epsilon^2), \nn \\
\Rightarrow -\gamma u_{3}\int_{\Omega}K\left(x-y\right)u_{1}^{2}\left(y\right)dy
& = & -\gamma C_{2}I_{20} A\left|A\right|^{4} \e^{ix}
+ \cdots + O(\epsilon^2). \nn
\ea
For the second nonlocal term in~(\ref{eq:nonlocal_e5}) we obtain
\ba
\int_{\Omega}K\left(x-y\right)u_{1}\left(y\right)u_{2}\left(y\right)dy
& = & \frac{b_0}{9}\left[19I_{10}A\left|A\right|^{2}\e^{ix} +I_{30}A^{3}\e^{3ix} \right. \nn \\
& & \left. +19I_{10}\bar{A}\left|A\right|^{2}\e^{-ix}
+I_{30}\bar{A}^3\e^{-3ix}\right],
\nn \\
\Rightarrow -2\gamma u_2 \int_{\Omega}K\left(x-y\right)u_1 \left(y\right)u_2\left(y\right)dy
& = & -\frac{2}{9}\gamma b_0^2\left(38I_{10}+\frac{19}{9}I_{10} \right. \nn \\
& & \left. +\frac{1}{9}I_{30}\right)
A\left|A\right|^{4} \e^{ix} + \cdots + O(\epsilon^2). \nn
\end{eqnarray}
For the third nonlocal term in~(\ref{eq:nonlocal_e5}) we obtain
\ba
\int_{\Omega}K\left(x-y\right)u_{1}\left(y\right)u_{3}\left(y\right)dy
& = & C_{2}\left[A^{4}\e^{4ix}I_{40} +A^{2}\left|A\right|^{2}\e^{2ix}I_{20} \right. \nn \\
& & \left. +\bar{A}^{4}\e^{-4ix}I_{40} +\bar{A}^2\left|A\right|^{2}\e^{-2ix}I_{20}\right],
\nn \\
\Rightarrow -2\gamma u_1\int_{\Omega}K\left(x-y\right)u_{1}\left(y\right)u_{3}\left(y\right)dy
& = & -2\gamma C_{2}\left(I_{20}\right) A\left|A\right|^4 \e^{ix} + \cdots + O(\epsilon^2). \nn
\ea
For the fourth nonlocal term in~(\ref{eq:nonlocal_e5}) we obtain
\ba
\int_{\Omega}K\left(x-y\right)u_{2}^{2}\left(y\right)dy & = & \frac{326}{81}b_0^2 \left|A\right|^{4}I_{00}+\frac{4}{9}b_0^2 I_{20}\left(A^{2}\left|A\right|^{2}\e^{2ix}+\bar{A}^{2}\left|A\right|^{2}\e^{-2ix}\right) \nn \\
& & +\frac{b_0^2}{81}I_{40}\left(A^{4}\e^{4ix}+\bar{A}^{4}\e^{-4ix}\right),
\nn \\
\Rightarrow -\gamma u_{1}\int_{\Omega}K\left(x-y\right)u_{2}^{2}\left(y\right)dy
& = & -\gamma b_0^2 \left(\frac{326}{81}I_{00}+\frac{4}{9}I_{20}\right) A\left|A\right|^{4}e^{ix}
+ \cdots + O(\epsilon^2). \nn
\ea
These four integral terms therefore contribute a term of the form
\begin{equation}
f_{5NL}^{(0)} =  -\gamma C_{7} A\left|A\right|^{4} \e^{ix}
+ \cdots ,
\end{equation}
to the right hand side of~(\ref{eq:nonlocal_e5}), where the coefficient
$C_7$ is defined to be
\ba
C_7 = 3C_{2}I_{20} +\frac{b_0^2}{81}\left(326I_{00} +722I_{10} +36I_{20} +2I_{30}\right).
\nn
\ea

\paragraph{Nonlocal terms that introduce higher order contributions at $O(\epsilon^5)$.}
At $O\left(\varepsilon^{5}\right)$ the second term in the series
expansion of $J_{11}$ contributes to the solvability condition, see~(\ref{eq:j11_2}).
Explicitly, we obtain
\ba
\int_{\Omega}K\left(x-y\right)u_{1}^{2}\left(y\right)dy & = & 2\int_{\Omega}K\left(x-y\right)\left|A\right|^{2}dy
+\int_{\Omega}K\left(x-y\right)A^2 \e^{2ix}dy \nn \\
& & +\int_{\Omega}K\left(x-y\right)\bar{A}^{2}\e^{-2ix}dy, \nn \\
& = & J_{11}^{(0)} +2\epsilon^2\left[\left(A_{X}\bar{A}+A\bar{A}_{X}\right)I_{01} +AA_{X}\e^{2ix}I_{21} +\bar{A}\bar{A}_{X}\e^{-2ix}I_{(-2)\, 1} \right]. \nn \\
& & \label{eq:nonlocal_e5_1}
\ea
The terms in $\e^{ix}$ from this expression that contribute to the solvability
condition are found by multiplying (\ref{eq:nonlocal_e5_1}) by $\gamma u_1$. This
gives
\begin{equation}
\gamma u_1\int_{\Omega}K\left(x-y\right)u_{1}^{2}\left(y\right)dy = \gamma u_1 J_{11}^{(0)} +\epsilon^2\gamma \e^{ix}\left[2\left(I_{01} +I_{21}\right)A_{X}\left|A\right|^2 +2I_{01}\bar{A}_X A^2\right]. \nn
\end{equation}
We remark that when the kernel $K(x)$ is real, as in all the cases we consider,
the coefficients $I_{m1}$ are purely imaginary.

\paragraph{Solvability condition at $O(\epsilon^5)$.}

Putting all these contributions together we obtain the following cubic--quintic
Ginzburg--Landau equation for $A(X,T)$ from the solvability condition at
fifth order in $\epsilon$:
\ba
A_T = \mu A +4A_{XX} +\frac{76}{9}b_0 b_2 A\left|A\right|^2
+\left(C_6-\gamma C_7\right)A\left|A\right|^{4}
+i\left[C_8 A_{X}\left|A\right|^2 +C_9\bar{A}_{X}A^{2}\right],
\label{eq:nlgl_e5}
\ea
where $b_0$ is defined in~(\ref{eq:b0nonlocal}) and
\ba
C_2 & = & -\frac{1}{76}+\frac{\gamma}{608}\left(I_{00}-9I_{20}\right), \nn \\
C_3 & = & b_0 \left[\frac{326}{81}b_0^2 -\frac{38}{3} -2\gamma\left(2I_{00}+\frac{38}{9}I_{10}+\frac{1}{9}I_{20}\right)\right], \nn \\
C_4 & = & b_0\left[\frac{4}{9}b_0^2 +2C_2 -\frac{20}{3} -\frac{2\gamma}{9}\left(I_{00}+19I_{10}+9I_{20}+I_{30}\right)\right], \nn \\
C_5 & = & b_0\left[\frac{b_0^2}{81} +2C_{2} -\frac{1}{3} -\frac{\gamma}{9}\left(I_{20} +2I_{30}\right)\right], \nn \\
C_6 & = & \frac{2}{9}\left(b_0^2 C_{2} +b_0 C_4\right) +2b_0 C_3 -3C_{2} -\frac{362}{27}b_0^2, \nn \\
C_7 & = & 3C_2 I_{20}+\frac{b_0^2}{81}\left(326I_{00}+722I_{10}+36I_{20}+2I_{30}\right), \nn \\
C_{8} & = & \frac{32}{27}b_0^2 +2\gamma \Im\left(I_{01}+I_{21}\right), \nn \\
C_{9} & = & 2\gamma \Im (I_{01}) . \nn
\ea
The coefficients $C_2,\ldots, C_9$ are real valued:
we introduce $\Im$ into $C_8$ and $C_9$ (assuming that the kernel is real), in order
to make this clear. In the case that the kernel $K$ is even-symmetric,
we see that the coefficients $I_{m1}$ will vanish, and so $b_0=27/38$,
$C_8=32 b_0^2 / 27=16/17$, and $C_9=0$.
As we expect, every term that contains a factor of $I_{mn}$ also contains 
at least one factor
of $\gamma$: in the case $\gamma=0$ the well-known solvability
condition (\ref{eq:gl5-23}) for the (2--3) Swift--Hohenberg
equation is recovered.

In the case of a real, even-symmetric kernel $K(x)$, the coefficients in
the amplitude
equation~(\ref{eq:nlgl_e5}) simplify a little to give the form
\ba
A_T = \mu A +4A_{XX} +\frac{76}{9}b_0b_2 A\left|A\right|^{2} +C_{10}A\left|A\right|^{4} +iC_{8}A_{X}\left|A\right|^2, \label{eq:nlgl_e5_001}
\nn
\ea
in which the coefficients, expressed directly in terms of values of the Fourier
transform $\hat K(k)$ and its derivatives, are
\ba
b_0 & = & \sqrt{\frac{27}{38}+\frac{9\gamma}{38}\left(2\hat{K}\left(0\right)+\hat{K}\left(2\right)\right)}, \nn \\
C_8 & = & \frac{32}{27}b_0^2 -2\gamma\left. \frac{d \hat{K}(k)}{dk} \right|_{k=2}, \nn \\
C_{10} & = & -\frac{8820}{361}-\gamma \bfv^{T}\mathbf{\mathbf{k}}-\gamma^{2}\mathbf{k}^{T}M\mathbf{k}, \nn
\ea
where
\ba
\bfv = \begin{bmatrix}
\frac{7998}{361}\\ 19 \\
\frac{2764}{361} \\
\frac{1}{19}
\end{bmatrix}, 
\qquad
M=\begin{bmatrix}\frac{67415}{17328} & \frac{19}{3} & \frac{14489}{17328} & \frac{1}{57}\\
\frac{19}{3} & 0 & \frac{19}{6} & 0\\
\frac{14489}{17328} & \frac{19}{6} & -\frac{1059}{5776} & \frac{1}{114}\\
\frac{1}{57} & 0 & \frac{1}{114} & 0
\end{bmatrix},
\qquad
\mathbf{k}=\begin{bmatrix}\hat{K}\left(0\right)\\
\hat{K}\left(1\right)\\
\hat{K}\left(2\right)\\
\hat{K}\left(3\right)
\end{bmatrix}.
\nn
\ea
The coefficients in the amplitude equation therefore depend only
on five pieces of data concerning $\hat K(k)$: the values at $k=0,1,2,3$ and
the derivative $\hat K'(2)$.

\begin{figure}[!ht]
\bc
\includegraphics[width=7.4cm]{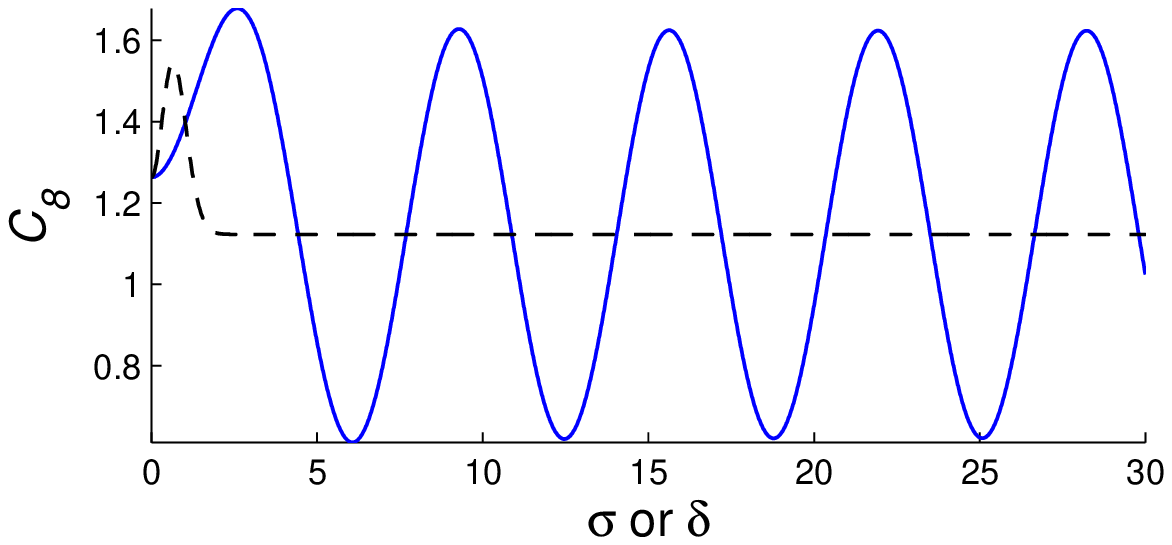}
\includegraphics[width=7.4cm]{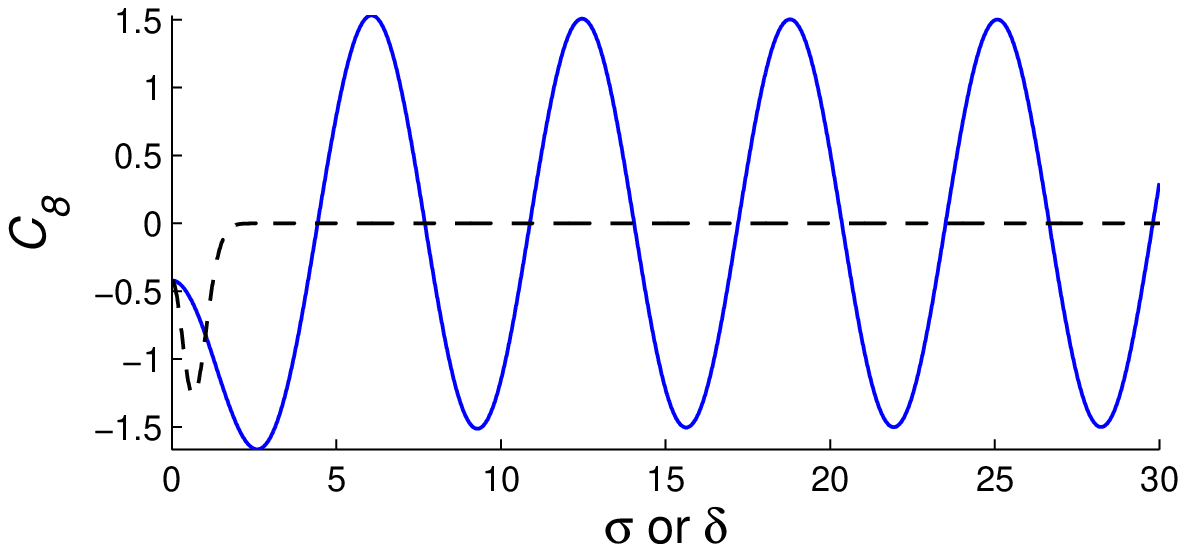}
\includegraphics[width=7.4cm]{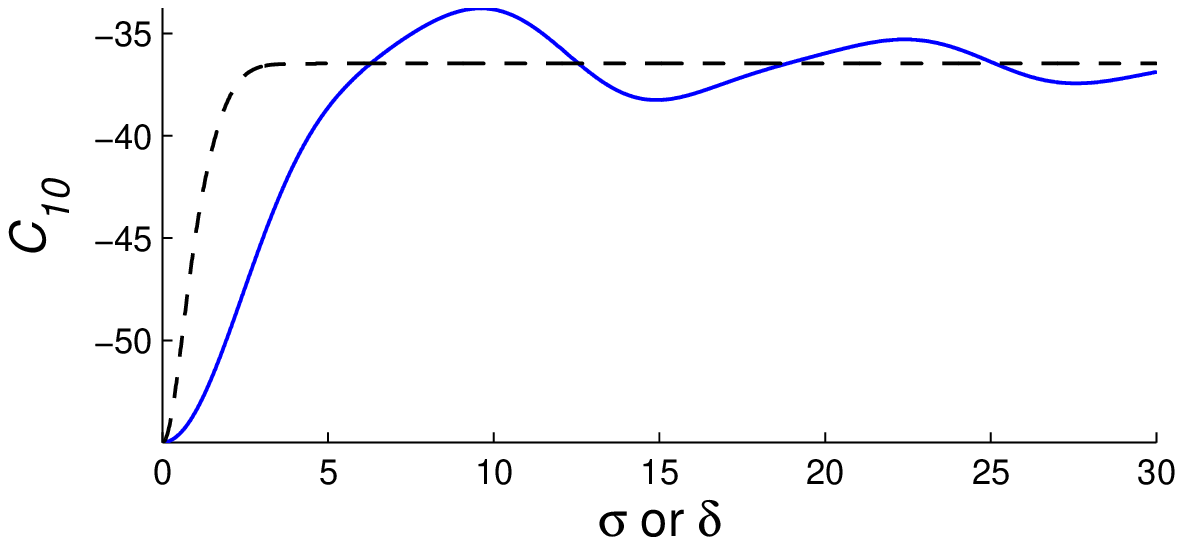}
\includegraphics[width=7.4cm]{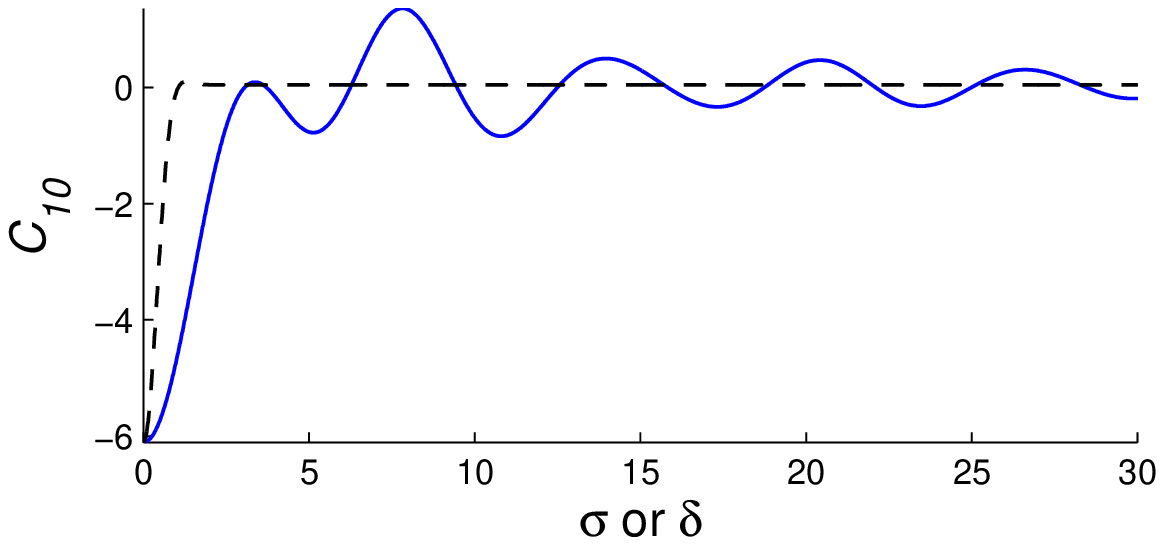}
\ec
\caption{\label{fig:coeffs}\small The coefficients $C_8$ and $C_{10}$
in the fifth order Ginzburg-Landau equation~(\ref{eq:nlgl_e5_001})
plotted as functions of the width parameters
$\delta$ (for the top hat kernel, solid blue lines)
and $\sigma$ (for the Gaussian kernel, dashed black lines).
Left-hand plots are for $\gamma=0.5$, right-hand plots are for
$\gamma=-1.5$. The remaining parameter, $b$, is determined
by the condition $q_2=0$, i.e. $b=b_0$ where $b_0$ is defined
in~(\ref{eq:b0nonlocal}).}
\end{figure}

Figure~\ref{fig:coeffs} shows the variation of the coefficients $C_8$ and
$C_{10}$ as functions of the kernel width for both the top hat (blue) and
Gaussian (black) cases. As anticipated, in the limit
of very small and very large width parameters these are indistinguishable,
which is consistent with the asymptotic solution of the $O\left(\varepsilon^{3}\right)$
nonlocal problem.

A final quantity of interest for the Ginzburg--Landau equation~(\ref{eq:nlgl_e5_001})
is the normal form coefficient $q_4$ that determines the dynamics near the codimension-two
point at which $\mu=q_2=0$. For a fuller discussion of the meaning of
the coefficient $q_4$ we refer the reader to the paper by
Woods and Champneys \cite{WC99}; brief discussion is contained also
in the paper by Burke \& Dawes \cite{BD12}. It is sufficient here to remark that homoclinic
snaking, and the consequent existence of localised states over an open interval in $\mu<0$ is typical in the case $q_2<0$ and $q_4>0$, whereas in the case
$q_4<0$ we anticipate the existence of a single localised state, spatially decaying
to zero at large $|x|$, that exists for $\mu<0$ and near zero, for $q_2$ of either sign.
As discussed in \cite{BD12}, $q_4$ can be computed as a combination of
coefficients in the Ginzburg--Landau equation~(\ref{eq:nlgl_e5_001}). For the problem
at hand we find
\ba
q_4 = -\frac{3}{256}C_8^2 -\frac{1}{4}C_{10}. \label{eq:q4}
\ea
This expression for $q_4$ is relevant in the normal form, strictly speaking, only when $q_2=0$; this
places an additional constraint on~(\ref{eq:q4}) so that, for the top hat
and Gaussian kernels discussed above, $b_0$, $\gamma$ and the
kernel width parameter ($\delta$ or $\sigma$, respectively) are not independent:
(\ref{eq:b0nonlocal}) must also be satisfied.

\begin{figure}
\bc
\includegraphics[width=7.4cm]{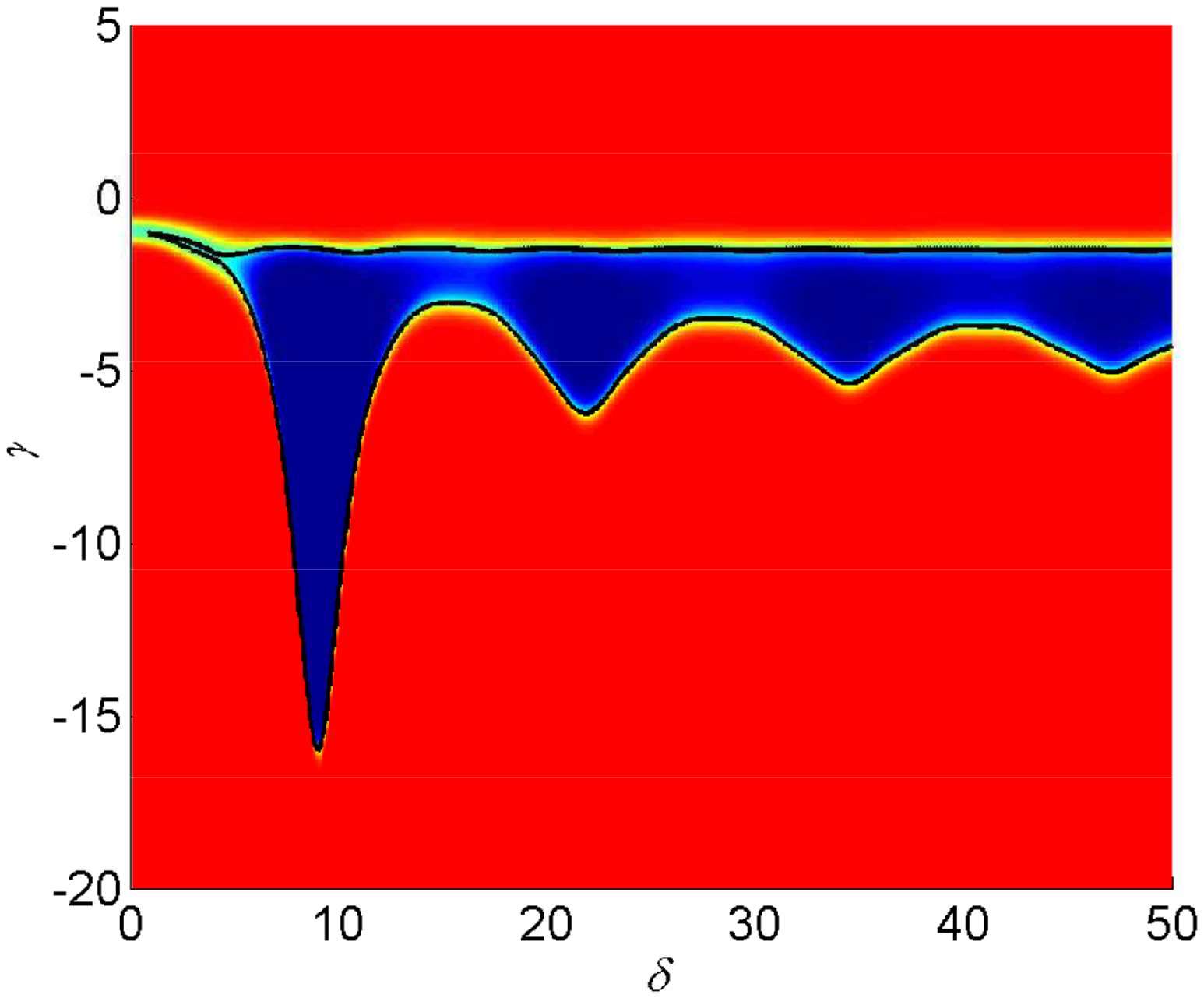}\includegraphics[width=7.4cm]{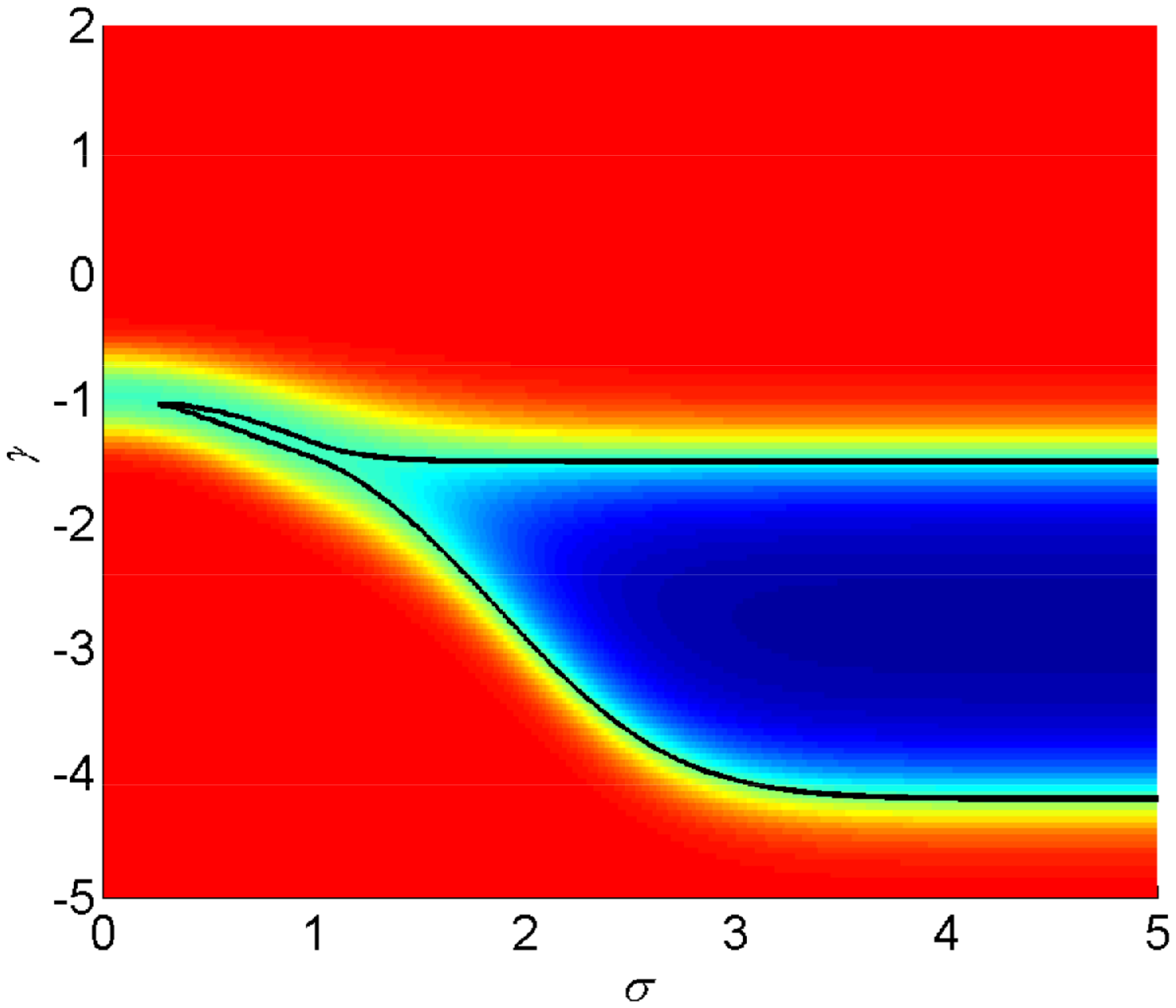}
\caption{\small Dependence of $q_4$ on $\gamma$ and the kernel width. Light red
regions correspond to $q_4>0$ and dark blue regions correspond to $q_4<0$. The
solid black line indicates the set $q_4=0$. Left: top hat kernel $K_{TH}(x)$.
Right: Gaussian kernel $K_G(x)$.
}
\label{fig:q4}
\ec
\end{figure}

The quadratic dependence of $q_4$ on $\gamma$ means that for $\gamma$ sufficiently
large, of either sign, $q_4$ is positive. Figure~\ref{fig:q4} illustrates the dependence
of $q_4$ on $\gamma$ for the top hat and Gaussian kernels, as functions of $\gamma$
and the width parameters $\delta$ and $\sigma$. In these plots, for a given point in
the parameter plane $b_0$ is chosen in order to fix $q_2=0$. Since both kernels
converge weakly to a delta function as the width parameters $\delta$ and $\sigma$
tend to zero, it is perhaps not surprising that the plots agree very closely
for small $\sigma$ and $\delta$.

\section{Longer range nonlocal interactions}
\label{sec:longrange}

In the previous section we computed quantitatively
how a nonlocal term with a short range kernel shifted the
codimension two point $q_2=0$ and the location of the snaking curves,
by changing the coefficients of the Ginburg--Landau equations that can be
derived at $O(\epsilon^3)$ or $O(\epsilon^5)$ that
provide leading-order descriptions of the weakly nonlinear dynamics.
In this section we make brief remarks about the existence of two
other distinguished limits in which
the width of the kernel can be rescaled so as to enable the derivation of
Ginzburg--Landau equations, at $O(\epsilon^5)$, that capture different
details of the effect of the nonlocal term. These distinguished limits
also connect the results of the previous section to the slanted snaking since
the new terms in the Ginzburg--Landau equations allow the branch of 
modulated states to deform in novel ways. For simplicity and brevity
we present results for the \Nqc Swift--Hohenberg equation and the Gaussian kernel.

In general, in moving away from the narrow kernel limit that we have implicitly
worked within so far, we anticipate deriving Ginzburg--Landau equations that contain
either higher-order derivatives, or integral terms, in order to capture the
nonlocal effects on the long length scale $X$. Since, as we have commented
above, it rapidly
becomes difficult to take more than the leading order contribution of the nonlocal term into account, we focus on the leading order term $\gamma u_1 J_{11}$. Using
the series expansion~(\ref{eq:FI_expand}) we have the expansion
\ba
\gamma u_1 J_{11} & = & 
-\epsilon \gamma \left(A\e^{ix}+\bar{A}\e^{-ix}\right)
\left[2\sum_{n=0}^{\infty}\frac{\left(\varepsilon^2 D_X\right)^n}{n!}I_{0n}|A|^2\left(X\right) \right. \nn \\
& & \left. +\e^{2ix}\sum_{n=0}^\infty \frac{\left(\varepsilon^2 D_X\right)^n}{n!}I_{2n}A^{2}(X)
+\e^{-2ix}\sum_{n=0}^{\infty}\frac{\left(\varepsilon^{2} D_X \right)^{n}}{n!}I_{\left(-2\right)n}\bar{A}^2(X)\right].
\label{eq:u1J11_002}
\ea
For the Gaussian kernel, for which $\hat{K}(k)=\exp(-\sigma^2 k^2/2)$, table~\ref{table:I_mn} lists the first few coefficients $I_{mn}$ for reference.
It is easily seen that in the Gaussian case, the coefficients $I_{mn}$ are
closely related to the Hermite polynomials $He_n(x)$. Precisely, we have
\ba
I_{mn} = (i\sigma)^n \ \e^{-(\sigma m)^2/2} \ He_n(\sigma m), \nn
\ea
where $He_n(x)$ is the Hermite polynomial of degree $n$, defined by the expression
\ba
He_n(x):= \e^{x^2/2} \ (-1)^n \ \left( \frac{d}{dx}\right)^n \e^{-x^2/2}, \nn
\ea
following the notation of Abramowitz \& Stegun \cite{AS64}, chapter 22.
\begin{table}
\bc
\begin{tabular}{lllll}
$m$       &  $n=0$              &  $n=1$ & $n=2$      &  $n=3$       \\ \hline 
$0$       &  $1$                &  $0$   & $\sigma^2$ &  $0$         \\ \hline
$m=\pm 1$ &  $\e^{-\sigma^2/2}$  &  $\pm i \sigma^2 \e^{-\sigma^2/2}$ &
$\sigma^2 (1-\sigma^2) \e^{-\sigma^2/2}$ & $\pm i \sigma^4 (3-\sigma^2) \e^{-\sigma^2/2}$ \\ \hline 
$m=\pm 2$ &  $\e^{-2\sigma^2}$  &  $\pm 2i\sigma^2\e^{-2\sigma^2}$  &
$\sigma^2(1-4\sigma^2) \e^{-2\sigma^2}$ & $\pm i \sigma^4(6-8\sigma^2)\e^{-2\sigma^2}$
\\ \hline 
$m=\pm 3$ &  $\e^{-9\sigma^2/2}$ & $\pm 3i\sigma^2\e^{-9\sigma^2/2}$ &
$\sigma^2(1-9\sigma^2) \e^{-9\sigma^2/2}$ & $\pm i \sigma^4(9-27\sigma^2)\e^{-9\sigma^2/2}$ \\ \hline 
$m=\pm 4$ &  $\e^{-8\sigma^2}$   & $\pm 4i\sigma^2 \e^{-8\sigma^2}$ &
$\sigma^2(1-16\sigma^2)\e^{-8\sigma^2}$ & $\pm i\sigma^4(12-64\sigma^2)\e^{-8\sigma^2}$ \\ \hline 
\end{tabular}
\caption{\label{table:I_mn}Coefficients $I_{mn}$ for the Gaussian kernel $K_G(x)$, for $|m|\leq 4$ and $0\leq n \leq 3$. $\pm$ signs in columns 1, 3 and 5 correspond to each other. 
}
\ec

\end{table}
To look for new distinguished limits we introduce the rescaled
kernel width parameter $\Sigma$, defined as
\ba
\Sigma = \varepsilon^\beta \sigma \label{eq:Sigma}
\ea
where $\beta>0$ is as yet undetermined. Since when $m>0$ we have that
$I_{mn} \propto \exp \left(-(m \Sigma)^2 \varepsilon^{2\beta} /2 \right)$,
terms with these coefficients are exponentially small
in the asymptotic limit $\varepsilon\rightarrow 0$ at fixed $\Sigma$
for any $\beta>0$, and so
the terms $I_{0n}$ might be expected to dominate the behaviour.
As a result, the
expressions for $I_{mn}$ can be simplified in the asymptotic limit, becoming
\ba
I_{mn} & \sim & \left\{ \begin{array}{cr}
\left(n-1\right)!! \ \Sigma^n \ \varepsilon^{-n\beta}
& \qquad m=0 \ \mathrm{and} \ n \ \mathrm{even},\\
O(\exp[-\Sigma^2 m^2 \epsilon^{-n\beta}/2]) & \mathrm{otherwise}.
\end{array}\right.
\label{eq:I_mn_002}
\ea
Substituting these limiting expressions
into~(\ref{eq:u1J11_002}) we see that the order of successive terms in
each of the summations is
lowered from $\varepsilon^{2n}$ to $\varepsilon^{\left(2-\beta\right)n}$.
The coefficients in the series still rapidly become small since
the factorial in the denominator grows more quickly than the
double factorial in the numerator of each term.

There are two choices of $\beta$ that allow the straightforward formation
of a solvability condition at $O\left(\varepsilon^5 \right)$. The first
of these is the choice $\beta=1$ which
leads to the amplitude equation
\ba
A_T & = & \mu A  +\frac{76}{9}b_0 b_2 A\left|A\right|^2 +C_{10}A\left|A\right|^4 +iC_{8}A_{X}|A|^2
\nn \\
& & -\gamma\Sigma^2\left(|A|^2 A_{XX} +2AA_X \bar{A}_X +A^{2}\bar{A}_{XX}\right)
+4A_{XX},
\label{eq:nlgl_beta1}
\ea
where the coefficients $b_0$, $C_8$ and $C_{10}$ are defined to be
\ba
b_0    & = & \sqrt{\frac{27}{38}+\frac{18\gamma}{38}},  \nn \\
C_8    & = & \frac{16}{19}+\frac{32\gamma}{57},         \nn \\
C_{10} & = & -\frac{8820}{361}-\frac{7998}{361}\gamma-\frac{67415}{17328}\gamma^2 . \nn
\ea
and $b_2$ is the departure from the codimension two point $b=b_0$, i.e. writing $b=b_0+\epsilon^2 b_2$ as in~(\ref{eq:b_scaling}).
The amplitude equation~(\ref{eq:nlgl_beta1}) is structurally distinct from~(\ref{eq:nlgl_e5}) which was obtained in the narrow kernel case due to
the presence of nonlinear cubic order
terms involving two derivatives. In some sense the appearance of these
new terms is balanced by a simplification of the dependence of the
coefficients on the kernel function:
in~(\ref{eq:nlgl_beta1}) the coefficients depend only $\gamma$
and not on the rescaled width $\Sigma$.

\begin{figure}
\bc
\includegraphics[width=7.0cm]{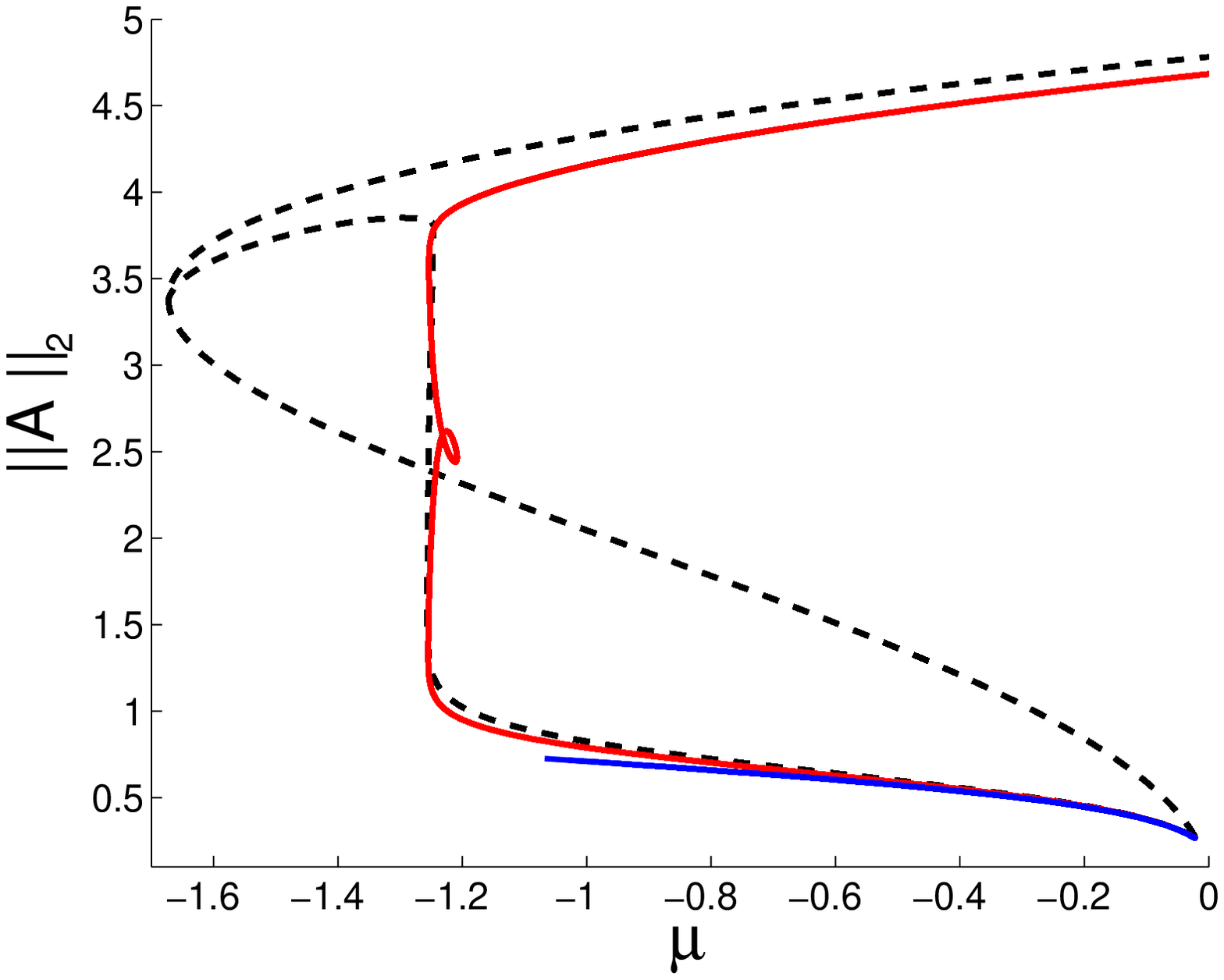}
\includegraphics[width=7.0cm]{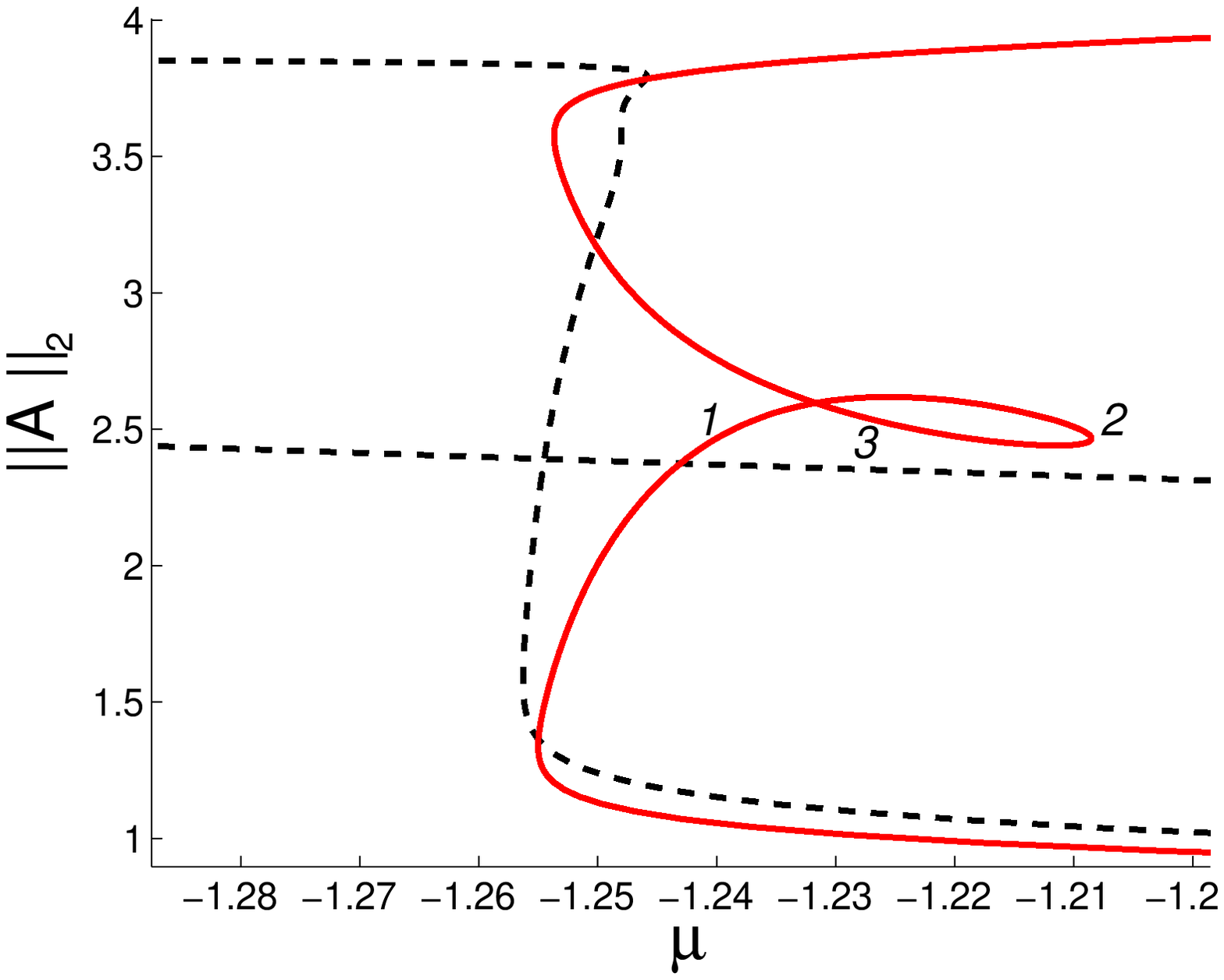}
\caption{\small Left: Bifurcation diagram for solutions of the nonlocal Ginzburg-Landau
equation~(\ref{eq:nlgl_beta1}) with a Gaussian kernel,
for different values of the width parameter:
$\Sigma= 1$ (black, dashed),
$\Sigma=2$ (red, solid) and $\Sigma=3$ (blue, solid, lowest curve).
Right: enlargement of left-hand figure showing the additional `loop'
that arises in the $\Sigma=2$ case. Labels 1, 2 and 3 correspond to
the profiles shown in figure~\ref{fig:gl_beta1_profiles}.
Other parameters are $\gamma=1.0$, $b_2=2.0$.
The domain size $L=|\Omega|=20\pi$.
}
\label{fig:gl_beta1}
\ec
\end{figure}

\begin{figure}
\bc
\includegraphics[width=7.0cm]{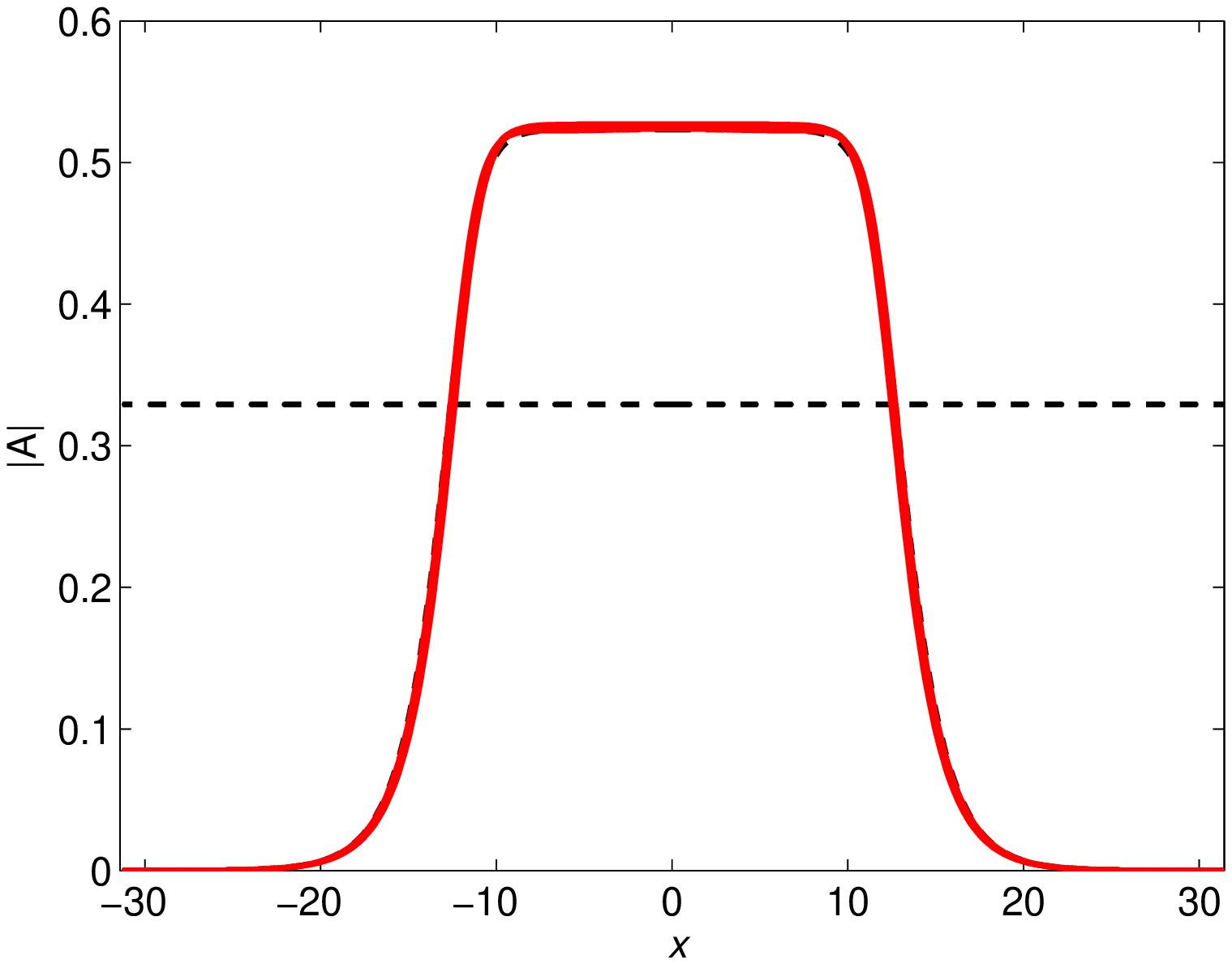}
\includegraphics[width=7.0cm]{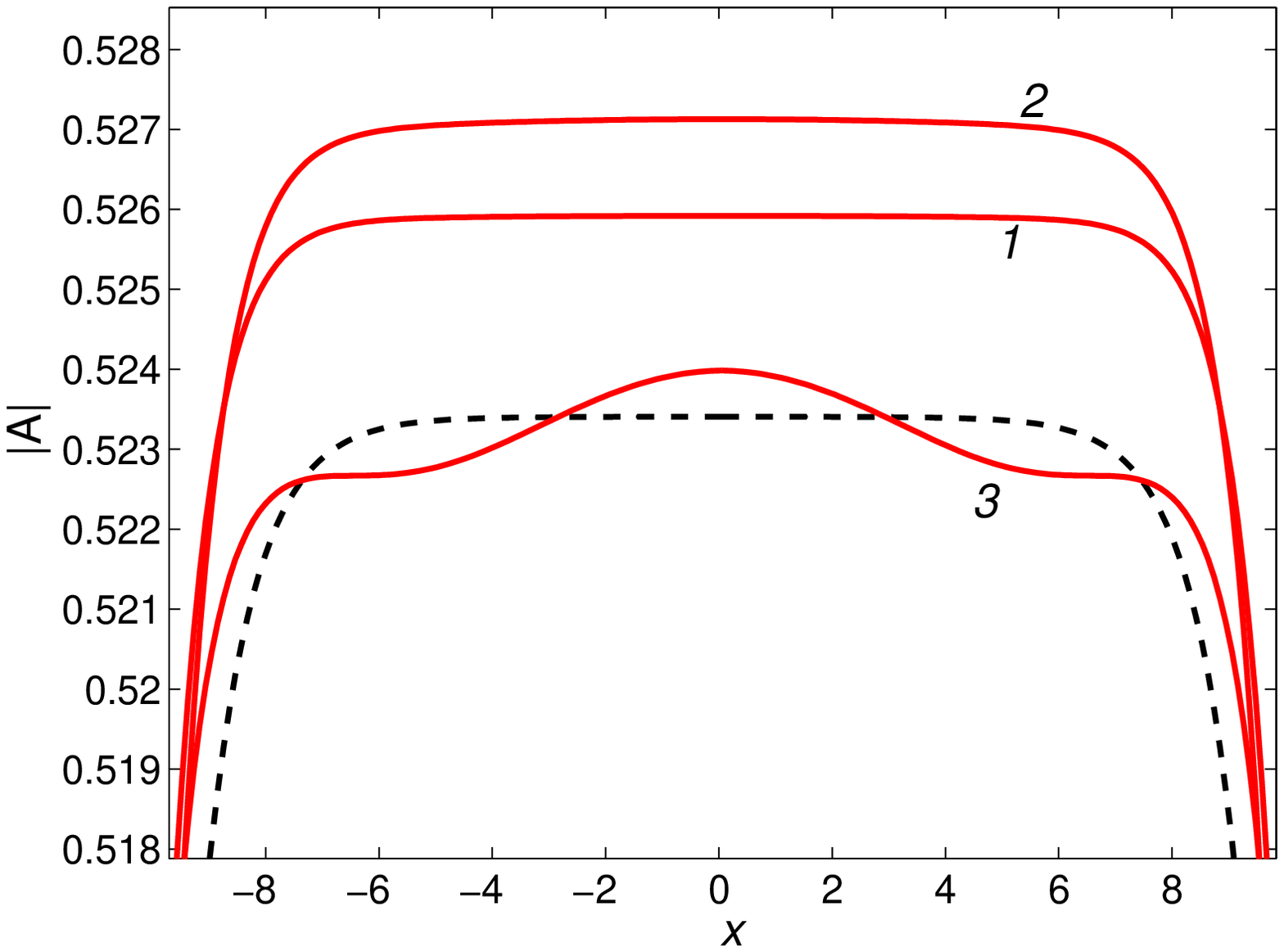}
\caption{\small Left: Solution profiles for the nonlocal Ginzburg-Landau
equation~(\ref{eq:nlgl_beta1}) with a Gaussian kernel,
for different values of the width parameter:
$\Sigma=1$ (black, dashed),
$\Sigma=2$ (red, solid), all at $\|A\|=2.5$. The horizontal
line is the uniform solution
Right: enlargement of left-hand figure showing the differences
between the three solutions for $\Sigma=2$ on the `loop'.
Continuing up the solid red curve in figure~\ref{fig:gl_beta1},
increasing $\|A\|$ the profiles at $\|A\|=2.5$ arise in
the order: middle (1), top (2), lower (3).
The black dashed curve is (as in the left-hand plot) for
$\Sigma=1$.
Other parameters are $\gamma=1.0$, $b_2=2.0$.
The domain size $L=|\Omega|=20\pi$.
}
\label{fig:gl_beta1_profiles}
\ec
\end{figure}

Figure~\ref{fig:gl_beta1} illustrates three typical bifurcation diagrams
obtained by continuation of solutions to~(\ref{eq:nlgl_beta1}). Solid and dashed lines do not indicate stability; they distinguish different parameter sets.
The dashed black curves correspond to $\Sigma=1$ and form the 
expected bifurcation diagram for the subcritical Ginzburg--Landau
equation: there is a branch of uniform (constant) solutions that
bifurcates subcritically and turns around at a fold bifurcation
close to $\mu=-1.7$. A branch of spatially non-constant solutions
bifurcates from the uniform branch near $\mu=0$ and rejoins it
near the saddle-node point. As the enlargement in
figure~\ref{fig:gl_beta1}(b) shows, in contrast to the case
$\Sigma=0$, this curve shows a small amount
of hysteresis close to the nearly-vertical section.
As $\Sigma$ is increased to $\Sigma=2$ (solid red curves) the
branch of spatially non-constant solutions develops an
intriguing additional loop in the centre of the 
`nearly-vertical' section before extending into positive $\mu$,
without reconnecting to the spatially uniform branch at
large amplitude. Solution profiles at $\|A\|=2.5$, a value that
cuts through the loop, are shown in figure~\ref{fig:gl_beta1_profiles}.
The profile becomes sharper, then develops additional undulations, as
the loop is traversed starting from small-amplitude solutions. At
larger amplitudes the solution resembles a sharp downwards-pointing
spike embedded in a uniform, but non-zero, background state.

\begin{figure}
\bc
\includegraphics[width=7.0cm]{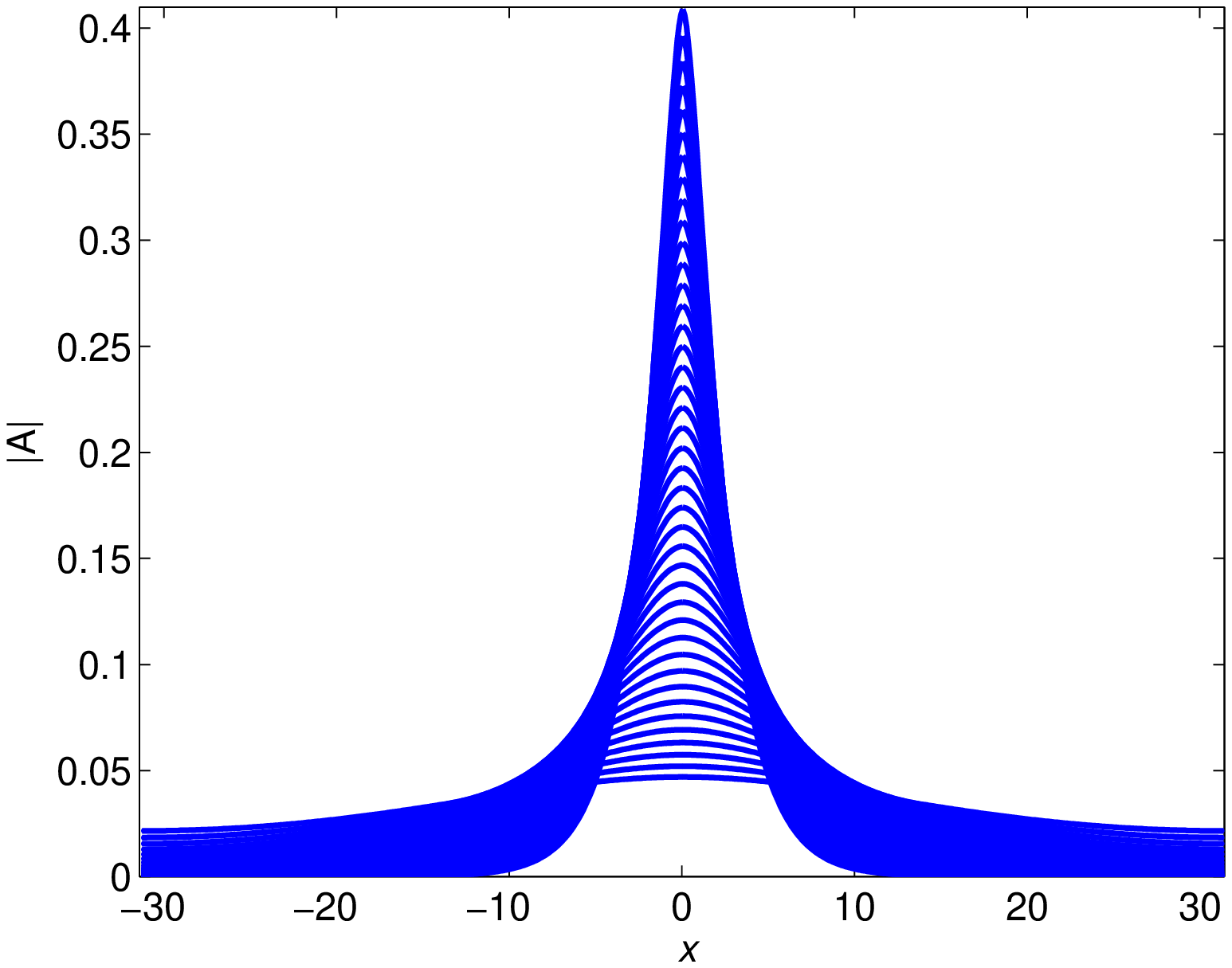}
\includegraphics[width=7.0cm]{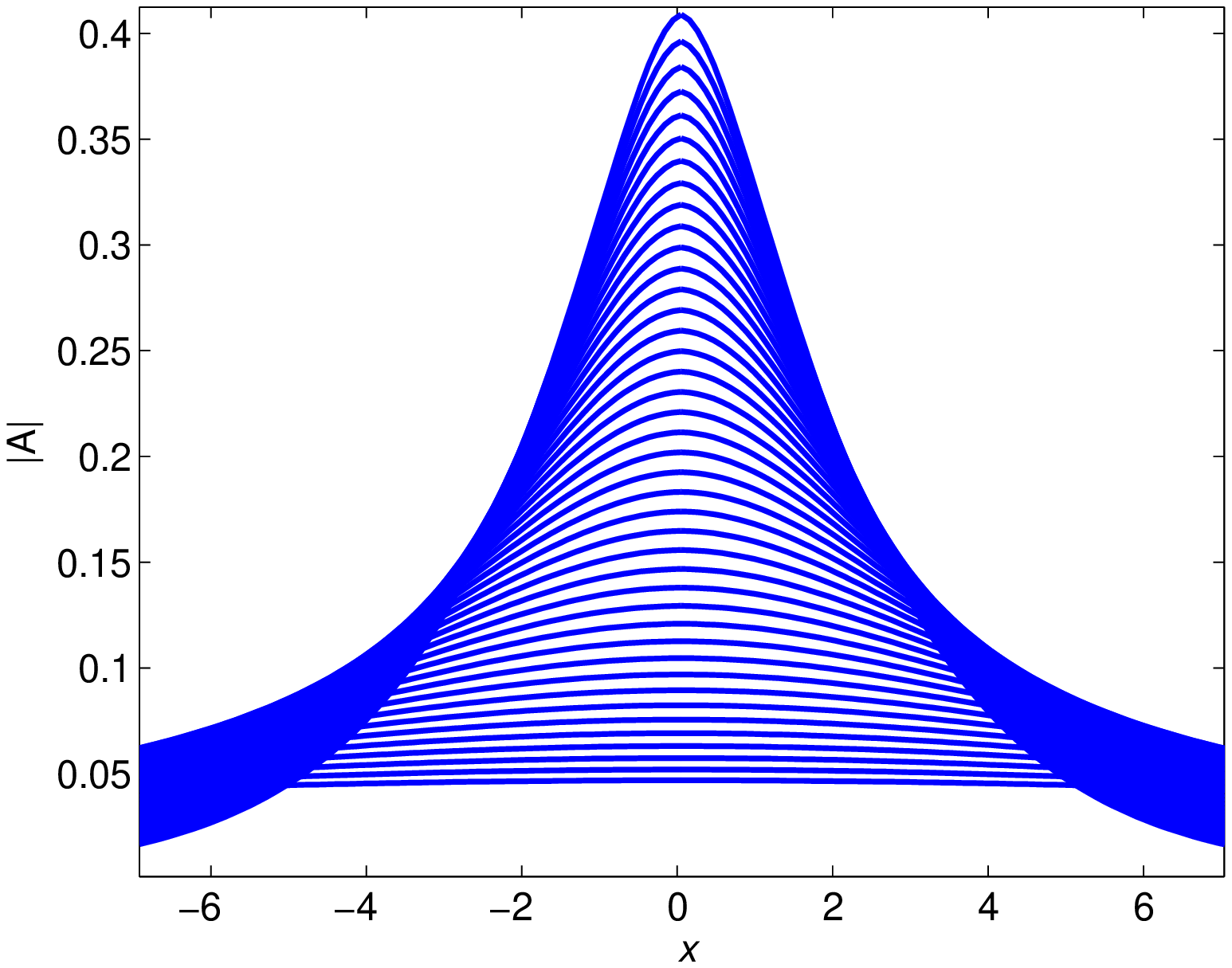}
\caption{\small Left: Solution profiles for the nonlocal Ginzburg-Landau
equation~(\ref{eq:nlgl_beta1}) with a Gaussian kernel
for $\Sigma=3$ along the lowest solution branch shown in
figure~\ref{fig:gl_beta1}.
Right: enlargement of left-hand figure showing the evolution
of the solutions near the point of blow-up.
Other parameters are $\gamma=1.0$, $b_2=2.0$.
The domain size $L=|\Omega|=20\pi$.
}
\label{fig:gl_beta1sigma3_profiles}
\ec
\end{figure}

In general, it is not {\it a priori} clear
that~(\ref{eq:nlgl_beta1}) remains well-posed 
as the amplitude of solutions $A(X)$ increases,
due to the additional terms and the likelihood of
singular behaviour for
$\gamma\Sigma^2 |A|^2 \approx 4$. This issue deserves
detailed investigation, and we will return to it in future work.
Its relevance to the formation of localised states is
indicated by the (numerical)
behaviour of solutions along the lowest (blue, solid)
curve in figure~\ref{fig:gl_beta1}. Solutions along this branch
are shown in figure~\ref{fig:gl_beta1sigma3_profiles}: solutions
become increasingly sharp as their amplitude increases until numerical
continuation fails to converge and the solution branch terminates. It
is possible that this blow-up arises in a self-similar
fashion.

The second natural distinguished limit that enables
formation of a solvability condition
at $O\left(\varepsilon^5\right)$ is the choice $\beta=2$. In order
to bring in the nonlocal terms at fifth order we combine this
with a rescaling of the coefficient $\gamma$, writing $\gamma = \epsilon^2 \gamma_2$.
In this distinguished limit the resulting envelope equation, again
choosing a Gaussian kernel, is
\ba
A_T & = & \mu A  +\frac{76}{9}b_0 b_2A|A|^2
-\frac{8820}{361}A|A|^4 +i\frac{16}{19}A_X |A|^2 \nn \\
& & -2\gamma_2 A\sum_{n=1}^\infty \frac{\left(2n-1\right)!!}{\left(2n\right)!} \left(\Sigma D_X\right)^{2n}|A|^2
+4A_{XX},
\label{eq:nlgl_beta2}
\ea
where again $b_0=\sqrt{27/38}$.
The choice $\beta=2$ effectively makes the kernel act entirely on the long
length scale $X$ and so every term in the series expansion~(\ref{eq:u1J11_002})
at the same
order in $\varepsilon$. Formally, the series expansion is equivalent to the integral
term, so that a more compact way of writing the amplitude equation is
\ba
A_{T} &= & \mu A  +\frac{76}{9}b_0 b_2A|A|^2
-\frac{8820}{361}A|A|^4 +i\frac{16}{19}A_X |A|^2 \nn \\
& & -2\gamma_2 \int_\Omega \e^{-(X-Y)^2/(2\Sigma^2)} |A(Y)|^2 \ dY
+4A_{XX} .
\label{eq:nlgl_beta2_002}
\ea
This nonlocal Ginzburg--Landau equation captures the nonlocal
dynamics sufficiently strongly to reproduce aspects of the
stretched and slanted snaking behaviour apparent in figures~\ref{fig:extreme} and~\ref{fig:nonlocal001}.
\begin{figure}
\bc
\includegraphics[width=7.4cm]{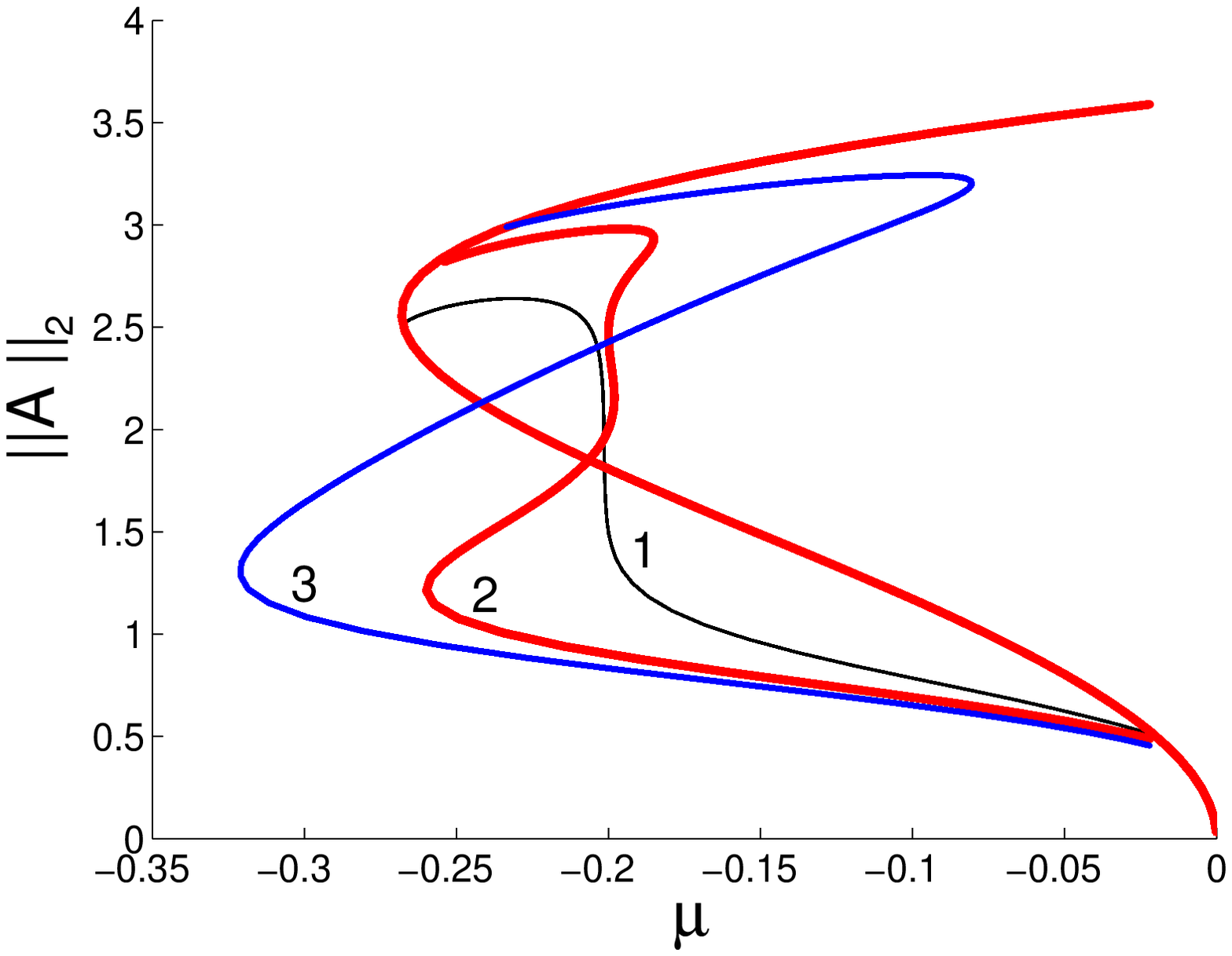}
\includegraphics[width=7.4cm]{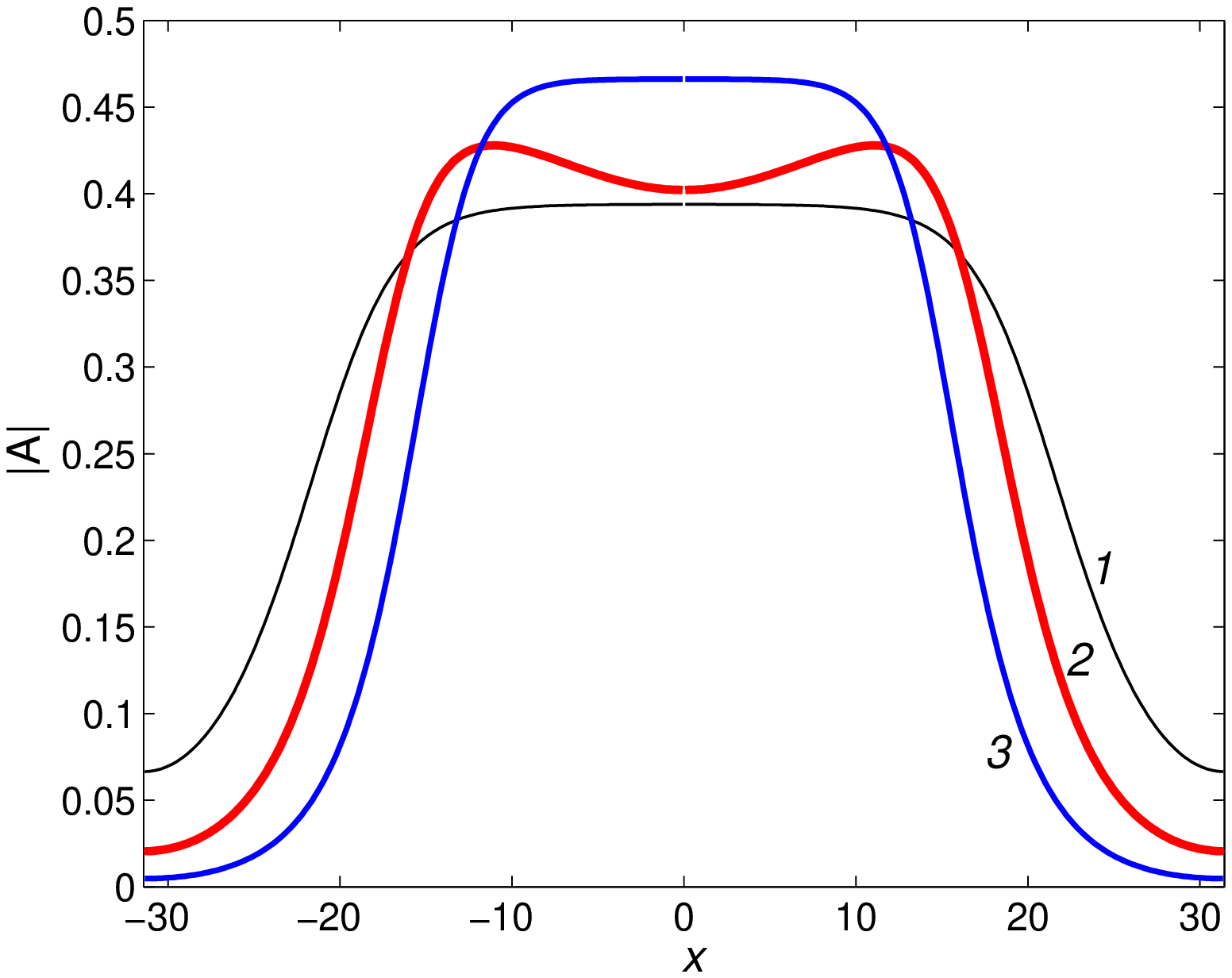}
\caption{\small Left: Bifurcation diagram for solutions of the nonlocal Ginzburg-Landau
equation~(\ref{eq:nlgl_beta2_002}) with a Gaussian kernel,
for different values of the width parameter: $\Sigma= 0$ (black, labelled 1),
$\Sigma=3\pi$ (red, labelled 2) and $\Sigma=10\pi$ (blue, labelled 3).
Right: solution
profiles for each curve of spatially modulated solutions at $\| A\|_2=2.5$, where
the curves in (a) are close to intersecting.
Note the central dip in the thick red curve (labelled 2).
Other parameters are $\gamma_2=1.0$, $b_2=1.0$. The domain size $L=|\Omega|=20\pi$.
}
\label{fig:int_gl}
\ec
\end{figure}
Figure~\ref{fig:int_gl} contains bifurcation diagrams and solution profiles
for three choices
of $\Sigma$, moving from the purely local problem ($\Sigma=0$, thin black curve)
to nearly the global one ($\Sigma=L/2$, medium thickness blue curve). For each
value of $\Sigma$ in figure~\ref{fig:int_gl}(a), the curves of uniform
constant solutions for $|A|$ are almost exactly superposed: of these
only the thickest (red) curve is visible. The thin black line for the $\Sigma=0$
case has a near-vertical central section, just as in figure~\ref{fig:gl5-23}. The
thick red curve indicates how this vertical curve is deformed as the width of
the nonlocal kernel is increased: the lower part moves to the left and the upper
part to the right. This is exactly the behaviour observed for the homoclinic
snaking in figures~\ref{fig:nonlocal001}. The central section remains
very close to the central section in the case $\Sigma=0$. As the kernel
width increases further we approach, smoothly, the slanted snaking limit in which
the curve assumes its characteristic `Z' shape. This transition takes place
monotonically and smoothly, in contrast to the many disconnection and
reconnection events associated with the homoclinic snaking curves themselves,
and illustrated, for the top hat kernel, in figures~\ref{fig:th_snake001} and~\ref{fig:th_snake002}. Figure~\ref{fig:int_gl}(b) shows solution profiles
$|A(X)|$ on the three bifurcation curves in~\ref{fig:int_gl}(a) at points close
to where all three curves of spatially modulated solutions intersect. We see
that the thickest (red) curve for intermediate kernel widths contains a central
trough and off-centre peaks that are not present in the other two cases. This
indicates the propensity of localised states for these intermediate kernel
widths (more precisely, kernel widths that are longer than the basic
wavelength of the pattern forming instability yet shorter than the width of the domain)
to prefer to form multipulse states rather than single-pulse states.
This therefore goes a little way towards justifying the collisions between
the primary snaking curves and those of multipulse states described
in figures~\ref{fig:th_snake001} and \ref{fig:th_snake002}.

\section{Conclusions}
\label{sec:discussion}

In this paper we have considered equilibrium solutions, in the form of spatially
localised states, to the one dimensional Swift--Hohenberg equation extended by a nonlocal term in convolution form.
Numerical results suggested that the well known `homoclinic snaking'
structure was deformed and perturbed in several different ways as a result
of the presence of nonlocality. We focussed our analytical efforts on
extending the multiple-scales analysis for the Swift--Hohenberg
equation to cope with the nonlocal term. Through (formal) expansions
and application of Fourier transforms we were able to
reduce the nonlocal integrodifferential Swift--Hohenberg equation into
one of three Ginzburg--Landau-type equations that applied in different
distinguished limits. These different limiting equations captured different
aspects of the problem. As a direction for future work it would be of
interest to link the results of Kao and Knobloch \cite{KK12}, which concern the
behaviour of solutions to the cubic-quintic Ginzburg--Landau equation,
to the parameter dependencies of the coefficients that the nonlocal
terms produce.

While we focussed on two particular kernels, the `top hat' composition of
Heaviside step functions, and a Gaussian kernel, we present analytic results 
in as much generality as possible, writing the coefficients in
the Ginzburg--Landau equations in terms of the Fourier transform of
the kernel function. The Gaussian case can be motivated straightforwardly in
applications since the convolution term is the solution of the initial value
problem for the diffusion
equation. The nonlocal term could then be thought of as directly
modelling the diffusive
spread of the solution, through interaction with another physical quantity, such as
temperature, over timescales short compared to the time evolution
of the Swift--Hohenberg equation itself.

On the mathematical side, it would be useful and interesting
to extend results obtained
recently by Achleitner and Kuehn \cite{Kuehn13} on the persistence of
solutions in the presence of convolution terms, to prove that for
kernels that are sufficiently narrow the homoclinic snaking bifurcation structure
persists. Numerical work to investigate exactly the sequence of collisions
with multipulse states for a specific choice of kernel would also be of interest.

Moreover, it would be desirable to extend
the analysis of this one dimensional problem to two or three space
dimensions, clearly of substantial physical relevance.
There is plenty of room for optimism since Fourier transform results
often extend easily to higher dimensions, so that such extensions may be achievable
with largely similar techniques to those presented here.
These and many other considerations in this complex problem are left to
be the subject of future work.

\section*{Acknowledgements}
JHPD gratefully acknowledges the support of the Royal Society through
a University Research Fellowship.



\end{document}